\documentclass[letterpaper,onecolumn,accepted=2026-01-11]{quantumarticle}
\pdfoutput=1

\usepackage{amsfonts, amsmath, amssymb, mathtools}
\usepackage[dvipsnames]{xcolor}
\usepackage{qcircuit}
\usepackage{tikz}
\usepackage[numbers]{natbib}
\usepackage[utf8]{inputenc}
\usepackage[english]{babel}
\usepackage[T1]{fontenc}
\usepackage{hyperref}

\newcommand{\maroon}[1]{{\color{Maroon}#1}}
\newcommand{\blue}[1]{{\color{RoyalBlue}#1}}
\newcommand{\green}[1]{{\color{Green}#1}}
\newcommand{\gray}[1]{{\color{gray}#1}}

\DeclareMathOperator{\tr}{tr}
\DeclareMathOperator{\diag}{diag}

\DeclareMathOperator{\sgn}{sgn}
\DeclareMathOperator{\Var}{Var}
\newcommand{\ket}[1]{{| {#1}  \rangle}}
\newcommand{\bra}[1]{{\langle {#1} |}}

\begin{document}

\title{Theory of quantum error mitigation for non-Clifford gates}

\author{David Layden}
\affiliation{IBM Research}

\author{Bradley Mitchell}
\affiliation{IBM Research}

\author{Karthik Siva}
\affiliation{IBM Research}

\begin{abstract}
Quantum error mitigation techniques mimic noiseless quantum circuits by running several related noisy circuits and combining their outputs in particular ways. How well such techniques work is thought to depend strongly on how noisy the underlying gates are. Weakly-entangling gates, like $R_{ZZ}(\theta)$ for small angles $\theta$, can be much less noisy than entangling Clifford gates, like CNOT and CZ, and they arise naturally in circuits used to simulate quantum dynamics. However, such weakly-entangling gates are non-Clifford, and are therefore incompatible with two of the most prominent error mitigation techniques to date: probabilistic error cancellation (PEC) and the related form of zero-noise extrapolation (ZNE). This paper generalizes these techniques to non-Clifford gates, and comprises two complementary parts. The first part shows how to effectively transform any given quantum channel into (almost) any desired channel, at the cost of a sampling overhead, by adding random Pauli gates and processing the measurement outcomes. This enables us to cancel or properly amplify noise in non-Clifford gates, provided we can first characterize such gates in detail. The second part therefore introduces techniques to do so for noisy $R_{ZZ}(\theta)$ gates. These techniques are robust to state preparation and measurement (SPAM) errors, and exhibit concentration and sensitivity---crucial statistical properties for many experiments. They are related to randomized benchmarking, and may also be of interest beyond the context of error mitigation. We find that while non-Clifford gates can be less noisy than related Cliffords, their noise is fundamentally more complex, which can lead to surprising and sometimes unwanted effects in error mitigation. Whether this trade-off can be broadly advantageous remains to be seen.
\end{abstract}

\maketitle

\tableofcontents

\section{Introduction}

While fault tolerance is essential for realizing the full potential of quantum computing, error mitigation may unlock some of this potential before the advent of large-scale fault tolerance \cite{cai:2022}. The most prominent error mitigation techniques seek to compute noiseless expectation values for a given quantum circuit by running several related circuits on a noisy quantum computer, then combining their measurement outcomes in nontrivial ways \cite{temme:2017, li:2017}. The resulting precision and/or accuracy (within a fixed running time) typically improves with increasing gate quality, but declines---often exponentially---with the size of the target circuit. In other words, better gates can enable these techniques on bigger circuits---a dynamic that may provide a continuous path towards fault tolerance \cite{bravyi:2022}.

Due to its strong dependence on circuit size, error mitigation is most promising for problems which admit an exponential quantum speedup in computing expectation values, and where the required gates closely match the connectivity between qubits in hardware. The most evident example is quantum simulation using Trotter/Floquet-type circuits \cite{lloyd:1996, childs:2018}. Such circuits implement repeated unitaries of the form $U_j = \exp(-i H_j \, \delta t)$ using one- and two-qubit gates, where $\{H_j\}$ are components of the Hamiltonian being simulated and $\delta t$ represents a timestep. One typically wants a small $\delta t$ to reduce Trotter error, which makes each $U_j$ (at most) weakly entangling \cite{childs:2021}. For example, alternating between $U_1$ and $U_2$ layers, generated by $H_1 = g \sum_i X_i$ and $H_2 = -J \sum_{\langle i, j \rangle} Z_i Z_j$ respectively, approximates evolution by the transverse-field Ising model $H_1+H_2$. Using the notation 
\begin{equation}
R_P(\theta)=e^{-i P \theta/2}
\end{equation}
where $P$ denotes a Pauli operator, $U_1$ comprises only single-qubit unitaries $R_X(\phi)$ with $\phi = 2 g \, \delta t$, while $U_2$ comprises two-qubit unitaries $R_{ZZ}(\theta)$ with $\theta= -2J \, \delta t$, which generate little entanglement per layer when $\delta t$ is small enough to give a reasonable Trotter error.

Such weakly-entangling unitaries can be realized through two main strategies, which use qualitatively different two-qubit gates \cite{clinton:2021}. The first uses fixed two-qubit Clifford gates, like CNOTs, regardless of $\delta t$, while the second uses weakly-entangling, non-Clifford two-qubit gates that approach $I$ as $\delta t \rightarrow 0$. (Both strategies can also use arbitrary single-qubit gates as needed.) Following the first strategy, for example, one might compile $R_{ZZ}(\theta)$ into two CNOTs and a single-qubit $R_Z(\theta)$ gate, where each CNOT is locally equivalent to $R_{ZZ}(\pi/2)$. Following the second strategy, one would instead implement $R_{ZZ}(\theta)$ directly, up to single-qubit gates, by shortening the control sequence used to perform CNOTs---effectively doing a fraction of a CNOT rather than two. We will call these two strategies \textit{digital} and \textit{semi-analog}, respectively. (The latter is not fully analog as it still discretizes the quantum dynamics of interest into circuits.) Since two-qubit gates are the dominant source of errors in most pre-fault-tolerant devices, these two strategies offer very different advantages. The digital strategy is manifestly compatible with the most prominent error mitigation techniques, which handle errors on two-qubit Clifford gates \cite{vandenberg:2023, kim:2023}. On the other hand, the semi-analog strategy can incur substantially fewer errors to begin with, by virtue of having faster, and sometimes also fewer, two-qubit gates \cite{earnest:2021, stenger:2021}. However, because these latter gates are non-Clifford, they have been largely incompatible with said error mitigation techniques to date.

\subsection{Contributions}

Motivated by this semi-analog strategy for quantum simulation, we introduce a broad error mitigation technique for non-Clifford gates. It comprises two interlocking, yet distinct parts: in Section~\ref{sec:mitigation} we develop a method for mitigating noise on arbitrary gates, given a detailed description of the noise. Then in Section~\ref{sec:learning}, we introduce a robust method to characterize said noise on common non-Clifford gates. Together, these two parts greatly generalize the most prominent error mitigation techniques, thereby making them compatible with the relatively low-noise, non-Clifford gates that arise naturally in many experiments.

More precisely, Section~\ref{sec:mitigation} extends probabilistic error cancellation (PEC) and zero-noise extrapolation (ZNE), two of the most successful error mitigation techniques to date, to generic noise on arbitrary gates. In their current form, these techniques can effectively transform one Pauli channel into another by inserting  single-qubit gates at random from some appropriate distribution. This suffices to mitigate multi-qubit Clifford gates because generic noise on such gates can always be transformed into a Pauli channel through Pauli-twirling, using single-qubit gates. The main obstacle to generalizing PEC and ZNE is that noise on non-Clifford gates cannot be twirled in this way, so it cannot generally be transformed into a Pauli channel. Our key insight here is to reframe the task: we effectively construct a ``super-channel'' (a linear map acting on quantum channels) that transforms a noisy non-Clifford gate into a desired channel---usually the ideal gate. We then show that this super-channel can be effectively realized by inserting random single-qubit gates into the circuit, as in Clifford PEC/ZNE, just with a different distribution. As usual, this distribution depends on the precise gate noise, which we assume is already known. Our technique applies to arbitrary gates, with only mild regularity requirements on their noise, and reduces to existing methods when applied to Cliffords. However, it behaves unexpectedly on some non-Clifford gates, in ways that have no clear analog in the Clifford case. Most notably, the sampling overhead required to effectively cancel noise on a non-Clifford gate can be discontinuous in the noise strength, and can remain large for arbitrarily weak noise, depending on subtle properties of the noise. We do not attempt to fully characterize the gates and noise structures where this effect occurs, but instead give three illustrative examples (two analytical and one numerical) with $R_{ZZ}(\theta)$ gates to inspire further study of this new effect.

In Section~\ref{sec:learning} we turn to the complementary problem: how to thoroughly characterize noise on non-Clifford gates to begin with. The first major hurdle is that the measurement outcomes in any experiment depend not only on gate noise, but also on state preparation and measurement (SPAM) errors, which can be significantly stronger and therefore make gate noise hard to isolate. This rules out quantum process tomography, which is not generally robust to SPAM errors. A potential alternative is gate set tomography, which learns gate and SPAM errors simultaneously and self-consistently \cite{merkel:2013, blume:2013, nielsen:2021, endo:2018}. While very general, it can be correspondingly expensive in terms of the experiments and computation required. Cycle benchmarking (CB) is therefore a more widely adopted technique, which avoids learning SPAM errors by channeling their effects into a single parameter that can ultimately be discarded. Our main contribution here is generalizing CB to non-Clifford gates. In doing so, one encounters the second major hurdle: standard CB is very robust to statistical fluctuations from running a finite number of random circuits finitely many times---a property that does not automatically carry over to the non-Clifford case. We show that our generalization is similarly robust by construction, while illustrating the pitfalls of more obvious approaches. How to ensure statistical robustness in the non-Clifford case depends significantly on the gate in question, since there is a wider variety of non-Clifford gates than of Cliffords. So for the sake of concreteness, we focus in this section on characterizing noisy $R_{ZZ}(\theta)$ gates, which are common across many experiments (unlike in Section~\ref{sec:mitigation}, which is completely general). Our technique learns all the noise parameters needed to effectively cancel generic noise using our method from Section~\ref{sec:mitigation}. We illustrate this numerically under strong SPAM errors and large statistical fluctuations. Finally, we find that, while not strictly necessary, learning certain noise parameters through CB-like methods seems fundamentally hard. Curiously, these are exactly the parameters that can cause unusual behavior in the sampling overhead in Section~\ref{sec:mitigation}, which suggests a novel connection between learnability and mitigation. 

We make several common mathematical assumptions about noise in this work. Specifically, we assume throughout that all noise is constant in time, or rather, that it drifts slowly enough to be treated as constant. We also assume that the noise on multi-qubit gates is Markovian, meaning that each noisy gate can be described by a fixed quantum channel regardless of when it occurs in a circuit. Moreover, we approximate single-qubit gates as being noiseless compared to multi-qubit gates, to reflect the fact that the latter are typically much noisier on pre-fault-tolerant devices. This is not strictly necessary, but it simplifies the presentation and keeps the focus on the main topics at hand. Finally, we assume in Section~\ref{sec:learning} that the gate noise is not so strong as to totally overwhelm the $R_{ZZ}(\theta)$ gate, as described by Eq.~\eqref{eq:eig_assumption}. From a theoretical perspective, we refer to this condition as the ``weak noise regime,'' although it encompasses noise that is very strong by experimental standards (e.g., 95\% average fidelity, as in our numerics), and is therefore a mild assumption in practice.

\section{Error Mitigation}
\label{sec:mitigation}

We begin this section by reviewing the essential mathematical background in Sec.~\ref{sec:math_background}, and the existing formalism of PEC and ZNE for Clifford gates in Sec.~\ref{sec:mitigation_clifford}, before finally introducing our more general mitigation scheme in Sec.~\ref{sec:mitigation_nonclifford}.

\subsection{Mathematical Background}
\label{sec:math_background}

A generic operation on $n$ qubits, unitary or not, can be described by a completely positive trace-preserving (CPTP\footnote{Many of our results apply more broadly to completely positive trace-non-increasing maps (which can describe leakage), but we will focus on CPTP maps here to simplify the presentation.}) map $\mathcal{G}$, also called a quantum channel, which maps an input state $\rho$ to an output state $\rho' = \mathcal{G}(\rho)$. There are several distinct matrix representations for a given quantum channel, but it will be convenient here to describe $\mathcal{G}$ by its Pauli transfer matrix (PTM) $\boldsymbol{G}$, a $4^n \times 4^n$ real matrix with elements
\begin{equation}
\boldsymbol{G}_{ij} = \tr \big[ P_i \, \mathcal{G}(P_j) \big]/2^n
\label{eq:G_ij_def}
\end{equation}
between $-1$ and $1$, where $P_i$ and $P_j$ are $n$-qubit Paulis \cite{greenbaum:2015}. ($\boldsymbol{G}$ is also known as a Liouville representation of $\mathcal{G}$.) We will denote PTMs and other matrices of the same size in bold to distinguish them from $2^n\times 2^n$-dimensional unitaries like $P_i$. Writing the input/output states above in the Pauli basis, $\rho= 2^{-n}\sum_i s_i P_i$ and $\rho' = 2^{-n} \sum_i s_i' P_i$, the PTM of $\mathcal{G}$ is the matrix relating their generalized Bloch vectors as $\vec{s} \, ' = \boldsymbol{G} \, \vec{s}$. For any two channels $\mathcal{G}^{(1)}$ and $\mathcal{G}^{(2)}$ with PTMs $\boldsymbol{G}^{(1)}$ and $\boldsymbol{G}^{(2)}$, the PTM of $\mathcal{G}^{(1)}$ followed by $\mathcal{G}^{(2)}$ is simply $\boldsymbol{G}^{(2)} \boldsymbol{G}^{(1)}$, and performing $\mathcal{G}^{(1)}$ or $\mathcal{G}^{(2)}$ with respective probabilities $p$ and $1-p$ gives a PTM of $p \, \boldsymbol{G}^{(1)} + (1-p) \boldsymbol{G}^{(2)}$.

A channel $\mathcal{P}$ is said to be a Pauli channel if it acts as $\mathcal{P}(\rho) = \sum_i p_i P_i \rho P_i$ for some probability distribution $\vec{p}=(p_i)_{i=0}^{4^n-1}$, meaning that it can be understood as a process where Pauli errors $P_i$ occur with probabilities $p_i$. Equivalently, a Pauli channel is one whose PTM is diagonal, $\boldsymbol{P}=\diag(\vec{f} \,)$, with elements $f_i$ that are sometimes called \textit{Pauli eigenvalues} \cite{flammia:2020} or \textit{Pauli fidelities} \cite{vandenberg:2023}. These are related to the error probabilities $p_i$ by $\vec{f} = \boldsymbol{W} \vec{p}$, where $\boldsymbol{W}$ is a $4^n \times 4^n$ Walsh-Hadamard matrix with elements
\begin{equation}
\label{eq:W}
\boldsymbol{W}_{ij} = 
\begin{cases}
+1, & [P_i, P_j] = 0 \\
-1, & \{ P_i, P_j \} = 0,
\end{cases}
\end{equation}
that describes a type of discrete Fourier transform and obeys $\boldsymbol{W} = \boldsymbol{W}^\top = 4^n \, \boldsymbol{W}^{-1}$.

Finally, a generic channel $\mathcal{G}$ can be transformed into a Pauli channel through Pauli-twirling \cite{dur:2005, dankert:2009}. Twirling means sampling a random unitary $V$ uniformly from some given set each time $\mathcal{G}$ is implemented, and applying $V$ and $V^\dag$ before and after $\mathcal{G}$, respectively. When $V$ is sampled from
\begin{equation}
\mathbb{P}
=
\Big \{ 
P^{(1)} \otimes \cdots \otimes P^{(n)} \,\Big|\, P^{(i)}\in \{I, X, Y, Z\} 
\Big \},
\end{equation}
the set of all $n$-qubit Paulis, i.e., when $V\sim \text{unif}(\mathbb{P})$, the process is called Pauli-twirling. For any $\mathcal{G}$, the resulting, overall channel
\begin{equation}
\bar{\mathcal{G}}(\rho)
= \frac{1}{4^n}\sum_{P_i \in \mathbb{P}} P_i \, \mathcal{G} \big( P_i \, \rho \, P_i \big) P_i
\end{equation}
is a Pauli channel, with Pauli fidelities $f_i = \boldsymbol{G}_{ii}$. All off-diagonal elements of $\boldsymbol{G}$ are averaged away by the twirling.

\subsection{Clifford Gates}
\label{sec:mitigation_clifford}

Suppose we want to implement a Clifford gate, or a layer of simultaneous Clifford gates, described by an $n$-qubit unitary $U$ and a corresponding quantum channel $\mathcal{U}(\rho) = U \rho \, U^\dag$, but we instead implement a slightly different channel $\mathcal{G}$ due to experimental imperfections. It is customary to factor this noisy gate into $\mathcal{G} = \mathcal{UN}$, where $\mathcal{N}=\mathcal{U}^{-1}\mathcal{G}$ describes the noise and $\mathcal{U}^{-1}(\rho) = U^\dag \rho \, U$. (We could equivalently factor it into $\mathcal{G}=\mathcal{N}' \, \mathcal{U}$ for noise $\mathcal{N}'=\mathcal{G} \,\mathcal{U}^{-1}$. Generically $\mathcal{N} \neq \mathcal{N}'$, although the choice of order is inconsequential, at least for PEC, as long as we are consistent.) PEC, and the related version of ZNE, both have two conceptual steps in terms of this factorization, as depicted in Fig.~\ref{fig:clifford}: 
\begin{description}
\item[Step 1] Pauli-twirl the noise channel $\mathcal{N}$ to simplify it, using only single-qubit gates.
\item[Step 2] Transform (i.e., cancel for PEC, or amplify for ZNE) the resulting Pauli noise channel $\bar{\mathcal{N}}$.
\end{description}
The noise $\mathcal{N}$ will not typically be a Pauli channel, but we can transform it into one through Pauli-twirling in Step 1. To do so while leaving $\mathcal{U}$ intact, we must apply a random Pauli $P_j \sim \text{unif}(\mathbb{P})$ before $\mathcal{G}$ and a corresponding unitary $U P_j U^\dag$ after, in order to reach $\mathcal{N}$ with $P_j$ from both sides \cite{knill:2004, wallman:2016}. It is essential that $U P_j U^\dag$ comprise only single-qubit gates, which are typically much less noisy than multi-qubit gates, so as not to introduce more noise comparable to $\mathcal{N}$ while trying to twirl $\mathcal{N}$. This locality is guaranteed when $U$ is Clifford, in which case $U P_j U^\dagger \propto P_i$ is an $n$-qubit Pauli (up to a global phase) for every $P_j \in \mathbb{P}$, by definition. Step 1 therefore consists of applying random Paulis $P_j$ and $P_i$ on either side of $\mathcal{G}$ according to the distribution described above. Under the approximation that Pauli gates are noiseless compared to multi-qubit gates, which we will make from now on, the resulting overall channel is $\bar{\mathcal{G}} = \mathcal{U} \bar{\mathcal{N}}$, where $\bar{\mathcal{N}}$ is a Pauli channel. Thus simplified, the noise is easily amenable to mitigation.

\begin{figure}[h]
\centering
\[
\Qcircuit @C=0.6em @R=.7em { 
& \gate{\mathcal{G}} & \qw & & = & & & \gate{\mathcal{N}} & \gate{U} & \qw
}
\vspace{1ex}
\]

{\color{gray} \hspace{6.4em} {\footnotesize Step 1} $\downarrow$  {\footnotesize (Pauli-twirl $\mathcal{N}$ into $\bar{\mathcal{N}}$)} }
\vspace{1ex}
\[
\Qcircuit @C=0.5em @R=1em { 
& \gate{P_j} & \gate{\mathcal{N}} & \gate{U} & \gate{P_i} & \qw 
& & = & & & \gate{P_j} & \gate{\mathcal{N}} & \gate{P_j} & \gate{U} & \qw & & = & &  & \gate{\bar{\mathcal{N}}} & \gate{U} & \qw  \\
&  \text{\scriptsize $\sim\,$unif($\mathbb{P}$) } & & & \text{\scriptsize $\!\! \propto \! U P_j U^\dag$} & & & & & & & \text{\scriptsize $\sim\,$unif($\mathbb{P}$) } & & & & & & & & \hspace{2em} \text{\scriptsize $\bar{\mathcal{G}}$}
 {\gategroup{1}{2}{1}{2}{.5em}{_\}}
 \gategroup{1}{5}{1}{5}{.5em}{_\}}
 \gategroup{1}{11}{1}{11}{.5em}{_\}}
 \gategroup{1}{13}{1}{13}{.5em}{_\}} 
 \gategroup{1}{20}{1}{21}{.5em}{.} }
}
\vspace{2ex}
\]
{\color{gray} \hspace{6em} {\footnotesize Step 2} $\downarrow$ {\footnotesize (cancel or amplify $\bar{\mathcal{N}}$)} }
\vspace{1ex}
\[
\Qcircuit @C=0.6em @R=1em { 
& \gate{P_k} & \gate{\bar{\mathcal{G}}} & \qw & & \stackrel{\text{avg}}{=} & & & \gate{\mathcal{M}} & \gate{\bar{\mathcal{N}}} & \gate{U} & \qw  & & = & & & \gate{\bar{\mathcal{N}}^{1+\alpha}} & \gate{U} & \qw \\
& \text{\scriptsize $\sim \! \vec{q}$}  & & & & & & & & \text{\scriptsize $\mathcal{A}$} &
 {\gategroup{1}{2}{1}{2}{.5em}{_\}}
 \gategroup{1}{9}{1}{11}{.5em}{.} }
}
\]
\caption{The two conceptual steps of PEC, and the related form of ZNE, for a noisy Clifford gate $\mathcal{G}$. Any $\mathcal{G}$ can described as a noise channel $\mathcal{N}$ followed by an ideal gate $U$. (We write the unitary $U$ in place of the channel $\mathcal{U}$ in the circuit above, and likewise for other noiseless gates, for simplicity.) In Step 1, one adds random Pauli gates on both sides of $\mathcal{G}$, chosen so as to twirl $\mathcal{N}$ into a Pauli channel $\bar{\mathcal{N}}$. We denote the resulting average channel as $\bar{\mathcal{G}}$. In Step 2, one then adds a random Pauli gate $P_k$ before $\bar{\mathcal{G}}$, sampled from a probability (for ZNE) or quasi-probability (for PEC) distribution $\vec{q}$, which is chosen to correctly amplify ($\alpha>0$) or invert ($\alpha=-1$) $\bar{\mathcal{N}}$, respectively. We denote the resulting aggregate channel as $\mathcal{A}$, and use the notation $\stackrel{\text{avg}}{=}$ to indicated that, for PEC, the relation only holds for expectation values.}
\label{fig:clifford}
\end{figure}

In Step 2, both PEC and the related version of ZNE perform operations of the form
\begin{equation}
\mathcal{M}(\rho) = \sum_k q_k P_k \rho P_k
\label{eq:M_superop}
\end{equation}
before $\bar{\mathcal{G}}$, leading to an aggregate channel of 
\begin{equation}
    \mathcal{A} = \bar{\mathcal{G}} \mathcal{M} = \mathcal{U}\,
    \bar{\mathcal{N}} \mathcal{M}
    \label{eq:A_clifford}
\end{equation} 
with a PTM of
\begin{equation}
\boldsymbol{A}
=
\boldsymbol{U} 
\; 
\overbrace{ \!\!
\begin{pmatrix}
\ddots &  & 0 \\
 & \vec{f} & \\
0 & & \ddots
\end{pmatrix}
\!\!}^{\bar{\boldsymbol{N}}=\diag(\vec{f} \,)}
\;
\overbrace{\!\!
\begin{pmatrix}
\ddots & & 0 \\
 & \boldsymbol{W} \vec{q} & \\
0 & & \ddots 
\end{pmatrix}
\!\! }^{\boldsymbol{M} = \diag(\boldsymbol{W} \vec{q} \,) } \;
,
\label{eq:clifford_mitigation}
\end{equation}
expressed in terms of the PTMs for $\mathcal{A}$, $\mathcal{U}$, $\bar{\mathcal{N}}$ and $\mathcal{M}$ respectively, where $f_i = \tr [P_i \, \mathcal{N}(P_i)]/2^n$. In other words, the aggregate channel behaves like an ideal gate $\mathcal{U}$ preceded by an adjustable Pauli noise channel $\bar{\mathcal{N}} \mathcal{M}$, which depends on $\vec{q}=(q_k)_{k=0}^{4^n-1}$ via $\mathcal{M}$. We now address ZNE and PEC in turn, which differ in their choice and their implementation of $\mathcal{M}$.

The idea of ZNE is to purposely increase the effective strength of gate noise so as to measure an expectation value of interest at multiple noise levels, then predict its zero-noise value by extrapolation. It is a heuristic technique whose performance depends on how exactly one increases the noise level. The most successful approach to date (sometimes called probabilistic error amplification, or PEA) seeks to effectively replace $\bar{\mathcal{N}}$ with $\bar{\mathcal{N}}^{1+\alpha}$ for different noise levels $1+\alpha \ge 1$ by picking $\mathcal{M}=\bar{\mathcal{N}}^\alpha$ \cite{kim:2023}. As Eq.~\eqref{eq:clifford_mitigation} shows, this can be done by applying random Paulis $P_k$ before $\bar{\mathcal{G}}$ with probabilities $q_k$ chosen so that $\boldsymbol{W} \vec{q} = \vec{f}^{\, \alpha}$, where $\vec{f}^{\, \alpha}$ is defined element-wise. The net effect is to amplify the twirled noise by a tunable amount $1+\alpha$ while preserving its structure. 

Rather than amplify the twirled noise, PEC seeks to cancel it by picking $\alpha=-1$ so that $\mathcal{M}=\bar{\mathcal{N}}^{-1}$ and therefore $\mathcal{A} = \mathcal{U}$. As per Eq.~\eqref{eq:clifford_mitigation}, this can be done by picking $\vec{q}$ such that $\boldsymbol{W}\vec{q} = \vec{f}^{\,-1}$, where $(\vec{f}^{\,-1})_i = 1/f_i$.  This $\vec{q}$, however, generally contains negative elements and is therefore not a valid probability distribution. In turn, $\mathcal{M}=\bar{\mathcal{N}}^{-1}$ is not a valid quantum channel, and cannot be implemented as described above in the context of ZNE. It is nonetheless possible to realize this $\mathcal{M}$ in effect, when measuring expectation values, by treating $\vec{q}$ as a quasi-probability distribution. That is, suppose we aim to measure $\langle P_m \rangle = \tr[ P_m \, \mathcal{U}( \rho) ]$ for some $n$-qubit Pauli $P_m$, but can only implement the noisy gate $\bar{\mathcal{G}}$ (with twirled noise) in place of $\mathcal{U}$. PEC provably recovers the noiseless expectation value, on average, by applying random Paulis $P_k$ before $\bar{\mathcal{G}}$ with probabilities $|q_k|/\gamma$, where $\gamma=\sum_k|q_k|$, then multiplying the $\pm 1$ measurement outcomes (corresponding to eigenspaces of $P_m$) by $\gamma \sgn(q_k)$ \cite{temme:2017}. Assuming noiseless readout, the expected value of these scaled outcomes is 
\begin{equation}
\sum_k \frac{|q_k|}{\gamma} \; \gamma \sgn(q_k) \tr \! \big[ P_m \, \bar{\mathcal{G}}(P_k \rho P_k) \big]
=
\tr \! \Big\{ P_m \, \bar{\mathcal{G}} \big[ \mathcal{N}^{-1} (\rho) \big] \Big \},
\end{equation}
which equals $\langle P_m \rangle$ as desired. However, because each shot returns $\pm \gamma$ rather than $\pm 1$, and $\gamma \ge 1$, one typically needs $\gamma^2$ times more shots to estimate $\langle P_m \rangle$ with a given precision than if the gate were noiseless \cite{temme:2017, vandenberg:2023}. Moreover, the $\gamma$ factors multiply when one does PEC for multiple gate layers within a circuit, resulting in a sampling overhead that (typically) grows exponentially in the number of noisy gates. The silver lining, however, is that $\gamma$ approaches 1  here as gate noise decreases, so PEC could be compatible with classically hard circuits despite this exponential overhead, provided the gate noise is sufficiently weak.

\subsection{Non-Clifford Gates}
\label{sec:mitigation_nonclifford}

The error mitigation schemes described above, which we will call \textit{Clifford PEC and ZNE}, rely critically on $\bar{\mathcal{N}}$ being a Pauli channel due to the Pauli-twirling in Step 1. In terms of PTMs, as in Eq.~\eqref{eq:clifford_mitigation}, both techniques  enact a diagonal matrix $\boldsymbol{M}$ before the noisy gate in Step 2. This suffices to amplify or cancel the twirled noise because the latter's PTM, $\bar{\boldsymbol{N}}$, is also diagonal, so we can set $ \boldsymbol{M} = \bar{\boldsymbol{N}}^\alpha$ for any $\alpha$. (We could equally well perform $\mathcal{M}$ after the noisy gate instead, with minor adjustments, following the alternate factorization $\mathcal{G}=\mathcal{N}' \, \mathcal{U}$.) If the effective noise channel were not Pauli, e.g., if we tried to implement $\boldsymbol{M}=\boldsymbol{N}^\alpha$ directly for a generic, non-diagonal $\boldsymbol{N}$, Step 2 would generally require multi-qubit non-Clifford gates, which would introduce more noise that also must be mitigated, leading to a circular situation. But this is exactly the problem we face when $U$ is non-Clifford, as it is not generally possible to Pauli-twirl the associated noise using only single-qubit gates since $U P_j U^\dag$ can be entangling, meaning Step 1 breaks down \cite{santos:2024}. More colloquially, there is no way to reach $\mathcal{N}$ from both sides with arbitrary Paulis without introducing more entangling gates, which are themselves noisy. Or alternatively, in terms of PTMs, there is no apparent way to twirl $\boldsymbol{N}$ into a diagonal matrix which can, in turn, be amplified or inverted by a diagonal $\boldsymbol{M}$. That means that Clifford PEC and ZNE cannot correctly amplify or cancel noise on non-Clifford gates.

\subsubsection{Formalism of Pauli Shaping}

We propose a simple generalization of these techniques that applies to both Clifford and non-Clifford gates. To motivate it, we begin with a thought-experiment: consider a noisy non-Clifford gate $\mathcal{G}= \mathcal{N}' \, \mathcal{U}$ shown in Fig.~\ref{fig:thought_exp}, where $\mathcal{U}(\rho)=U\rho\,U^\dag$ is the intended (non-Clifford) unitary and $\mathcal{N}'$ happens to be a Pauli noise channel from the outset. Since there is no need to twirl such noise, we could still do Clifford PEC or ZNE (skipping Step 1) by inserting $\mathcal{M}=(\mathcal{N}')^\alpha$, as in Eq.~\eqref{eq:M_superop}, \textit{after} $\mathcal{G}$. Notice, however, that if we factored the same noisy gate in the order $\mathcal{G}=\mathcal{UN}$ instead, the resulting noise channel $\mathcal{N} = \mathcal{U}^{-1} \mathcal{N}' \, \mathcal{U}$ would generally be non-Pauli (since $U$ is non-Clifford), and could not be amplified or inverted by inserting an $\mathcal{M}$ \textit{before} $\mathcal{G}$, as described in the previous section. Moreover, there is no apparent way to twirl $\mathcal{N}$ into a Pauli channel using single-qubit gates since $U$ is not Clifford. This thought-experiment suggests two conclusions about mitigating errors on non-Clifford gates. First, whether we insert $\mathcal{M}$ before or after $\mathcal{G}$ can make a critical difference for PEC and ZNE, unlike in the Clifford case. We should therefore seek a formalism that finds the right placement automatically. Second, while it is always possible to factor a noisy gate $\mathcal{G}$ into noise and an ideal gate, doing so for non-Clifford gates can give qualitatively different noise channels (e.g., Pauli or non-Pauli) depending on the factorization order, which is an arbitrary mathematical choice of no physical significance. It can therefore be more informative to think in terms of the noisy gate $\mathcal{G}$ directly, rather than factoring out a noise channel.

\begin{figure}[h]
\[
\Qcircuit @C=0.75em @R=1.3em { 
& \gate{\mathcal{G}} & \qw & & := & & & \gate{U} & \gate{\mathcal{N}'} & \qw & & = & & & \gate{\mathcal{N}} & \gate{U} & \qw \\
& & & & & & & & \parbox{2cm}{\scriptsize \centering \vspace{0.5ex} Pauli channel \\[-0.2ex] by definition} & & & & & & \parbox{2cm}{\scriptsize \centering not generally a \\[-0.2ex] Pauli channel} 
\gategroup{1}{9}{1}{9}{.6em}{_\}}
\gategroup{1}{15}{1}{15}{.6em}{_\}}
} 
\]
\caption{A thought-experiment where a noisy non-Clifford gate $\mathcal{G}$ happens to factorize into an ideal gate $U$ followed by a Pauli noise channel $\mathcal{N}'$, as in the middle circuit. $\mathcal{N}'$ can therefore be amplified or inverted by inserting random Paulis after the noisy gate. However, if we factorized $\mathcal{G}$ as shown on the right, the resulting noise channel $\mathcal{N}$ would generally be non-Pauli, and could not be amplified or inverted by inserting random Paulis before $\mathcal{G}$. This is a fundamental difference between Clifford and non-Clifford gates.}
\label{fig:thought_exp}
\end{figure}

In light of these conclusions, we propose the following scheme. Consider performing $P_j$, then a noisy gate $\mathcal{G}$, then $P_i$, as shown in Fig.~\ref{fig:pauli_shaping}, where $P_i$ and $P_j$ are arbitrary $n$-qubit Paulis. The resulting channel
\begin{equation}
\mathcal{A}^{(ij)} (\rho) = P_i \, \mathcal{G} (P_j \rho P_j) P_i
\end{equation}
has a PTM $\boldsymbol{A}^{(ij)}$ whose $(k,\ell)^\text{th}$ element is
\begin{equation}
\boldsymbol{A}_{k\ell}^{(ij)}
=
\tr \big[ P_k \, \mathcal{A}^{(ij)} (P_\ell) \big]/2^n = \boldsymbol{W}_{ki} \boldsymbol{W}_{j \ell} \, \boldsymbol{G}_{k\ell},
\label{eq:single_PTM}
\end{equation}
where we have used the trace's cyclic property, the fact that $P_a P_b P_a = \boldsymbol{W}_{ab} \, P_b$ for all $P_a, P_b \in \mathbb{P}$, and the definition of the PTM elements $\boldsymbol{G}_{k\ell}$ of $\mathcal{G}$ from Eq.~\eqref{eq:G_ij_def}. Now consider the aggregate channel $\mathcal{A}$ that is a linear combination of $\mathcal{A}^{(ij)}$, weighted by real coefficients $\boldsymbol{Q}_{ij}$ forming a $4^n \times 4^n$ matrix $\boldsymbol{Q}$ of our choice:
\begin{equation}
\mathcal{A}(\rho)
=
\sum_{ij} \boldsymbol{Q}_{ij} \, \mathcal{A}^{(ij)} (\rho)
=
\sum_{ij} \boldsymbol{Q}_{ij} P_i \, \mathcal{G} (P_j  \rho P_j) P_i.
\end{equation}
Using Eq.~\eqref{eq:single_PTM}, the $(k,\ell)^\text{th}$ PTM element of $\mathcal{A}$ is
\begin{equation}
\boldsymbol{A}_{k\ell} 
=
\sum_{ij} \boldsymbol{W}_{ki} \, \boldsymbol{Q}_{ij} \boldsymbol{W}_{j \ell} \, \boldsymbol{G}_{k\ell},
\end{equation}
so its PTM, in matrix form, is
\begin{equation}
\boldsymbol{A}
=
(\boldsymbol{WQW}) \odot \boldsymbol{G}
=
\boldsymbol{C} \odot \boldsymbol{G},
\label{eq:pauli_shaping}
\end{equation}
where $\odot$ denotes an element-wise (i.e., Hadamard) product. It is convenient to define $\boldsymbol{C} =\boldsymbol{WQW}$ in Eq.~\eqref{eq:pauli_shaping}, which we call a \textit{characteristic matrix} in analogy to characteristic functions from probability theory, which are Fourier transforms of probability density functions \cite{casella:2002}. Intuitively, $\boldsymbol{Q}$ and $\boldsymbol{C}$ can be understood as Fourier transforms of one another (since $\boldsymbol{W}$ describes a type of discrete Fourier transform), so the element-wise product in Eq.~\eqref{eq:pauli_shaping} is reminiscent of a convolution in some ``frequency domain.''

\begin{figure}
\[
\Qcircuit @C=0.75em @R=.7em { 
& & \text{\hspace{2ex} \scriptsize $\mathcal{A}^{(ij)}$} & & & & & & \\
& \gate{P_j} & \gate{\mathcal{G}} & \gate{P_i} & \qw & & \stackrel{\text{avg}}{=} & & & \gate{\mathcal{A}} & \qw & \\
 & {\underbrace{\hspace{2em}}} & \raisebox{-2.5em}{\text{\scriptsize \hspace{-0.ex} jointly $\sim \! \boldsymbol{Q}$}} & {\underbrace{\hspace{2em}}} & & & & \\ 
{\gategroup{2}{2}{2}{4}{.4em}{.}}
} 
\] 
\caption{A circuit description of Pauli shaping. To shape a given channel $\mathcal{G}$ into a desired channel $\mathcal{A}$, one can insert random $n$-qubit Paulis $P_j$ and $P_i$ before and after $\mathcal{G}$, respectively, to form a channel $\mathcal{A}^{(ij)}$. One picks $P_i$ and $P_j$ randomly with probability $|\boldsymbol{Q}_{ij}|/\gamma$ in each shot, then multiplies the measurement outcomes by $\gamma \sgn(\boldsymbol{Q}_{ij})$ to realize $\mathcal{A}$, for $\boldsymbol{Q}$ and $\gamma$ from Eqs.~\eqref{eq:pauli_shaping} and \eqref{eq:pauli_shaping_gamma} respectively. When all $\boldsymbol{Q}_{ij} \ge 0$, i.e., when $\boldsymbol{Q}$ is a valid probability distribution of pairs of Paulis, the last step is trivial and the channel $\mathcal{A}$ is actually realized. When some $\boldsymbol{Q}_{ij} < 0$, $\boldsymbol{Q}$ is instead a quasi-probability distribution, and $\mathcal{A}$ is only realized in terms of expectation values, as indicated by the notation $\stackrel{\text{avg}}{=}$.
}
\label{fig:pauli_shaping}
\end{figure}

More concretely, Eq.~\eqref{eq:pauli_shaping} shows that through an appropriate choice of $\boldsymbol{Q}$, the aggregate channel $\mathcal{A}$ can be chosen almost arbitrarily, at least in terms of expectation values, at the cost of a potential sampling overhead. That is, to realize a desired $\mathcal{A}$ with PTM elements $\boldsymbol{A}_{ij}$, it suffices to pick characteristic matrix elements $\boldsymbol{C}_{ij} = \boldsymbol{A}_{ij}/\boldsymbol{G}_{ij}$. (Anytime $\boldsymbol{A}_{ij} = \boldsymbol{G}_{ij} = 0$, the corresponding $\boldsymbol{C}_{ij}$ can be chosen arbitrarily.) The resulting $\boldsymbol{C}$ then corresponds to a unique $\boldsymbol{Q} = \boldsymbol{WCW}/2^{4n}$. If all $\boldsymbol{Q}_{ij} \ge 0$ then $\boldsymbol{Q}$ can be interpreted as a probability distribution over \textit{pairs} of Paulis $(P_j, P_i) \in \mathbb{P} \times \mathbb{P}$, rather than over individual Paulis like $\vec{q}$ in Eq.~\eqref{eq:M_superop}. In other words, the corresponding $\mathcal{A}$ can be realized, with no sampling overhead, by performing $P_j$, then $\mathcal{G}$, then $P_i$ with probability $\boldsymbol{Q}_{ij}$, as in Fig.~\ref{fig:pauli_shaping}. (Normalization, $\sum_{ij} \boldsymbol{Q}_{ij}=1$, is guaranteed if $\mathcal{A}$ and $\mathcal{G}$ are both trace-preserving.) Much like in the Clifford case, we can still realize a desired $\mathcal{A}$ in expectation when the corresponding $\boldsymbol{Q}$ contains negative elements by treating $\boldsymbol{Q}$ as a quasi-probability distribution. That is, we can insert $P_j$ and $P_i$ before and after $\mathcal{G}$ respectively with probability $|\boldsymbol{Q}_{ij}|/\gamma$, where 
\begin{equation}
\gamma = \sum_{ij}|\boldsymbol{Q}_{ij}|,
\label{eq:pauli_shaping_gamma}
\end{equation} 
then multiply the measurement outcomes (the measured eigenvalue of the observable in question) by $\gamma \sgn (\boldsymbol{Q}_{ij})$. The proof is almost identical to that of the Clifford case (see Appendix~\ref{secA:exp_vals}), and the meaning of $\gamma$ is the same: $\gamma^2$ is a sampling overhead that combines multiplicatively with that from other gate layers, leading to an exponential overhead. We are not aware of any prior name for this technique, which is encapsulated by Eq.~\eqref{eq:pauli_shaping} and Fig.~\ref{fig:pauli_shaping}, so we will refer to it here as \textit{Pauli shaping}. Moreover, we will generally refer to $\boldsymbol{Q}$ as a quasi-probability matrix, and similarly for its elements, even though it can also describe a true probability distribution.

Rather than just invert or amplify Pauli noise, Pauli shaping effectively transforms any implemented channel $\mathcal{G}$ into (almost) any desired channel $\mathcal{A}$. It applies to both Clifford and non-Clifford gates. (The only minor limitation is that it requires $\boldsymbol{G}_{ij} \neq 0$ in order to achieve $\boldsymbol{A}_{ij} \neq 0$ for any Paulis $P_i, P_j \in \mathbb{P}$, otherwise no choice of characteristic matrix will satisfy $\boldsymbol{A}_{ij} = \boldsymbol{C}_{ij} \, \boldsymbol{G}_{ij}$. While it is unclear how to rephrase this condition in more conventional terms, we expect it to be easily satisfied outside of pathological scenarios involving fine-tuned or infinite noise, as illustrated in Sec.~\ref{sec:examples}.) For instance, one can use Pauli shaping to do PEC for arbitrary gates by choosing $\boldsymbol{C}$ so that $\mathcal{A} = \mathcal{U}$, i.e., by demanding $\boldsymbol{A}_{ij} = \boldsymbol{U}_{ij}$. Similarly, one could do several forms of ZNE for different notions of noise amplification. One possibility would be to mimic the Clifford case by picking $\boldsymbol{C}$ such that $\mathcal{A} = \mathcal{U N}^{1+\alpha}$ for different noise levels $1+\alpha\ge1$, even though $\mathcal{N} = \mathcal{U}^{-1}\mathcal{G}$ need not be a Pauli channel. (The other noise factorization order works too.) Alternatively, one could aim to find a Lindbladian $\mathcal{L}$ such that $\mathcal{G} = \exp(-i\mathcal{H} +\mathcal{L})$, where $\mathcal{H} = [H,\, \cdot \,]$ is an effective Hamiltonian superoperator (with $H=H^\dag$) that generates the intended gate $\mathcal{U} = e^{-i \mathcal{H}}$, then similarly implement $\mathcal{A} = \exp(-i\mathcal{H} + \alpha \mathcal{L})$ for different $\alpha$. Variants of these schemes where the noise is twirled over a subset of $\mathbb{P}$, depending on the intended gate, are also possible.

Pauli shaping reduces to Clifford PEC/ZNE when the target gate $U$ is Clifford. This may not be obvious since the latter is typically broken into two conceptual steps (as in Sec.~\ref{sec:mitigation_clifford}), which can obscure the full picture: while Step 2 only adds random Paulis on one side of the noisy gate $\mathcal{G}$, Step 1 adds them on both sides to twirl the noise. By combining any adjacent Paulis as shown in Fig.~\ref{fig:clifford_sandwich}, these two steps can be jointly described as inserting random Paulis $P_j$ and $P_i$ before and after a noisy gate $\mathcal{G}$, respectively, with quasi-probability
\begin{equation}
\boldsymbol{Q}_{ij} = 4^{-n} \, q_{ \, \sigma(i) \oplus j},
\label{eq:Clifford_Q}
\end{equation}
where $\vec{q}=(q_k)_{k=0}^{4^n-1}$ are the coefficients appearing throughout Sec.~\ref{sec:mitigation_clifford}, the function $\sigma$ is defined by $P_{\sigma(i)} \propto U^\dag P_i U$, and we write $k = i \oplus j$ when $P_k \propto P_i P_j$. These are the same probabilities, or quasi-probabilities, as one gets by starting with the formalism of Pauli shaping (see Appendix~\ref{secA:clifford_reduction}). The key difference between Pauli shaping and Clifford PEC/ZNE, then, is that the former allows more general correlations between the random Paulis flanking $\mathcal{G}$, giving it a much broader scope without requiring deeper circuits. More precisely, Pauli shaping uses a quasi-probability matrix $\boldsymbol{Q}$ with up to $O(4^{2n})$ distinct elements, whereas Clifford PEC/ZNE implicitly uses a $\boldsymbol{Q}$ with only $O(4^n)$ distinct elements, repeated according to Eq.~\eqref{eq:Clifford_Q}. Of course, it is not feasible to compute $2^{O(n)}$ quasi-probabilities in either case for modern quantum devices. In Clifford PEC/ZNE, ``sparse'' noise models, which approximate $\bar{\mathcal{N}}$, or equivalently $\bar{\mathcal{G}}$, using only $\text{poly}(n)$ parameters based on hardware connectivity, have been empirically successful \cite{vandenberg:2023, kim:2023}. (See also \cite{flammia:2020, harper:2021} for related theoretical results.) We expect a similar approach to be possible for non-Clifford gates, on similar grounds, but we leave such modeling to follow-up work.

\begin{figure}[h]
\centering
\[
\Qcircuit @C=0.65em @R=1em { & & & & \text{\hspace{5ex} \raisebox{-1.25em}{\scriptsize $\mathcal{G}$}} \\ 
& \gate{P_k} & \qw & \gate{P_j} & \gate{\mathcal{N}} & \gate{U} & \gate{P_i} & \qw 
& & = & & & \gate{P_{k \oplus \sigma(i)}} & \gate{\mathcal{G}} & \gate{P_i} & \qw \\
& \text{\scriptsize Step 2} & & & \text{\hspace{1.6em} \scriptsize Step 1}
\gategroup{2}{2}{2}{2}{0.75em}{_\}}
\gategroup{2}{4}{2}{7}{.75em}{_\}}
{\gategroup{2}{5}{2}{6}{.4em}{.}}
}
\]
\caption{Two equivalent ways to view Clifford PEC/ZNE. Left: the conceptual steps are shown separately, where the random Paulis $P_i\sim \text{unif}(\mathbb{P})$ and $P_j =  P_{\sigma(i)} \propto U^\dag P_i U$ serve to twirl the factored noise channel $\mathcal{N}$, then a random Pauli $P_k$ is added with quasi-probability $k \sim \vec{q}$ to amplify or invert the twirled noise. Right: $P_k$ and $P_j$ are combined into a single Pauli $P_\ell = P_{k \oplus \sigma(i)}$, so $P_\ell$ and $P_i$ are added before and after $\mathcal{G}$, respectively, with quasi-probability $q_k / |\mathbb{P}| = 4^{-n} \, q_{\sigma(i) \oplus \ell}$.}
\label{fig:clifford_sandwich}
\end{figure}

\subsubsection{Examples of Pauli Shaping}
\label{sec:examples}

To illustrate the scope and behavior of Pauli shaping more broadly, we present three concrete examples in this section---two analytical and one numerical. While Pauli shaping can be applied to arbitrary gates, these examples focus on a 2-qubit gate that is common in many experiments \cite{krantz:2019, moses:2023, evered:2023}:
\begin{equation}
U=R_{ZZ}(\theta),
\end{equation}
with a generic angle $\theta \notin \{0,\, \pm \pi/2,\, \pi\}$ for which the gate is non-Clifford. When constructing PTMs and the like related to this gate, it will be convenient to order the 2-qubit Paulis as
\begin{align}\label{eq:pauli_order}
\mathbb{P} = \big( & II, \, ZZ,\, XX,\, YY, \,IZ,\, ZI, \,YX,\, XY, \;\;
 XI, \,YZ,\, XZ, \,YI, \,IX, \,ZY,\, ZX,\, IY  \big),
\end{align}
so that those in the first/second half commute/anti-commute with $ZZ$, and each Pauli is related to one of its neighbors through multiplication by $ZZ$. (Of course, the choice of basis ordering is arbitrary and has no impact on the results that follow.) Expressed in the ordered basis \eqref{eq:pauli_order}, the PTM of the channel $\mathcal{U}(\rho) = U\rho \, U^\dag$ describing the ideal gate is block-diagonal:
\begin{equation}
\boldsymbol{U}
=
\begin{pmatrix}
I\quad & & & & & & & 0 \\
& I\quad & & & & & & \\
& & I\quad & & & & & \\ 
& & & I\quad & & & & \\
& & & & R^{(2)}\, & & & \\ 
& & & & & R^{(2)}\, & & \\ 
& & & & & & R^{(2)}\, & \\
0\quad & & & & & & & R^{(2)}
\end{pmatrix} ,
\label{eq:U_PTM}
\end{equation}
with $2\times 2$ identity matrices $I$ in the upper-left (acting trivially on Paulis that commute with $ZZ$) and $2 \times 2$ rotation matrices
\begin{equation}
R^{(2)} = 
\begin{pmatrix}
\cos \theta & -\sin \theta \\
\sin \theta & \,\,\,\,\, \cos \theta 
\end{pmatrix}
\label{eq:R2}
\end{equation}
in the lower-right (mixing together Paulis that anti-commute with $ZZ$). 

\textit{Example 1-} To illustrate the enhanced scope of Pauli shaping, suppose the only error in implementing $U$ is a coherent over- or under-rotation by an angle $\epsilon \in \mathbb{R}$, that is:
\begin{equation}
\mathcal{G}(\rho) = R_{ZZ}(\theta + \epsilon) \, \rho \,  R_{ZZ}(\theta + \epsilon)^\dag.
\end{equation}
This is a paradigmatic non-Pauli error \cite{santos:2024}, which cannot be twirled into a Pauli error channel because $U$ is non-Clifford. Pauli shaping lets us recover the ideal gate\footnote{The only mathematical requirement for Pauli shaping, namely that $\boldsymbol{G}_{ij} \neq 0$ wherever $\boldsymbol{A}_{ij} \neq 0$, is met here for all values of $\theta$ and $\epsilon$ except those where $2(\theta+\epsilon)$ is an integer multiple of $\pi$; that is, the miscalibrated gate $R_{ZZ}(\theta+\epsilon)$ must not be exactly Clifford. This is more of a theoretical concern than a practical one.}, in terms of expectation values, by using a block-diagonal characteristic matrix (also expressed in the ordered basis \eqref{eq:pauli_order}):
\begin{equation}
\boldsymbol{C} = \text{diag}(C_1, \, C_1, \, C_1, \, C_1, \, C_2, \, C_2, \, C_2, \, C_2),
\label{eq:T_coherent}
\end{equation}
where the four bottom blocks must be
\begin{equation}
C_2 = 
\begin{pmatrix}
\frac{\cos (\theta)}{\cos(\theta+\epsilon)} & \frac{\sin (\theta)}{\sin(\theta+\epsilon)} \\[1ex]
\frac{\sin (\theta)}{\sin(\theta+\epsilon)} & \frac{\cos (\theta)}{\cos(\theta+\epsilon)}
\end{pmatrix}
\end{equation}
in order to satisfy Eq.~\eqref{eq:pauli_shaping}. Similarly, the four top blocks of $\boldsymbol{C}$ must have the form
\begin{equation}
C_1 = 
\begin{pmatrix}
1 & x \\ 
y & 1
\end{pmatrix},
\label{eq:Ta_coherent}
\end{equation}
where the parameters $x$ and $y$ can be chosen arbitrarily since both get multiplied by zero in Eq.~\eqref{eq:pauli_shaping}, so their values do not affect the resulting aggregate channel. They do, however, generally affect the required sampling overhead, in potentially complicated ways. A judicious choice of $x$ and $y$ (see Appendix~\ref{secA:examples} for details) cancels the coherent error, in expectation, at the cost of a $\gamma^2$ sampling overhead, where
\begin{equation}
\gamma = \max \left\{ 
\left| \frac{\cos (\theta)}{\cos(\theta+\epsilon)} \right|, \, \left| \frac{\sin (\theta)}{\sin(\theta+\epsilon)} \right|
\right \}
=
1+O(\epsilon).
\label{eq:gamma_overrot}
\end{equation}
In other words, the sampling overhead (per gate) vanishes linearly as the calibration error goes to zero. 

\textit{Example 2-} Unfortunately, while Pauli shaping applies to arbitrary gates, it can be prohibitively expensive on non-Clifford gates depending on the noise in question. To illustrate the issue, consider a different noisy implementation of $U=R_{ZZ}(\theta)$ with a block-diagonal PTM of
\begin{equation}
\boldsymbol{G}
= \!
\scalebox{0.78}{$
\begin{pmatrix}
\!\begin{pmatrix}
1 & 0 \\
2 \epsilon & 1 - 2\epsilon
\end{pmatrix} &&&&&&& 0\\ 
& \!\!\! \begin{pmatrix}
1 - \epsilon & -\epsilon \\
-\epsilon & 1- \epsilon
\end{pmatrix} \\
&& \!\!\! \begin{pmatrix}
1 - \epsilon & \epsilon \\
\epsilon & 1- \epsilon
\end{pmatrix} \\
&&& \!\!\! \begin{pmatrix}
1 - \epsilon & \epsilon \\
\epsilon & 1- \epsilon
\end{pmatrix}\\ 
&&&& \sqrt{1\!-\!2\epsilon} \, R^{(2)} \\
&&&&& \sqrt{1\!-\!2\epsilon} \, R^{(2)} \\
&&&&&& \sqrt{1\!-\!2\epsilon} \, R^{(2)} \\ 
0 &&&&&&& \sqrt{1\!-\!2\epsilon} \, R^{(2)}
\end{pmatrix}
$},
\label{eq:bad_PTM}
\end{equation}
where $\epsilon \in [0, \frac{1}{2})$ describes the noise strength\footnote{This example can be derived from a Lindblad equation (see Appendix~\ref{secA:examples}) in which $\epsilon \rightarrow \frac{1}{2}$ describes the limit of infinite noise strength. The mathematical requirement for Pauli shaping that $\boldsymbol{G}_{ij} \neq 0$ wherever $\boldsymbol{A}_{ij} \neq 0$, is met here for all for all $\epsilon < \frac{1}{2}$, i.e., all finite noise strengths.} ($\epsilon=0$ for the ideal gate) and $R^{(2)}$ is the $2\times 2$ rotation matrix from Eq.~\eqref{eq:R2}. The optimal characteristic matrix (in terms of $\gamma$) that recovers the ideal gate, in expectation, is similarly block-diagonal (see Appendix~\ref{secA:examples}). To satisfy $\boldsymbol{U} = \boldsymbol{C} \odot \boldsymbol{G}$, it must equal
\begin{equation}
\boldsymbol{C} = \diag(C_1, \, C_2,\,  C_2,\,  C_2, \, C_3,\,  C_3,\, C_3,\, C_3 ),
\label{eq:T_bad}
\end{equation}
where 
\begin{equation}
C_1 = 
\begin{pmatrix}
1 & x\\
0 & (1-2\epsilon)^{-1}
\end{pmatrix}
\end{equation}
for some parameter $x$ that we are free to choose, and 
\begin{equation}
C_2 = \frac{1}{1-\epsilon} 
\begin{pmatrix}
1 & 0 \\ 0 & 1
\end{pmatrix}
\qquad 
C_3 = \frac{1}{\sqrt{1-2\epsilon}} 
\begin{pmatrix}
1 & 1 \\ 1 & 1
\end{pmatrix}.
\label{eq:G_ex3}
\end{equation}
The resulting $\gamma$ is a complicated function of $\epsilon$ and $x$ in general (see Appendix~\ref{secA:examples}), but its noiseless limit is
\begin{equation}
\lim_{\epsilon \rightarrow 0} \gamma 
=
\frac{1}{32}
\Big( 
28\,  |x| + 3\,|x-8| + |x + 24|
\Big),
\label{eq:gamma_limit}
\end{equation}
which attains a minimum value of $\gamma = 1.5$ at $x=0$. 

While contrived, this second example illustrates a stark difference between Pauli shaping on Clifford and non-Clifford gates. With Clifford gates, the sampling overhead required to cancel any (twirled) noise, in expectation, approaches 1 in the noiseless limit. The same is not true, in general, for non-Clifford gates, where canceling certain types of noise can require a large sampling overhead, regardless of how weak the noise is. (Since the overheads from different gates combine multiplicatively, $\gamma \approx 1.5$ is effectively much larger than $\gamma \approx 1$ since it leads to a total overhead that grows much faster with circuit size.) In the case of Eq.~\eqref{eq:bad_PTM}, the issue, mathematically, comes from the $O(\epsilon)$ off-diagonal elements in the upper-left part of $\boldsymbol{G}$, which cannot be twirled away without spoiling the action of the gate. Clifford gates have no such elements---rather, all elements of their PTMs that should be zero (for noiseless gates) can be twirled away in noisy gates. In this second non-Clifford example, however, the only way to suppress the aforementioned PTM elements is to set the corresponding elements of $\boldsymbol{C}$ to zero, leading to a large $\gamma$. (This is true whenever these PTM elements are nonzero, regardless of their exact values, i.e., even if they did not all equal $\pm \epsilon$ as in Eq.~\eqref{eq:bad_PTM}.) Because Eq.~\eqref{eq:pauli_shaping} involves an element-wise product of $\boldsymbol{C}$ and $\boldsymbol{G}$, instead of a typical matrix multiplication, this holds true even for arbitrarily small $\epsilon > 0$. In other words, the only number $\boldsymbol{C}_{ij}$ for which $\boldsymbol{C}_{ij} \times \epsilon = 0$ is $\boldsymbol{C}_{ij} = 0$, no matter how small $\epsilon$ might be, which can impose a prohibitive lower bound on $\gamma$ even when the noise is weak. This issue is illustrated numerically in the next example.

\begin{figure}
    \centering
    \includegraphics[width=\textwidth]{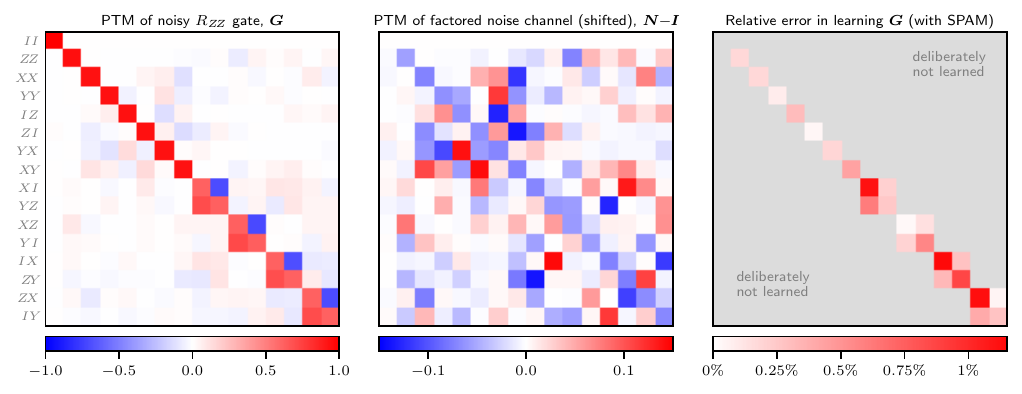} \\[-1em]
    \raggedright \hspace*{7.7em} \textbf{a.} \hspace{12.8em} \textbf{b.} \hspace{12.6em} \textbf{c.} \\ 
    \caption{A noisy $R_{ZZ}(\theta)$ gate used as a numerical example. \textit{Left:} The PTM of the channel $\mathcal{G}$ constructed in Example 3 of Sec.~\ref{sec:examples}, which describes a noisy $U= R_{ZZ}(47^\circ)$ gate. \textit{Center:} The PTM of the associated noise channel $\mathcal{N} = \mathcal{U}^{-1} \mathcal{G}$, with the identity matrix subtracted to reduce the range of values and thereby improve visibility. \textit{Right:} The relative error in learning the PTM elements of this $\mathcal{G}$ that are essential for Pauli shaping, using the schemes developed in Sec.~\ref{sec:learning}. The learned elements are all accurate to within $\lesssim 1\%$ (with room for further optimization), even though the underlying simulation includes state preparation and measurement (SPAM) errors, each with only 90\% average fidelity \cite{nielsen:2002}. The elements in gray were not learned, as their values are not needed to cancel the noise through Pauli shaping, which requires one to find numbers $\{ \boldsymbol{C}_{ij} \}_{i,j=0}^{15}$ such that $\boldsymbol{U}_{ij} = \boldsymbol{C}_{ij} \boldsymbol{G}_{ij}$ (as in Eq.~\eqref{eq:pauli_shaping}, with $\boldsymbol{A}=\boldsymbol{U}$). There is no need to learn $\boldsymbol{G}_{0,0}$, as it equals 1 for any CPTP map, which fixes $\boldsymbol{C}_{0,0}=1$. For all other grayed out indices, $\boldsymbol{U}_{ij}=0$ (see Eqs.~\eqref{eq:U_PTM} and \eqref{eq:R2}), so can we can pick $\boldsymbol{C}_{ij}=0$ regardless of $\boldsymbol{G}_{ij}$.
    }
    \label{fig:PTMs}
\end{figure}

\textit{Example 3-} The last two examples considered error models with very particular structures in order to be analytically tractable. In this one we take the opposite approach by choosing a complicated error model, generated at random with no special structure, and analyzing it numerically. Specifically, suppose one intends to perform a $U=R_{ZZ}(\theta)$ gate with $\theta=47^\circ$, a nondescript angle chosen for illustration so that $U$ is neither Clifford, nor even close to a Clifford gate. Suppose also that, in place of $U$, one can only implement a noisy version described by the channel $\mathcal{G} = \exp(-i \mathcal{H} + \mathcal{L})$ where the first term, defined by $\mathcal{H}(\rho) = \frac{\theta}{2}[ZZ, \rho]$, describes the ideal gate, and the second term is a perturbation of general Lindblad form (to ensure that $\mathcal{G}$ remains CPTP):
\begin{equation}
\mathcal{L}(\rho)
=
-i \, \epsilon_\textsc{h} [H_\mathsf{err},\rho] + \epsilon_\textsc{d} \! \sum_{j,k=1}^{15} \Gamma_{jk} \Big(
P_j \rho P_k - \frac{1}{2} \big \{ P_k P_j, \, \rho \big \} \Big).
\end{equation}
To form $\mathcal{G}$, we picked generic matrices $H_\mathsf{err}=H_\mathsf{err}^\dag$ and $\Gamma = \Gamma^\dag \ge 0$ at random by sampling elements from standard normal distributions. We then set the strengths of the Hamiltonian and dissipative noise terms, $\epsilon_\textsc{h}$ and $\epsilon_\textsc{d}$ respectively, through bisection so that $\mathcal{G}$ has an average fidelity \cite{nielsen:2002} of $F_\text{avg} = 0.95$ with respect to $U$, and the noise unitarity \cite{wallman:2015} is $u=0.9$. These two numbers were chosen for illustration, to give strong noise that is somewhat non-unital, and partway between the fully incoherent and fully coherent. The PTM of the resulting $\mathcal{G}$ is shown in Fig.~\ref{fig:PTMs}a, and that of the corresponding non-Pauli noise channel $\mathcal{N} = \mathcal{U}^{-1} \mathcal{G}$ is depicted in Fig.~\ref{fig:PTMs}b.

\begin{figure}
\hspace*{2em} 
\Qcircuit @C=0.6em @R=0.6em { 
\lstick{\textbf{a.} \;\;}  & & & & \text{\hspace{0em} \scriptsize 10 identical layers} \\ 
\lstick{\ket{y_+}} & \gate{R_X(\phi)} & \multigate{1}{R_{ZZ}(47^\circ)}  & \qw & \raisebox{-3em}{\,\dots} & & \gate{R_X(\phi)} & \multigate{1}{R_{ZZ}(47^\circ)} & \meterB{Z} \\ 
\lstick{\ket{y_+}} & \gate{R_X(\phi)} & \ghost{R_{ZZ}(47^\circ)} & \qw & & & \gate{R_X(\phi)} & \ghost{R_{ZZ}(47^\circ)} & \meterB{X}
\gategroup{2}{2}{2}{8}{.75em}{^)} 
}
\vspace{1.5em}
\centering
\begin{tikzpicture}
    \node[inner sep=0, anchor=south west] (image) at (0, 0) {\includegraphics[width=3.65in]{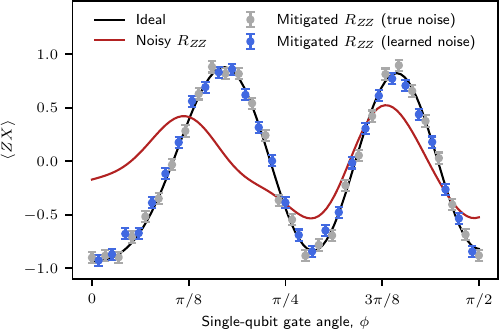}};
    \node[anchor=south west] at (0.03, 5.3) {\textbf{b.}};
 \end{tikzpicture}
\caption{A numerical example of noise cancellation through Pauli shaping. \textit{Top:} An illustrative circuit used to estimate $\langle Z X \rangle$ as a function of $\phi$, where $\ket{y_+}$ denotes the $+1$ eigenstate of $Y$. \textit{Bottom:} Implementing $\mathcal{G}$, the noisy gate depicted in Fig.~\ref{fig:PTMs}a, in place of the ideal $U=R_{ZZ}(47^\circ)$ gates generally yields incorrect values of $\langle ZX \rangle$, as shown by the two curves. (Both use perfect initial states and measurements.) However, Pauli shaping can recover the correct expectation values despite using $\mathcal{G}$. Every data point is the mean from $10^7$ shots, each from an independent noisy circuit drawn from the quasi-probability distribution given by Eq.~\eqref{eq:pauli_shaping} with $\boldsymbol{A}=\boldsymbol{U}$. The gray points use a distribution derived from the true PTM $\boldsymbol{G}$, whereas the blue points use one derived from an imperfect version thereof, which is learned using the schemes from Sec.~\ref{sec:learning}, and described in Fig.~\ref{fig:PTMs}c. Error bars show the estimated standard error of the mean.}
\label{fig:mitigation_numerics}
\end{figure}

To demonstrate the broad applicability of Pauli shaping, we used it to mitigate repeated $\mathcal{G}$ gates in the circuit from the top panel of Fig.~\ref{fig:mitigation_numerics}a. As shown in Fig.~\ref{fig:mitigation_numerics}b, this gives the correct observable expectation values\footnote{The mathematical requirement for Pauli shaping that $\boldsymbol{G}_{ij} \neq 0$ wherever $\boldsymbol{A}_{ij}\neq 0$ is manifestly met for $\boldsymbol{G}$ in Fig.~\ref{fig:PTMs}a when $\boldsymbol{A}=\boldsymbol{U}$.}. However, because $\boldsymbol{G}_{i,i+1}$ and $\boldsymbol{G}_{i+1,i}$ are mostly nonzero for $0 \le i \le 7$, as in Example 2, this noise cancellation requires $\gamma \approx 1.67$. Note that while Example 1 features coherent noise and small $\gamma$, whereas Examples 2 and 3 feature (at least partly) incoherent noise and large $\gamma$, one should not infer a trend here. That is, noise coherence is not what causes $\gamma$ to be large or small. As a simple counter-example, consider an $R_{ZZ}(\theta)$ gate with weak Pauli noise---the resulting $\gamma$ is small like in Example 1, even though the noise is purely incoherent. The same is true for a mixture of over/under-rotation and Pauli noise.

We will discuss the phenomenon of large $\gamma$ despite weak noise more broadly in Sec.~\ref{sec:discussion}. For now, however, we turn our attention to the challenge of accurately measuring the PTM $\boldsymbol{G}$ of a noisy non-Clifford gate. Since the characteristic matrix $\boldsymbol{C}$, and ultimately, the quasi-probability matrix $\boldsymbol{Q}$ in Eq.~\eqref{eq:pauli_shaping}, depend on $\boldsymbol{G}$, addressing this challenge is essential for Pauli shaping.

\section{Characterizing Noisy Gates}
\label{sec:learning}

The performance of both PEC and ZNE depends on how accurately one can characterize, or ``learn,'' the noisy gate in question, or equivalently, the factored noise channel. This is true for both Clifford and non-Clifford gates. A notable obstacle, however, is that the measurement outcomes from any circuit meant to characterize a noisy gate will also depend on state preparation and measurement (SPAM) errors, which can be significant \cite{bravyi:2021}. It is therefore desirable to use learning schemes that can distinguish gate and SPAM errors as much as possible, so as to accurately isolate the former. Here we review such learning schemes for Clifford gates in Sec.~\ref{sec:clifford_learning}, then introduce new ones suited for non-Clifford $R_{ZZ}(\theta)$ gates in Sec.~\ref{sec:non-clifford_learning}. But first we discuss readout twirling, which is a common component throughout both sections \cite{vandenberg:2022}.

Readout twirling consists of applying an $X$ gate with 50\% probability independently on each qubit before readout, then flipping the measured bits wherever such gates were applied. (We use ``measurement'' and ``readout'' synonymously.) In principle, these random $X$ gates should be sampled independently in each circuit execution, i.e., each ``shot.'' Suppose we want to measure an $n$-qubit Pauli expectation value $\langle P_i \rangle = \tr(P_i \, \rho)$ for $P_i \in \mathbb{P}$ and some state $\rho$, and we denote our estimate thereof after a finite number $N_\text{tot}$ of independent shots as
\begin{equation} 
\hat{\mu} = 
\frac{(\texttt{\# of +1 outcomes}) - (\texttt{\# of -1 outcomes})}{N_\text{tot}},
\label{eq:mu_hat}
\end{equation}
where $\pm 1$ outcomes refer to the observed eigenvalues of $P_i$. Because $\hat{\mu}$ will vary from one experiment to the next, even under identical conditions, we can treat it as a random variable whose expectation value, $\mathbb{E}(\hat{\mu})$, describes an average over many hypothetical experiments. Absent any measurement errors, $\mathbb{E}(\hat{\mu}) = \langle P_i \rangle$, i.e., $\hat{\mu}$ is an unbiased estimate of $\langle P_i \rangle$, which means we can get it arbitrarily close to the true value $\langle P_i \rangle$, with arbitrarily high probability, by simply taking enough shots. However, measurement errors could bias $\hat{\mu}$ in complicated ways, such that $\mathbb{E}(\hat{\mu})$ bears no simple relation to $\langle P_i \rangle$. This issue is partially remedied through readout twirling, which ensures that $\mathbb{E}(\hat{\mu}) = m_i \langle P_i \rangle$ for some coefficient $m_i$ that depends on the statistics of the readout noise but not on the measured quantum state $\rho$. ($m_i=1$ for ideal readout---see Appendix~\ref{secA:readout_twirling}.) In other words, readout twirling still gives a biased estimate for $\langle P_i \rangle$, but the bias has a predictable form that will let us distinguish gate errors from SPAM errors.

\subsection{Clifford Gates}
\label{sec:clifford_learning}

Consider a Clifford unitary $U$ whose noisy implementation is described by the channel $\mathcal{G} = \mathcal{UN}$, where $\mathcal{N}$ describes generic noise and $\mathcal{U}(\rho) = U\rho \, U^\dag$, as in Sec.~\ref{sec:mitigation_clifford}. We will assume throughout that the noise does not vary with time, and is independent of any previous gates. Since we can Pauli-twirl $\mathcal{N}$ using only single-qubit gates, it suffices to learn the twirled noise channel $\bar{\mathcal{N}}$. In the language of PTMs, we only need to learn $\bar{\boldsymbol{N}} = \diag(\vec{f} \, )$, since the Pauli fidelities $f_i = \boldsymbol{N}_{ii}$ are the only components of the noise that figure in Clifford PEC/ZNE. It is possible to learn these (at least in part) in a way that is robust to SPAM errors through cycle benchmarking (CB) \cite{erhard:2019}, a variant of randomized benchmarking.

To introduce CB, we will begin with the simple case where $U=I$. This means that every Pauli $P_i \in \mathbb{P}$ is an eigenvector of the twirled, noisy identity gate $\bar{\mathcal{G}} = \mathcal{U} \bar{\mathcal{N}} = \bar{\mathcal{N}}$ with eigenvalue $f_i$---that is:
\begin{equation}
\bar{\mathcal{G}} (P_i) = f_i \, P_i.
\end{equation}
Applying $\bar{\mathcal{G}}$ $d$ times to an initial state $\rho = 2^{-n} \sum_j s_j P_j$ therefore leads to a final state of
\begin{equation}
    \rho' = \bar{\mathcal{G}}^d (\rho) = \frac{1}{2^n} \sum_j s_j f_j^d P_j
\end{equation}
with Pauli expectation values $ \langle P_i \rangle = \tr(P_i \, \rho') =  s_i \, f_i^d$ that decay exponentially in the circuit depth $d$ at rates $f_i$. CB exploits this relation by performing the following steps to estimate each Pauli fidelity $f_i$:
\begin{enumerate}
\item Prepare an initial state $\rho = 2^{-n} \sum_j s_j P_j$ for which $s_i = \tr(P_i \, \rho)$ is as large as possible (to maximize the eventual signal-to-noise ratio). E.g., attempt to prepare $\rho=\ket{\psi} \! \bra{\psi}$ where $\ket{\psi}$ is a separable $+1$ eigenstate of $P_i \in \mathbb{P}$, so $s_i = 1$ ideally.
\item Apply $\bar{\mathcal{G}}$ $d$ times to $\rho$ for varying depths $d$.
\item Estimate $\langle P_i \rangle$ for the resulting state $\bar{\mathcal{G}}^d(\rho)$ as in Eq.~\eqref{eq:mu_hat} using readout twirling, denoting the result by $\hat{\mu}$.
\end{enumerate}
The expected value of $\hat{\mu}$ (which we will denote as $\mu$), i.e., the average estimate of $\langle P_i \rangle$ for a circuit depth $d$ from noisy experimental data, is
\begin{equation}
\mu \; := \; \mathbb{E}(\hat{\mu}) = s_i \, m_i \times f_i^d,
\label{eq:exp_decay}
\end{equation}
where the coefficients $s_i$ and $m_i$ depend on state preparation and measurement errors, respectively, but not on $f_i$ or $d$. Therefore, even though $\mu \neq \langle P_i \rangle$ in general, CB obtains an estimate of $f_i$ that is robust to SPAM by fitting the tuples $( d, \, \hat{\mu} )$ to a function $d \mapsto A r^d$ and extracting $r$. These steps are then repeated for all desired Pauli fidelities $f_i$ (although it is possible to re-use some of the same experimental data to get different Pauli fidelities, as we discuss later).

While there are more sophisticated alternatives with substantially better scaling \cite{chen:2022, chen:2024, king:2024}, CB's relative simplicity and robustness make it especially practical on large pre-fault-tolerant quantum devices. In fact, besides being robust to SPAM, CB is well-behaved in two other ways that are critical in many experiments but less often discussed theoretically. First, it is \textit{sensitive}, in that a small change in $f_i$ produces a large change in the data, so the number of shots needed is reasonably small. Second, it \textit{concentrates}, meaning that the different random circuits that arise due to twirling give similar expectation values, so few of them are needed \cite{wallman:2014, helsen:2019b}. We now elaborate on both points in turn.

\textit{Sensitivity:} Suppose we estimate $\langle P_i \rangle$ for a circuit depth $d$, as per the CB steps described above. Each $\pm 1$ measurement outcome follows a Bernoulli distribution\footnote{We use the term ``Bernoulli'' loosely here since this distribution is supported on $\{1,-1\}$ rather than $\{0,1\}$.} with mean $\mu$ from Eq.~\eqref{eq:exp_decay}. One way to quantify how much information about $f_i$ is conveyed by each such outcome is through its (classical) Fisher information \cite{casella:2002}
\begin{equation}
\mathcal{I}(f_i)
= \frac{1}{1-\mu^2} \left( \frac{\partial \mu}{\partial f_i} \right)^2
=
 \frac{\mu^2}{1-\mu^2}  \frac{d^2}{f_i^2},
\label{eq:fisher_info}
\end{equation}
which diverges for large $d$ in the noiseless limit (see Appendix~\ref{secA:sensitivity}).  Why consider this limit? Because we need $f_i \approx 1$ for error mitigation to be feasible in the first place, so while the exact value of $f_i$, and therefore also of $\mathcal{I}(f_i)$, is unknown \textit{a priori}, a learning scheme that works well as $f_i \rightarrow 1$ should also work well in the relevant regime of $f_i$, by continuity. (Also, crucially, this limit is analytically tractable.) Concretely, $\mu \rightarrow f_i^d$ in the absence of SPAM errors, so we can re-write Eq.~\eqref{eq:fisher_info} as
\begin{equation}
\mathcal{I}(f_i)
\;
\mathop{\longrightarrow}^{\text{no SPAM}}_{\text{errors}}
\;
\frac{1}{f_i^2 \ln(f_i)^2} 
\frac{x^2}{4 (e^x - 1)}
\lesssim 
\frac{0.162}{f_i^2 \ln(f_i)^2},
\label{eq:fisher_info_max}
\end{equation}
where we maximized over $x=2 d \ln(1/f_i)$ numerically in the last step, which is a convenient way to maximize over $d$ in effect (see Appendix~\ref{secA:sensitivity}). This maximum value of $\mathcal{I}(f_i)$ then diverges as $f_i \rightarrow 1$, thanks to the slow decay of $\mu$ versus $d$ when $f_i\approx 1$, wherein a small change in $f_i$ produces a big change in $\mu$ at large depths $d$, as quantified by $\partial \mu / \partial f_i$. Consequently, CB does not require an exorbitant number of shots to precisely estimate large Pauli fidelities, since each shot can be very informative.

\textit{Concentration:} CB, as described above, uses a new random circuit for each shot because it must twirl the gate and measurement noise. For any given depth, this leads to independent and identically distributed (IID) measurement outcomes, which are relatively simple to analyze. For instance, the standard error (i.e., the standard deviation of $\hat{\mu}$, a widely-used measure of statistical uncertainty) in estimating $\mu = \mathbb{E}(\hat{\mu})$ with $N_\text{tot}$ shots has the familiar $O(1/\sqrt{N_\text{tot}})$ scaling, specifically:
\begin{equation}
\sqrt{\Var(\hat{\mu})} = \sqrt{\frac{1-\mu^2}{N_\text{tot}}}.
\label{eq:Var_mu}
\end{equation}

In many experiments, however, loading a new circuit into the control electronics used to implement gates is slow compared to running an already-loaded circuit \cite{wack:2021}. It is therefore common practice to use a modified version of CB where, for any given depth, $N_\text{c}$ different random circuits are chosen, each of which is run $N_\text{s/c} \ge 1$ times (the subscript stands for ``shots per circuit'') for a total of $N_\text{tot} = N_\text{c} \, N_\text{s/c}$ shots. The resulting estimate of $\langle P_i \rangle$ (i.e., the sample average), denoted $\hat{\mu}'$, has the same mean as $\hat{\mu}$, and reduces to $\hat{\mu}$ when $N_\text{s/c}=1$ as in proper CB. In general, however, it suffers from larger statistical fluctuations than $\hat{\mu}$, which do not generally scale as $O(1/\sqrt{N_\text{tot}})$ since the measurement outcomes are not IID. Rather, $\hat{\mu}'$ has a standard error of
\begin{equation}
    \sqrt{ \Var(\hat{\mu}') }
    =
    \sqrt{ \frac{1-\mu^2}{N_\text{tot}}
    +
    \left(  \frac{N_\text{s/c}-1}{N_\text{s/c}}  \right) 
    \frac{ \Delta^2}{N_{\text{c}}}
    },
    \label{eq:Var_mu'}
\end{equation}
where 
\begin{equation}
\Delta^2 
=
\Var \{ \, \tr[P_i \,\mathcal{T}(\rho)] \,\} 
= \mathbb{E} \big\{ \tr[P_i \,\mathcal{T}(\rho)]^2 \big\}
-
\mu^2
\label{eq:circ_variance}
\end{equation}
is the variance in expectation values over different random, noisy circuits $\mathcal{T}$. (See Appendix~\ref{secA:concentration} for details.) That is, CB defines many random noisy circuits $\mathcal{T}$, sometimes called ``twirl circuits'' or ``twirl instances,'' comprising $d$ sequential noisy gates $\mathcal{G}$ interleaved by random Paulis, as depicted in Fig.~\ref{fig:T_variance}a. Each such circuit can have a different expectation value, and $\Delta^2$ is the variance thereof, as shown in Fig.~\ref{fig:T_variance}b. It is an important, albeit rarely analyzed quantity, since the standard error of $\hat{\mu}'$ approaches $\Delta/\sqrt{N_\text{c}}$ as $N_\text{s/c} \rightarrow \infty$. A large $\Delta$ would therefore necessitate many different twirl circuits ($N_\text{c} \gg 1$) to precisely estimate $\mu$ for any given depth $d$, which can be very slow (in terms of wall-clock time). Fortunately, $\Delta \rightarrow 0$ in the noiseless limit for CB. This can be seen from Eq.~\eqref{eq:circ_variance}, since $\tr[P_i \, \mathcal{T}(\rho)] \rightarrow 1$ for all $\mathcal{T}$ in this limit. We will not seek a formal bound on $\Delta$ more broadly; rather, we simply note that the common practice of reusing a small number of random circuits many times is well-justified for CB when the noise is weak.

\begin{figure}[h]

    \raggedright \hspace{8.5em} \textbf{a.} \vspace{-1.5ex}
\[
\Qcircuit @C=0.75em @R=1.3em { 
& \gate{\mathcal{T}} & \qw 
} 
\; = \; 
\raisebox{2.2em}{
\Qcircuit @C=0.75em @R=0.5em { 
& & & & & \raisebox{1.5em}{\text{\footnotesize random circuit with $d$ noisy gates $\mathcal{G}$}} \\ 
& \gate{X} & \multigate{1}{\mathcal{G}} & \gate{Y} & \qw & \raisebox{-1.8em}{\,\dots} & & \gate{I} & \multigate{1}{\mathcal{G}} & \gate{Z} & \multigate{1}{\parbox{2.7em}{\scriptsize \centering twirled\\readout\\noise}} & \qw \\ 
& \gate{Y} & \ghost{\mathcal{G}} & \gate{Z} & \qw & & & \gate{Z} & \ghost{\mathcal{G}} & \gate{Z}  & \ghost{ \parbox{2.7em}{\scriptsize twirled\\readout\\noise}} & \qw
\gategroup{2}{2}{2}{10}{.7em}{^\}}
}}
\]

\vspace{0.5em}
\raggedright \hspace{8.5em} \textbf{b.}
\vspace{-1em}

\hspace*{3em}
\centering
\begin{tikzpicture}
\draw[->, thick] (-3, 0) -- (3, 0) node[right] {\scriptsize $\tr[P_i \mathcal{T}(\rho)]$};
\draw[->, thick] (0, 0) -- (0, 2) node[right] {\scriptsize probability};

\node at (-2.8, -0.2) {\scriptsize -1};
\node at (2.8, -0.2) {\scriptsize 1};

\draw[-, densely dotted] (-0.6, 0) -- (-0.6, 1.8);
\node at (-0.6, -0.2) {\scriptsize $\mu$};
\draw[|-|] (-1.4, 0.8) -- (-0.63, 0.8);
\node at (-1.015, 0.65) {\scriptsize $\Delta$};

\draw[domain=-2.8:2.8, smooth, variable=\x, gray, samples=1000, thick] plot ({\x}, { 1.8*exp(-(\x+0.6)*(\x+0.6) * 1)  });
\end{tikzpicture}   

\caption{Cycle benchmarking, like the non-Clifford learning schemes introduced below, involves running random circuits comprising $d$ noisy gates $\mathcal{G}$ interleaved with random Paulis drawn from some prescribed distribution. These gates are then followed by a readout noise channel flanked by random bit flips \cite{vandenberg:2022} (see Appendix~\ref{secA:readout_twirling}). We describe such a noisy circuit by a channel $\mathcal{T}$, as shown in the top panel. Each of these possible circuits can give a different Pauli expectation value $\tr[P_i \mathcal{T}(\rho)]$, and some values arise more often than others, as illustrated in the bottom panel. (This is not quantum randomness from measurement outcomes, but rather, classical randomness from the choice of circuit.) The mean of this distribution, $\mu$, describes our average estimate of $\langle P_i \rangle$, and the standard deviation $\Delta$ determines how many random circuits are needed to do so accurately.}
\label{fig:T_variance}
\end{figure} 

\newpage

These two properties, namely sensitivity and concentration for weak noise, do not arise automatically. As we will see in the next section, special care must be taken to maintain them when generalizing CB for non-Clifford gates.

Before moving on, however, we must consider non-trivial Clifford gates $U \neq I$. CB works very similarly in this setting---the key difference being that $U$ now maps a generic Pauli $P_i$ to a potentially different one $P_j = \pm U P_i U^\dag$, so
\begin{equation}
\bar{\mathcal{G}}(P_i) = \pm f_i P_j.
\end{equation}
(In other words, $P_i$ is now a generalized eigenvector of $\bar{\mathcal{G}}$ \cite{flammia:2021}.) Since the right-hand side is proportional to $P_j$, another application of $\bar{\mathcal{G}}$ will introduce a factor of $f_j$ rather than another $f_i$. However, repeated applications of $\bar{\mathcal{G}}$ will eventually yield a term proportional to $P_i$, ultimately leading to an exponential decay similar to that in Eq.~\eqref{eq:exp_decay}. Consider $U=\text{CNOT}$ and $P_i =IZ$ for illustration. This $U$ maps $P_i$ to $P_j=ZZ$ and vice versa, meaning: 
\begin{equation}
\bar{\mathcal{G}}^d(P_i)
=
(f_i f_j )^{d/2} P_i
\end{equation}
for even depths $d$. Therefore, by measuring $\langle P_i \rangle$ for various (even) circuit depths, as described earlier, one can obtain an estimate of $f_i f_j$ that is robust to SPAM errors. Unfortunately, this product does not uniquely specify $f_i$ or $f_j$. One can partially sidestep this ``degeneracy'' and isolate certain Pauli fidelities by interleaving other single-qubit gates between applications of $\bar{\mathcal{G}}$. However, some $f_i$ fundamentally cannot be learned in a SPAM-robust way for general Clifford gates \cite{chen:2023}. Perhaps unsurprisingly, a similar issue will arise with non-Clifford gates as well. For CNOTs, principled guesses for $f_i$ and $f_j$, given a measured value of $f_i f_j$, have led to very effective error mitigation \cite{vandenberg:2023, kim:2023}. We will assume here that similar heuristics may work for non-Clifford gates too.

\subsection{Non-Clifford Gates}
\label{sec:non-clifford_learning}

While there exist some SPAM-robust methods for learning general, noisy non-Clifford gates (at least in part) \cite{merkel:2013, blume:2013, nielsen:2021, endo:2018, kimmel:2014, helsen:2019}, it is easier to ensure sensitivity and concentration by devising schemes that are tailored for specific gates of interest. In particular, we will consider 2-qubit gates of the form $U=R_{PQ}(\theta)$, where $P,Q \in \{X,Y,Z\}$ and $\theta \notin \{0,\, \pm \pi/2,\, \pi\}$ is a generic rotation angle for which the gate is non-Clifford. Such gates arise naturally in several types of qubits \cite{krantz:2019, moses:2023, evered:2023}. For concreteness, we will focus on $U = R_{ZZ}(\theta)$ gates, from which the more general $R_{PQ}(\theta)$ case follows through a change of basis. In contrast with the CB formalism for Clifford gates from the previous section, it will be more natural here to learn PTM elements of the noisy gate $\mathcal{G}=\mathcal{UN}$ itself, rather than those of the factored noise channel $\mathcal{N}$. Of course, learning $\mathcal{G}$ is equivalent to learning $\mathcal{N}$, although the former approach is especially convenient for Pauli shaping, which deals directly with $\mathcal{G}$, not $\mathcal{N}$ (see, e.g., Eq.~\eqref{eq:pauli_shaping}).

For mathematical convenience, we will express all PTMs here in the same ordered basis as in Sec.~\ref{sec:mitigation}, i.e., we will order the set of 2-qubit Paulis $\mathbb{P}$ as in Eq.~\eqref{eq:pauli_order}. It will also be convenient to define the subsets
\begin{align}
\mathbb{P}_\textsc{c} = \big \{ II, \, ZZ,\, XX,\, YY, \,IZ,\, ZI, \,YX,\, XY  \big \} 
\qquad 
\mathbb{P}_\textsc{a} = \big \{  XI, \,YZ,\, XZ, \,YI, \,IX, \,ZY,\, ZX,\, IY  \big \} 
\end{align}
of $\mathbb{P} = \mathbb{P}_\textsc{c} \cup \mathbb{P}_\textsc{a}$, comprising the eight Paulis that commute and anti-commute with $ZZ$ respectively. The PTM $\boldsymbol{U}$ of the ideal gate, given in Eqs.~\eqref{eq:U_PTM} and \eqref{eq:R2}, is then block-diagonal with $2 \times 2$ blocks. By contrast, the PTM $\boldsymbol{G}$ of the noisy gate has no such special structure in general; rather, almost all its elements could be nonzero due to noise. Learning $\boldsymbol{G}$ in full would be a daunting task. To cancel arbitrary noise through Pauli shaping, however, we need only form a characteristic matrix $\boldsymbol{C}$ such that $\boldsymbol{U} = \boldsymbol{C} \odot \boldsymbol{G}$ (see Eq.~\eqref{eq:pauli_shaping}). Notice that since $\boldsymbol{U}_{ij}=0$ for most indices $i$ and $j$, we can simply set these $\boldsymbol{C}_{ij}=0$ and avoid learning the corresponding $\boldsymbol{G}_{ij}$. This argument does not necessarily hold if we instead want to amplify the noise in $\boldsymbol{G}$ for ZNE, though. Rather, we observe that arbitrary noise can be greatly simplified (though not to the point of being Pauli noise), without affecting the logical action of the gate, by twirling $\mathcal{G}$ over $\mathbb{P}_\textsc{c}$. The resulting channel
\begin{equation}
\bar{\mathcal{G}}_\textsc{c}(\rho) = \frac{1}{8} \sum_{P_i \in \mathbb{P}_\textsc{c}} 
P_i \, \mathcal{G} \big( P_i \, \rho \, P_i \big) P_i,
\end{equation}
has a PTM of
\begin{equation}
\bar{\boldsymbol{G}}_\textsc{c}
=
\begin{pmatrix}
\maroon{\boldsymbol{G}_{0,0}} & \gray{\boldsymbol{G}_{0,1}} & & & & & & & & 0 \\
\gray{\boldsymbol{G}_{1,0}} & \maroon{\boldsymbol{G}_{1,1}}  \\ 
& & \maroon{\ddots} \\
& & & \maroon{\boldsymbol{G}_{6,6}} & \gray{\boldsymbol{G}_{6,7}} \\
& & & \gray{\boldsymbol{G}_{7,6}} & \maroon{\boldsymbol{G}_{7,7}} \\
& & & & & \blue{\boldsymbol{G}_{8,8}} & \green{\boldsymbol{G}_{8,9}} \\
& & & & & \green{\boldsymbol{G}_{9,8}} & \blue{\boldsymbol{G}_{9,9}} \\
& & & & & & & \blue{\ddots} \\ 
& & & & & & & & \blue{\boldsymbol{G}_{14,14}} & \green{\boldsymbol{G}_{14,15}} \\
0 & & & & & & & & \green{\boldsymbol{G}_{15,14}} & \blue{\boldsymbol{G}_{15,15}}
\end{pmatrix},
\label{eq:G_color}
\end{equation}
which is $2 \times 2$ block-diagonal regardless of $\boldsymbol{G}$, and whose nonzero elements are equal to the corresponding ones of $\boldsymbol{G}$ (see Appendix~\ref{secA:examples}). This ``partially'' twirled (over $\mathbb{P}_\textsc{c}$ rather than $\mathbb{P}$, hence the subscript $\textsc{c}$) noisy gate is therefore a simpler starting point for ZNE, and it also includes all the $\boldsymbol{G}_{ij}$ that are relevant for noise cancellation. Consequently, we will focus here on learning only the elements of $\bar{\boldsymbol{G}}_\textsc{c}$, i.e., those on the $2 \times 2$ diagonal blocks of $\boldsymbol{G}$. In principle, they  could take (almost) arbitrary values. But in practice, if the noise is weak enough to be amenable to mitigation, then these elements should be close to their ideal values. This leads to four qualitatively different types of PTM elements to be learned, which are color-coded in Eq.~\eqref{eq:G_color}:
\begin{description}
\item[Type 1 \maroon{($\boldsymbol{G}_{ii} \approx 1$)}] These top-left, diagonal elements describe how $\mathcal{G}$ preserves Paulis that commute with $ZZ$, and should be close to 1. They must be learned in order to cancel the gate noise.
\item[Type 2 \blue{($\boldsymbol{G}_{ii} \approx \cos \theta$)}] These bottom-right, diagonal elements describe how $\mathcal{G}$ preserves Paulis that anti-commute with $ZZ$, and should be close to $\cos \theta$. They must be learned in order to cancel the gate noise.
\item[Type 3 \green{($\boldsymbol{G}_{ij}\approx \pm \sin \theta$)}] These bottom-right, off-diagonal elements describe how $\mathcal{G}$ mixes Paulis that anti-commute with $ZZ$, and should be close to $ \pm \sin \theta$. They must be learned in order to cancel the gate noise.
\item[Type 4 \gray{($\boldsymbol{G}_{ij} \approx 0$)}] These top-left, off-diagonal elements describe how $\mathcal{G}$ mixes Paulis that commute with $ZZ$, and should be close to 0. They need not necessarily be learned in order to cancel gate noise, since the corresponding elements of $\boldsymbol{U}$ equal zero. However, they could be necessary for ZNE.
\end{description}
These distinct types stand in contrast with the Clifford case, where all Pauli fidelities approach 1 in the noiseless limit (so there is only one type). We now introduce learning schemes tailored for each different type of PTM element.

\subsubsection{Type 1 Elements}

\begin{figure}
    \centering
    \begin{tikzpicture}
    \node[inner sep=0, anchor=south west] (image) at (0, 0) {\includegraphics[width=3.8in]{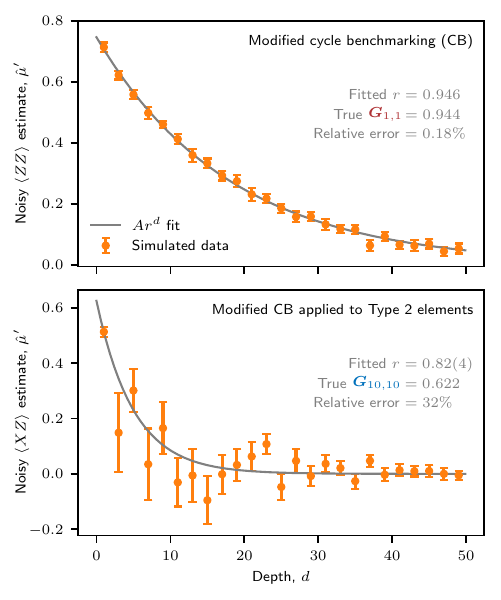}};
    \node[anchor=south west] at (3.4, 10.57) {\textbf{a.}};
    \node[anchor=south west] at (3.4, 5.37) {\textbf{b.}};
 \end{tikzpicture}
    \caption{A numerical example of how modified CB learns the Type 1 elements of $\boldsymbol{G}$ from Fig.~\ref{fig:PTMs}a, but fails on the Type 2 elements. \textit{Top:} To illustrate how modified CB concentrates for Type 1 elements, we estimated $\langle ZZ \rangle$ at each depth $d$ using only $N_\textsc{c}=10$ random circuits, repeated for $N_\textsc{s/c}=10^3$ shots each, and denote the resulting means by $\hat{\mu}'$. The error bars, which show the standard error of $\hat{\mu}'$ estimated through Eq.~\eqref{eq:Var_mu'}, are small because the random circuits for a given $d$ have similar $ZZ$ expectation values (i.e., $\Delta$ is small). The slow decay enables a good fit and ultimately a good estimate of $\boldsymbol{G}_{1,1}$, despite strong SPAM errors, which we model by inserting random error channels with average fidelity $F_\text{avg}=0.9$ at the beginning and end of each circuit. The relative error in learning this, and other Type 1 elements, is shown in Fig.~\ref{fig:PTMs}c. \textit{Bottom:} Analogous data for a Type 2 element of this $\boldsymbol{G}$ using the same settings ($N_\text{c}$, $N_\text{s/c}$, and SPAM error strength). The learning error is much worse because the decay is faster, and because the error bars/scatter are larger due to different twirl circuits having very different $XZ$ expectation values (i.e., $\Delta$ is large). This data is not included in Fig.~\ref{fig:PTMs}c; rather, it only motivates the two other learning schemes that follow. 
    }
    \label{fig:modified_CB}
\end{figure}

\begin{figure*}

\includegraphics[width=\textwidth]{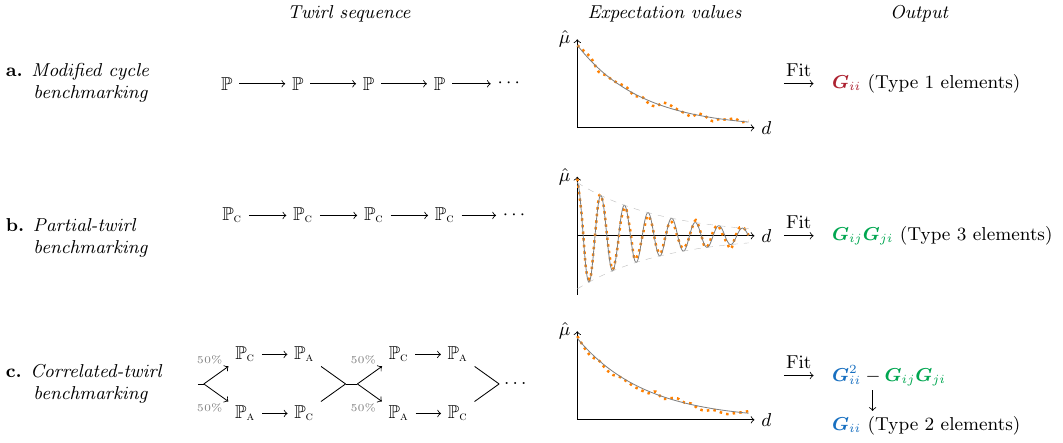}

\caption{A summary of the non-Clifford learning schemes introduced in this paper. All three schemes work by repeating the noisy gate $\mathcal{G}$ $d$ times for various depths $d$, twirling each one in a prescribed way, then estimating Pauli expectation values for the resulting quantum state. Fitting these estimated values, $\hat{\mu}$, versus $d$ to a prescribed function then gives SPAM-robust estimates of the various PTM elements of $\mathcal{G}$ (or products thereof). Each row above depicts a different, complementary, learning scheme. The first column (\textit{Twirl sequence}) shows the sets of Paulis over which each $\mathcal{G}$ is twirled. The branching in the bottom row indicates that either path is taken with 50\% probability. The middle column (\textit{Expectation values}) illustrates typical results, with orange dots depicting empirical estimates $\hat{\mu}$ of the true expectation values, and solid gray curves depicting the resulting fits.  (The dashed curves in the middle plot show the exponential envelope $d \mapsto \pm r^d$, and are meant to guide the eye.) Finally, the right column (\textit{Output}) lists the PTM elements, or products thereof, that can be extracted by fitting the data.}
\label{fig:learning_sequences}
\end{figure*}

Type 1 elements behave much like Pauli fidelities. We can therefore learn them using \textit{modified cycle benchmarking}, in which we Pauli-twirl $\mathcal{G}$ (or equivalently $\bar{\mathcal{G}}_\textsc{c}$), then follow the steps of standard CB for the resulting channel. Note that this is different than standard CB for nontrivial (i.e., $U \neq I$) Clifford gates. There, one inserts random Paulis on either side of $\mathcal{G}$ so as to Pauli-twirl the associated \textit{noise channel} $\mathcal{N}$ without changing the logical effect of the gate, described by $\mathcal{U}$. Here, we instead Pauli-twirl \textit{the noisy gate itself}, thus intentionally spoiling its logical effect---in a particular way---and turning it into a Pauli channel with PTM
\begin{equation}
\bar{\boldsymbol{G}}
=
\text{diag}\big(
\maroon{\boldsymbol{G}_{0,0}}, \dots, \maroon{\boldsymbol{G}_{7,7}}, \, \blue{\boldsymbol{G}_{8,8}}, \dots, \blue{\boldsymbol{G}_{15,15}} \big).
\label{eq:G_full_twirl}
\end{equation}
This channel is not unitary, even in the noiseless limit. However, because $\bar{\boldsymbol{G}}$ is diagonal, we can learn the elements $\maroon{\boldsymbol{G}_{ii}}$, for $1 \le i \le 7$, by preparing an eigenstate of $P_i$, applying this Pauli channel $d$ times for various depths $d$, estimating $\langle P_i \rangle$ for each using readout twirling, and fitting the results to $d \mapsto A r^d$, from which $r$ gives a SPAM-robust estimate of $\maroon{\boldsymbol{G}_{ii}}$.  ($\maroon{\boldsymbol{G}_{00}}=1$ assuming the noisy gate is CPTP.) Moreover, this scheme is sensitive and it concentrates, much like standard CB; that is, $\mathcal{I}(\maroon{\boldsymbol{G}_{ii}}) \rightarrow \infty$ and $\Delta \rightarrow 0$ in the noiseless limit. (See Appendices~\ref{secA:sensitivity} and \ref{secA:concentration}.) It is illustrated numerically in Fig.~\ref{fig:modified_CB}a and summarized in Fig.~\ref{fig:learning_sequences}a.

\subsubsection{Type 2 and 3 Elements}

It may seem from Eq.~\eqref{eq:G_full_twirl} that we could learn the Type 2 elements ($\blue{\boldsymbol{G}_{ii}}$ for $8 \le i \le 15$) in the same way. Indeed, this approach would formally give SPAM-robust estimates of said elements---but it would not be sensitive nor would it concentrate, thus making it of limited practical use. These issues are illustrated numerically in Fig.~\ref{fig:modified_CB}b. In more mathematical terms, Eq.~\eqref{eq:fisher_info_max} implies that
\begin{equation}
\mathcal{I}( \blue{\boldsymbol{G}_{ii}} )
\lesssim
\frac{0.162}{\cos(\theta)^2 \ln[ \cos(\theta)]^2}
< \infty
\end{equation}
in the noiseless limit, meaning that each shot would give relatively little information about $\blue{\boldsymbol{G}_{ii}}$, so far more shots would be needed than for Type 1 elements (or Pauli fidelities in the Clifford case). Intuitively, the problem is that $\blue{\boldsymbol{G}_{ii}}^d \approx \cos(\theta)^d$ generally decays quickly with $d$, so the measured expectation values $\langle P_i \rangle$ would quickly approach zero regardless of $\blue{\boldsymbol{G}_{ii}}$'s exact value, and resolving them to within a reasonable relative error would require many shots. To make matters worse, the expectation values of different random circuits would not concentrate; rather, they would have a variance of
\begin{equation}
\Delta^2
\rightarrow
\frac{1}{2} \Big[ 
1 + \cos(2\theta)^d - 2 \cos(\theta)^{2d} 
\Big]
\label{eq:Delta}
\end{equation}
in the noiseless limit, which quickly asymptotes to $1/2$ with growing depth $d$. (See Appendix~\ref{secA:concentration}.) The issue is that the lower-right blocks of the ideal PTM $\boldsymbol{U}$ (see Eqs.~\eqref{eq:U_PTM} and \eqref{eq:R2}) are 2-dimensional rotation matrices, so Pauli-twirling the ideal gate implements rotations over $\mathbb{P}_\textsc{a}$ by a uniformly random angle of $\pm \theta$. Repeating such twirled gates therefore produces a random walk with a rapidly growing variance given by Eq.~\eqref{eq:Delta}. Ultimately, this means that modified CB is impractical for Type 2 elements, since it would require many more shots from many more random circuits (compared to Type 1 elements). These issues highlight the importance of grouping PTM elements into distinct types based on their approximate values---Type 1 and Type 2 elements may appear similarly in Eq.~\eqref{eq:G_full_twirl}, but they can behave very differently since the latter can be much smaller.

Instead, we introduce two other learning schemes which, together, satisfy all of our desiderata. The first of these schemes (called \textit{partial-twirl benchmarking}) yields some information about the Type 3 elements, which we then use, together with the second scheme (called \textit{correlated-twirl benchmarking}), to get the Type 2 elements.

\textit{Partial-twirl benchmarking:} The main idea of this scheme is to apply $\bar{\mathcal{G}}_\textsc{c}^d$ for various depths $d$, i.e., to apply the noisy gate $\mathcal{G}$ $d$ times, twirling each one independently over $\mathbb{P}_\textsc{c}$, the set of Paulis that commute with $ZZ$, rather than over all Paulis. Estimating $\langle P_i \rangle$ for $P_i \in \mathbb{P}_\textsc{a}$ at different depths and fitting the results will then give SPAM-robust estimates of certain PTM elements. 

Since the PTM of the partially-twirled gate, $\bar{\boldsymbol{G}}_\textsc{c}$, is block-diagonal (see Eq.~\eqref{eq:G_color}), we can find $\bar{\boldsymbol{G}}_\textsc{c}^d$ by simply taking the $d^{\,\text{th}}$ power of each $2\times 2$ block. Consider one such block from the bottom-right of $\bar{G}_\textsc{c}$, which we will denote as $B$:
\begin{equation}
B :=
\begin{pmatrix}
\blue{\boldsymbol{G}_{ii}} & \green{\boldsymbol{G}_{ij}} \\ 
\green{\boldsymbol{G}_{ji}} & \blue{\boldsymbol{G}_{jj}}
\end{pmatrix},
\end{equation}
where $i \in \{8, 10, 12, 14\}$ and $j=i+1$. The form of $B^d$ depends qualitatively on the eigenvalues of $B$, which are:
\begin{equation}
\lambda_\pm
=
\frac{1}{2} \left[ 
\blue{\boldsymbol{G}_{ii}} + \blue{\boldsymbol{G}_{jj}}
\pm 
\sqrt{(\blue{\boldsymbol{G}_{ii}}-\blue{\boldsymbol{G}_{jj}})^2 
+ 
4 \, \green{\boldsymbol{G}_{ij} \boldsymbol{G}_{ji}}}
\right].
\end{equation}
Intuitively, the elements of $B^d$ are functions of $\lambda_\pm^d$, so there can be two distinct cases: If $\lambda_\pm$ are real, they will produce exponential decays. If $\lambda_\pm = r e^{\pm i \omega}$ are complex, because the term in the square root is negative, they will instead produce exponentially-damped oscillations with some frequency $\omega$ and decay rate $r$. We will call these two cases, namely when $\text{Im}(\lambda_\pm)=0$ and $\text{Im}(\lambda_\pm) \neq 0$, the strong and weak noise regimes respectively. (They are analogous to over/critically-damped and under-damped classical harmonic oscillators, respectively.) In the strong noise regime, applying $\bar{\mathcal{G}}_\textsc{c}$ repeatedly to a generic initial state $\rho$ and measuring $\langle P_i \rangle$ for $P_i \in \mathbb{P}_\textsc{a}$ will give expectation values that decay steadily towards their asymptotic values with growing circuit depth $d$. In the weak noise regime, these expectation values will instead oscillate with $d$ as they gradually decay, much like Rabi oscillations. (In the noiseless limit there is no decay and the oscillations persist as $d \rightarrow \infty$.) The two regimes should therefore typically be easy to distinguish experimentally. This scheme assumes that the gate is in the weak noise regime, i.e., that:
\begin{equation}
(\blue{\boldsymbol{G}_{ii}} - \blue{\boldsymbol{G}_{jj}} )^2 
\; \stackrel{\text{assumed}}{<} \;
-4 \, \green{\boldsymbol{G}_{ij} \boldsymbol{G}_{ji}}
\label{eq:eig_assumption}
\end{equation}
for all $2 \times 2$ blocks of Type 2 and 3 elements. This is our only assumption about these PTM elements, and it amounts to assuming that the gate's logical effect is not overwhelmed by noise. In the noiseless limit, the left and right hand sides of \eqref{eq:eig_assumption} approach 0 and $4 \sin(\theta)^2$ respectively. More broadly, we expect this condition to be easily satisfied on modern quantum processors, provided the chosen $\theta$ is not unreasonably small.

Assuming condition \eqref{eq:eig_assumption}, we can write $B^d$ in the form
\begin{equation}
B^d
=
\begin{pmatrix}
a r^d \cos(\omega d - \delta) & \cdots \; \\
\cdots & \cdots \;
\end{pmatrix},
\label{eq:block_powers}
\end{equation}
where
\begin{align}
a &= 2 \sqrt{
\frac{ \green{ \boldsymbol{G}_{ij} \boldsymbol{G}_{ji} } }
{(\blue{\boldsymbol{G}_{ii}}- \blue{\boldsymbol{G}_{jj}} )^2 + 4 \, \green{ \boldsymbol{G}_{ij} \boldsymbol{G}_{ji}}
}
} \nonumber \\[2ex]
r &= \sqrt{
\blue{\boldsymbol{G}_{ii} \boldsymbol{G}_{jj}} - \green{\boldsymbol{G}_{ij} \boldsymbol{G}_{ji} }
} 
\label{eq:params_to_PTM}
\end{align}
\vspace{-3ex}
\begin{align}
\omega &= \text{arctan2} \! \left(
\sqrt{
-(\blue{\boldsymbol{G}_{ii}} - \blue{\boldsymbol{G}_{jj}} )^2 - 4 \, \green{\boldsymbol{G}_{ij} \boldsymbol{G}_{ji}}
}
\, ,\;
\blue{\boldsymbol{G}_{ii}} + \blue{\boldsymbol{G}_{jj}}
\right) \nonumber \\ 
\delta &= \text{arctan2} \! \left(
\blue{\boldsymbol{G}_{ii}} - \blue{\boldsymbol{G}_{jj}}
\, ,\;
\sqrt{
-(\blue{\boldsymbol{G}_{ii}}-\blue{\boldsymbol{G}_{jj}})^2 - 4 \, \green{\boldsymbol{G}_{ij} \boldsymbol{G}_{ji}}
} \nonumber
\right),
\end{align}
and ellipses denote different matrix elements that are generally nonzero (see Appendix~\ref{secA:partial_twirl_benchmarking}). We also use the notation $\text{arctan2}(y,x)$ to denote $\arctan(y/x)$ with an appropriate quadrant correction, as in many programming languages. In the noiseless limit\footnote{This can be a useful initial guess for curve fitting.} $(a, r, \omega, \delta) \rightarrow (1,1,\theta,0)$. More generally, Eqs.~\eqref{eq:block_powers} and \eqref{eq:params_to_PTM} suggest that one could perhaps follow steps akin to CB, but with measured expectation values at different $d$ forming a decaying sinusoid $\propto r^d \cos(\omega d - \delta)$ rather than a pure exponential decay. Fitting the data to this curve would then give a decay rate $r$, a frequency $\omega$, and a phase $\delta$, from which one could learn about the relevant PTM elements by inverting Eq.~\eqref{eq:params_to_PTM}.

There remains one problem, however. In the other learning schemes discussed so far, the PTM of interest was diagonal, so an expectation value $\langle P_i \rangle$ at depth $d$ only depended on one component $s_i = \tr(\rho P_i)$ of the initial state $\rho=\frac{1}{4} \sum_k s_k P_k$. Here, however, because $B^d$ is not diagonal, $\langle P_i \rangle$ will be a linear combination of $s_i$ and $s_j$, with weights that vary with $d$. That is, assuming ideal measurements for the moment (for simplicity):
\begin{equation}
\langle P_i \rangle 
=
\tr[ \, \bar{\mathcal{G}}_\textsc{c}^d (\rho) \, P_i]
=
s_i \; a r^d \cos(\omega d - \delta)   +  s_j \; (B^d)_{01},
\label{eq:state_prep_issue}
\end{equation}
where $(B^d)_{01}$ is the top-right element of $B^d$ from Eq.~\eqref{eq:block_powers}, which is generally nonzero and depends on $d$. This means that state preparation errors, which can cause $s_i$ and $s_j$ to deviate from their intended values independently, can impact our estimates of $\langle P_i \rangle$ in more pernicious ways than before, i.e., not just as a constant scale factor that can be absorbed into the amplitude of the fitted curve and ignored. To sidestep this issue, we prose applying $P_i$ to $\rho$ with $50\%$ probability before applying $\bar{\mathcal{G}}_\textsc{c}$. The resulting state 
\begin{equation}
\rho' = \frac{1}{2} \big(\rho + P_i \rho P_i \big) = \frac{1}{4} \sum_k s_k' P_k
\end{equation}
has 
\begin{equation}
s_i' = \tr(\rho' P_i) = s_i
\qquad \quad 
s_j' =  \tr(\rho' P_j) = 0,
\end{equation}
i.e., it has the same $P_i$ component as $\rho$ but no $P_j$ component, because $\{P_i, P_j\}=0$. We will refer to this step as \textit{state-prep twirling}, in analogy to readout twirling, since it uses randomization to make state preparation errors better behaved. By suppressing the second term in Eq.~\eqref{eq:state_prep_issue} (effectively replacing $s_j$ with $s_j'=0$), state-prep twirling ensures that state preparation errors only contribute a constant scale factor in our estimates of $\langle P_i \rangle$, as in CB. This allows us to fit our estimates of $\langle P_i \rangle$ versus $d$ and extract values of $r$, $\omega$, and $\delta$ that are robust to SPAM errors.

\begin{figure}[h]
    \centering
    \vspace{1em}
    \includegraphics[width=3.7in]{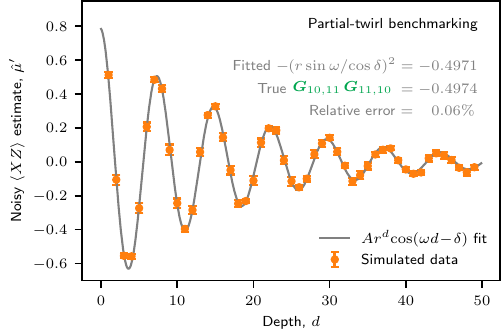}
    \caption{A numerical example of how partial-twirl benchmarking learns the Type 3 elements of $\boldsymbol{G}$ from Fig.~\ref{fig:PTMs}a. We estimated $\langle XZ \rangle$ at each depth $d$ using the same settings as in Fig.~\ref{fig:modified_CB} ($N_\text{c}=10$, $N_\text{s/c}=10^3$, and state preparation and measurement errors with $F_\text{avg}=0.9$ each), and denote the resulting means by $\hat{\mu}'$. By design, the decay is slow and the error bars, which show the standard error of $\hat{\mu}'$ estimated through Eq.~\eqref{eq:Var_mu'}, are small for the same reason as in Fig.~\ref{fig:modified_CB}a (small $\Delta$). The resulting fit therefore gives a good estimate of $\green{\boldsymbol{G}_{10,11}} \green{\boldsymbol{G}_{11,10}}$. To fully isolate these and other Type 3 elements, we assumed $\green{\boldsymbol{G}_{11,10}} = \green{-\boldsymbol{G}_{10,11}} > 0$ as in the noiseless limit, leading to the sub-$1\%$ errors shown in Fig.~\ref{fig:PTMs}c. Notice from the oscillations that this example is clearly in the required ``weak noise regime,'' defined by Eq.~\eqref{eq:eig_assumption}, despite $\boldsymbol{G}$ describing much stronger noise than is in most experiments.}
    \label{fig:PTB}
\end{figure}

Ultimately, then, partial-twirl benchmarking comprises the following steps. For each $i \in \{8, 10, 12, 14\}$ and $j=i+1$ (indices that label 2-qubit Paulis according to Eq.~\eqref{eq:pauli_order}):
\begin{enumerate}
    \item Prepare an initial state $\rho = \frac{1}{4} \sum_k s_k P_k$ for which $s_i = \tr(P_i \, \rho)$ is as large as possible. E.g., attempt to prepare $\rho=\ket{\psi} \! \bra{\psi}$ where $\ket{\psi}$ is a separable $+1$ eigenstate of $P_i$, so $s_i = 1$ ideally.
    \item (State-prep twirling.) Apply $P_i$ with $50\%$ probability to $\rho$, independently in each shot, to produce an average state $\rho'$.
    \item Apply $\bar{\mathcal{G}}_\textsc{c}$ $d$ times to $\rho'$ for varying depths $d$, where $\bar{\mathcal{G}}_\textsc{c}$ denotes the noisy $R_{ZZ}(\theta)$ gate $\mathcal{G}$ twirled over $\mathbb{P}_\textsc{c}$, the eight Paulis that commute with $ZZ$.
    \item Estimate $\langle P_i \rangle$ for the resulting state $\bar{\mathcal{G}}^d(\rho')$ as in Eq.~\eqref{eq:mu_hat} using readout twirling, denoting the result by $\hat{\mu}$.
\end{enumerate}
The expected value of $\hat{\mu}$ for a circuit depth $d$ is then
\begin{equation}
\mu \; := \;
\mathbb{E}(\hat{\mu})
=
s_i \, m_i \, a \times r^d \cos (\omega d - \delta),
\label{eq:oscillating_decay}
\end{equation}
where $a$, $r$, $\omega$ and $\delta$ are given by Eq.~\eqref{eq:params_to_PTM}, and the coefficient $m_i$ describes the measurement errors, as introduced at the start of Sec.~\ref{sec:learning}. (Cf. the equivalent expression for cycle benchmarking in Eq.~\eqref{eq:exp_decay}.) Note that SPAM errors only affect the amplitude of this decaying sinusoid. We can therefore estimate its decay rate, frequency and phase in a SPAM-robust way by fitting the tuples $(d, \hat{\mu})$ to $d \mapsto A r^d \cos(\omega d - \delta)$ and extracting $r$, $\omega$ and $\delta$, respectively. We can then solve Eq.~\eqref{eq:params_to_PTM} for the underlying PTM elements to get:
\begin{align}
\green{\boldsymbol{G}_{ij} \boldsymbol{G}_{ji}}
&=
-\left(  \frac{r \sin (\omega)}{\cos(\delta)} \right)^2 \label{eq:type3_prod} \\ 
\blue{ \boldsymbol{G}_{ii} }
&=
r \Big[
\cos(\omega) + \sin(\omega) \tan(\delta)
\Big] \label{eq:G_ii} \\ 
\blue{ \boldsymbol{G}_{jj} }
&=
r \Big[
\cos(\omega) - \sin(\omega) \tan(\delta)
\Big], \label{eq:G_jj}
\end{align}
which do not depend on the amplitude $A$. The procedure is illustrated numerically in Fig.~\ref{fig:PTB} and summarized in Fig.~\ref{fig:learning_sequences}b.

The expectation values from this scheme concentrate as desired, i.e., $\Delta \rightarrow 0$ in the noiseless limit (see Appendix~\ref{secA:concentration}). Unfortunately, the steps above only give us products $\green{\boldsymbol{G}_{ij} \boldsymbol{G}_{ji}}$ of Type 3 elements, rather than their isolated values. While there are partial workarounds\footnote{One can interleave single-qubit Cliffords between the noisy gates, for instance \cite{chen:2023}.}, we suspect this is a fundamental limitation like the CB ``degeneracy'' arising in the Clifford case \cite{chen:2023}. If so, one could rely on similar approximations here to isolate the Type 3 elements, e.g., assume that $\green{\boldsymbol{G}_{ij}} = -\green{\boldsymbol{G}_{ji}}$ as in the noiseless limit.

There is one remaining issue with this scheme: the measurement results are highly sensitive to the decay rate and oscillation frequency, $r$ and $\omega$ respectively, but not to the phase $\delta$ (see Appendix~\ref{secA:sensitivity}). Intuitively, a small change in $r$ or $\omega$ leads to a big change in $\mu$ at large depths, as quantified by $\partial \mu /\partial r$ and $\partial \mu /\partial \omega$. In contrast, a small change in $\delta$ only produces a small offset in $\mu$ regardless of the depth. In other words, the phase is typically harder to fit precisely than the other two parameters. This is a minor issue for Type 3 elements, since $\delta \approx 0$ for weak noise, and Eq.~\eqref{eq:type3_prod} only depends on $\delta$ to order $O(\delta^2)$. However, Eqs.~\eqref{eq:G_ii} and \eqref{eq:G_jj} both depend on it more strongly, namely to order $O(\delta)$, so our inability to precisely fit $\delta$ can lead to poor estimates of Type 2 elements ($\blue{\boldsymbol{G}_{ii}}$ and $\blue{\boldsymbol{G}_{jj}}$) using this method. We therefore introduce one final scheme to more accurately estimate these latter elements.

\textit{Correlated-twirl benchmarking:} Due to the above difficulty in fitting $\delta$, partial-twirl benchmarking should only be used to learn the Type 3 elements---a different scheme can then be used to learn the Type 2 elements. The key insight underpinning this final scheme is that twirling $\mathcal{G}$ over $\mathbb{P}_\textsc{a}$, the 8 Paulis that anti-commute with $ZZ$, leads to a block-diagonal PTM $\bar{\boldsymbol{G}}_\textsc{a}$ that resembles $\bar{\boldsymbol{G}}_\textsc{c}$ in Eq.~\eqref{eq:G_color}, but with all the off-diagonal components negated\footnote{In other words, $\frac{1}{2} \bar{\boldsymbol{G}}_\textsc{c} + \frac{1}{2} \bar{\boldsymbol{G}}_\textsc{a} = \bar{\boldsymbol{G}}$, since twirling over $\mathbb{P}_\textsc{c}$ or $\mathbb{P}_\textsc{a}$, each with 50\% probability, amounts to a full Pauli-twirl.}. Suppose we apply $\mathcal{G}$ twice, and with equal probability we either twirl the first instance over $\mathbb{P}_\textsc{c}$ then the second over $\mathbb{P}_\textsc{a}$, or twirl the first over $\mathbb{P}_\textsc{a}$ then the second over $\mathbb{P}_\textsc{c}$. We refer to this as \textit{correlated twirling}, since the second gate is twirled in a manner that depends on how the first gate was twirled. It results in a Pauli channel, but not the same one as if we had simply Pauli-twirled $\mathcal{G}$ directly. As in the previous section, all PTMs in question are block-diagonal, so it suffices to consider a generic $2\times 2$ PTM block. Specifically, the overall PTM that describes correlated twirling has blocks:
\begin{equation}
\frac{1}{2} \!
\underbrace{
\begin{pmatrix}
\blue{\boldsymbol{G}_{ii}} & \green{\boldsymbol{G}_{ij}} \\ 
\green{\boldsymbol{G}_{ji}} & \blue{\boldsymbol{G}_{jj}}
\end{pmatrix}
}_\text{block from $\bar{\boldsymbol{G}}_\textsc{c}$}
\!
\underbrace{
\begin{pmatrix}
\blue{\boldsymbol{G}_{ii}} & \green{-\boldsymbol{G}_{ij}} \\ 
\green{-\boldsymbol{G}_{ji}} & \blue{\boldsymbol{G}_{jj}}
\end{pmatrix}
}_\text{block from $\bar{\boldsymbol{G}}_\textsc{a}$}
+
\frac{1}{2} \!
\underbrace{
\begin{pmatrix}
\blue{\boldsymbol{G}_{ii}} & \green{-\boldsymbol{G}_{ij}} \\ 
\green{-\boldsymbol{G}_{ji}} & \blue{\boldsymbol{G}_{jj}}
\end{pmatrix}
}_\text{block from $\bar{\boldsymbol{G}}_\textsc{a}$}
\!
\underbrace{
\begin{pmatrix}
\blue{\boldsymbol{G}_{ii}} & \green{\boldsymbol{G}_{ij}} \\ 
\green{\boldsymbol{G}_{ji}} & \blue{\boldsymbol{G}_{jj}}
\end{pmatrix}
}_\text{block from $\bar{\boldsymbol{G}}_\textsc{c}$}
=
\begin{pmatrix}
\blue{\boldsymbol{G}_{ii}^2} - \green{\boldsymbol{G}_{ij}  \boldsymbol{G}_{ji}} & 0 \\
0 & \!\! \blue{\boldsymbol{G}_{jj}^2} - \green{\boldsymbol{G}_{ij}  \boldsymbol{G}_{ji}}
\end{pmatrix}.
\label{eq:alternating_twirl_block}
\end{equation}
The resulting PTM is therefore diagonal, but with elements that depend non-trivially on the elements of $\boldsymbol{G}$. (We use the colors for Type 2 and 3 elements in Eq.~\eqref{eq:alternating_twirl_block}, but it applies also to Type 1 and Type 4 elements.) Therefore, by repeating this sequence for varying depths, as in cycle benchmarking, we can learn $\blue{\boldsymbol{G}}^2_\blue{ii} - \green{\boldsymbol{G}_{ij} \boldsymbol{G}_{ji}}$ and $\blue{\boldsymbol{G}}^2_\blue{jj} - \green{\boldsymbol{G}_{ij} \boldsymbol{G}_{ji}}$ in a SPAM-robust way. And since we have already learned $\green{\boldsymbol{G}_{ij} \boldsymbol{G}_{ji}}$ through partial-twirl benchmarking, we can therefore isolate the Type 2 elements $\blue{\boldsymbol{G}_{ii}}$ and $\blue{\boldsymbol{G}_{jj}}$.

\begin{figure}
    \centering
    \includegraphics[width=3.5in]{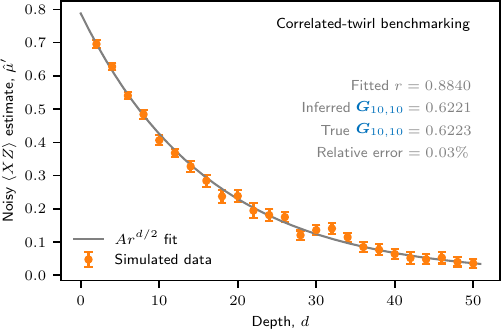}
    \caption{A numerical example of how correlated-twirl benchmarking learns the Type 2 elements of $\boldsymbol{G}$ from Fig.~\ref{fig:PTMs}a. We estimated $\langle XZ \rangle$ at each depth $d$ using the same settings as in Figs.~\ref{fig:modified_CB} and \ref{fig:PTB} ($N_\text{c}=10$, $N_\text{s/c}=10^3$, and state preparation and measurement errors with $F_\text{avg}=0.9$ each), and denote the resulting means by $\hat{\mu}'$. This data contrasts sharply with Fig.~\ref{fig:modified_CB}b by design: the decay is slow and the error bars, which show the standard error of $\hat{\mu}'$ estimated through Eq.~\eqref{eq:Var_mu'}, are small for the same reason as in Fig.~\ref{fig:modified_CB}a (small $\Delta$). The resulting fit therefore gives a good estimate of $\blue{\boldsymbol{G}_{10,10}^2} - \green{\boldsymbol{G}_{10,11}  \boldsymbol{G}_{11,10}}$, and in turn of $\blue{\boldsymbol{G}_{10,10}}$. The relative error in learning this, and other Type 2 elements, is shown in Fig.~\ref{fig:PTMs}c.}
    \label{fig:CTB}
\end{figure}

We refer to this scheme as \textit{correlated-twirl benchmarking}. It uses the same steps as CB, but with different---and to the best of our knowledge, unusual---twirling. That is, for each $i \in \{8, 10, 12, 14 \}$ and $j=i+1$ (indices that label 2-qubit Paulis according to Eq.~\eqref{eq:pauli_order}):
\begin{enumerate}
    \item Prepare an initial state $\rho = \frac{1}{4} \sum_k s_k P_k$ for which $s_i = \tr(P_i \, \rho)$ is as large as possible. E.g., attempt to prepare $\rho=\ket{\psi} \! \bra{\psi}$ where $\ket{\psi}$ is a separable $+1$ eigenstate of $P_i$, so $s_i = 1$ ideally.
    \item Apply $\bar{\mathcal{G}}_\textsc{c}$ then $\bar{\mathcal{G}}_\textsc{a}$, or $\bar{\mathcal{G}}_\textsc{a}$ then $\bar{\mathcal{G}}_\textsc{c}$, each with 50\% probability, where $\bar{\mathcal{G}}_\textsc{c}$ and $\bar{\mathcal{G}}_\textsc{a}$ denote the noisy gate $\mathcal{G}$ twirled over $\mathbb{P}_\textsc{c}$ or $\mathbb{P}_\textsc{a}$ respectively. Repeat this process independently $d/2$ times for varying (even) depths $d$, as shown in Fig.~\ref{fig:learning_sequences}c.
    \item Estimate $\langle P_i \rangle$ for the resulting state as in Eq.~\eqref{eq:mu_hat} using readout twirling, denoting the result $\hat{\mu}$.
\end{enumerate} 
The resulting expectation values decay exponentially with depth, with no oscillations, as in CB. Concretely, the expected value of $\hat{\mu}$ is
\begin{equation}
\mu \; := \; \mathbb{E}(\hat{\mu}) = s_i \, m_i \times (\blue{\boldsymbol{G}}^2_\blue{ii} - \green{\boldsymbol{G}_{ij} \boldsymbol{G}_{ji}} )^{d/2},
\label{eq:exp_decay_correlated}
\end{equation}
where the coefficients $s_i$ and $m_i$ depend on state preparation and measurement errors respectively, as in CB, but not on the noisy gate in question. So by fitting the tuples $(d, \hat{\mu})$ to $d \mapsto A r^{d/2}$, the resulting $r$ gives a SPAM-robust estimate of  $\blue{\boldsymbol{G}}^2_\blue{ii} - \green{\boldsymbol{G}_{ij} \boldsymbol{G}_{ji}}$. Like CB, this scheme is sensitive and it concentrates (see Appendices~\ref{secA:sensitivity} and \ref{secA:concentration}). Finally, by adding the learned value of $\green{\boldsymbol{G}_{ij} \boldsymbol{G}_{ji}}$ from partial-twirl benchmarking, we obtain an estimate of the Type 2 element $\blue{\boldsymbol{G}_{ii}}$. These same steps can be repeated with $i \leftrightarrow j$ to learn $\blue{\boldsymbol{G}_{jj}}$. The procedure is illustrated numerically in Fig.~\ref{fig:CTB} and summarized in Fig.~\ref{fig:learning_sequences}c.

While the three learning schemes we have introduced may seem more complicated than CB, they actually have similar experimental requirements. Not only do they involve the same kinds of circuits (just drawn from different distributions), but they require a similar number of distinct experiments. In particular, a 2-qubit Pauli channel has 15 non-trivial Pauli fidelities\footnote{There are 16 diagonal elements, but the top-left PTM element of a CPTP map always equals 1.}. One might therefore expect that it takes 15 distinct experiments to learn these with CB, where each ``experiment'' consists of estimating some Pauli expectation value $\langle P_i \rangle$ for various depths $d$. However, it is possible to recycle data and use the same measurement outcomes to estimate $\langle ZX \rangle$ and $\langle IX \rangle$ simultaneously, for instance, since $[I,Z]=0$. In fact, 6 different expectation values (for the weight-1 Paulis, namely $IX$, $IY$, \dots, $ZI$) can be found for free in this way, reducing the required number of distinct experiments to 9. The same trick applies to all the schemes we have introduced, which therefore require only 11 distinct experiments in total. Specifically, the Type 1 elements require 5 distinct experiments (modified CB), the Type 3 elements require just 2 experiments (partial-twirl benchmarking), and the Type 2 elements require 4 experiments (correlated-twirl benchmarking). 

\subsubsection{Type 4 Elements}

The only potentially nonzero PTM elements in Eq.~\eqref{eq:G_color} not learned so far are those of Type 4. Unfortunately, we do not yet know of a good way to learn these. The issue is that, like the Type 3 elements, all SPAM-robust learning schemes we have found\footnote{For instance, modified CB followed by correlated-twirl benchmarking.} let us measure products $\gray{\boldsymbol{G}_{ij} \boldsymbol{G}_{ji}}$, rather than isolated elements $\gray{\boldsymbol{G}_{ij}}$ and  $\gray{\boldsymbol{G}_{ji}}$. We suspect this to be a fundamental limitation, analogous to that for Clifford gates \cite{chen:2023}. And since these Type 4 elements should be close to zero for low-noise gates, we expect their products to be extremely small in practice, and therefore very challenging to resolve. For now, we propose to simply bound them using our knowledge of the other PTM elements, by demanding that the learned channel be CPTP \cite{greenbaum:2015}. 

Of course, it is possible to cancel arbitrary noise through Pauli shaping without learning these Type 4 elements, as mentioned earlier, by simply setting the corresponding characteristic matrix elements $\boldsymbol{C}_{ij}$ to be zero. This is what we did in Examples 2 and 3 of Sec.~\ref{sec:examples} after all. However, this is what led to a pathologically large values $\gamma$. If we could learn these Type 4 elements, we could perhaps amplify them correctly and do ZNE, or simply ignore them (and allow $\boldsymbol{C}_{ij} \neq 0$) with confidence, and thereby get a much lower $\gamma$. As it stands, however, these small---but potentially nonzero---elements seem both hard to learn and expensive to mitigate (specifically, to cancel). Such elements are unique to non-Clifford gates like $R_{ZZ}(\theta)$, since any PTM element of a noisy Clifford that should be zero (ideally) can be made zero through twirling. The same is not true of non-Clifford gates, whose noise generally cannot be twirled over the full Pauli group without spoiling the effect of the gate. It is interesting to note that the average fidelity between the ideal gate $\mathcal{U}$ and its noisy realization $\mathcal{G}$ depends on the Type 1--3 elements, but not on the Type 4 elements.

\section{Discussion \& Outlook}
\label{sec:discussion}

This work was motivated by the prospect of using error mitigation to simulate quantum dynamics on pre-fault-tolerant quantum computers in a semi-analog way. More specifically, the Trotter/Floquet circuits arising in quantum simulation can be realized using weakly-entangling gates (e.g., $R_{ZZ}(\theta)$ for small angles $\theta$), which can be performed faster and with higher fidelity in some experiments than can entangling Clifford gates (e.g., CNOTs). However, the current prevailing machinery of error mitigation---including both the noise learning and noise cancellation or amplification components---relies critically on the gate(s) of interest being Clifford, and is therefore incompatible with such non-Clifford, weakly-entangling gates. We have shown how to generalize both components of error mitigation to non-Clifford gates. Specifically, we introduced the framework of \textit{Pauli shaping}, which transforms any quantum channel into almost any other channel in expectation, at the cost of a sampling overhead, and which reduces to earlier methods when applied to Clifford gates. As this technique relies on detailed knowledge of the former channel, we also introduced three schemes to characterize noisy $R_{ZZ}(\theta)$ gates, which are natural in many experiments, in a SPAM-robust way. In doing so, however, we uncovered several new challenges that do not arise with Clifford gates.

Clifford gates have a simple structure by definition, in that they map every Pauli operator (in a density matrix) to some other Pauli, rather than to a mixture thereof. This makes it possible to twirl the noise in an imperfect Clifford gate over the full set of Paulis using only single-qubit gates, leading to relatively simple noise in effect. In this sense, non-Clifford gates have more complicated structures that admit less twirling, so mitigating them presents a trade-off: the associated noise is potentially weaker, but more complex. While this upside is substantial, the cost can be serious, leading to unwanted effects in error mitigation that have no analogue in Clifford gates. The main examples we encountered involve PTM elements of noisy $R_{ZZ}(\theta)$ gates that describe how a Pauli $P_i$ gets mapped to a different one $P_j$, where $[P_i, \, ZZ]=0$ and $P_j \propto P_i \, ZZ$ (e.g., $P_i=ZI$ and $P_j=IZ$). We called these Type 4 elements in Sec.~\ref{sec:learning}. They equal zero in noiseless gates, but can be slightly nonzero in practice due to experimental imperfections. Because they belong to a non-Clifford gate, we know of no good way to eliminate them through twirling. And while it is possible to do so through Pauli shaping, the resulting sampling overhead is impractically large, no matter how weak the noise. This would not be an issue if these Type 4 PTM elements happened to be negligibly small and could simply be ignored. However, they also seem particularly difficult to measure in a SPAM-robust way, making it hard to know precisely how small they are in experiments. There are no such troublesome PTM elements in Clifford gates, because these gates' simpler structure enables more twirling, which leads to simpler noise.

We do not yet know how common such small-but-not-easily-eliminated PTM elements are for other non-Clifford gates, although they appear to be quite generic. However, we can imagine several tentative ways to sidestep them. One way could be at the device physics level, by designing gates whose errors overwhelmingly affect the larger PTM elements. Another would be to synthesize a noiseless non-Clifford gate by Pauli-shaping a probabilistic mixture of Cliffords in which entangling gates rarely arise. For instance, rather than mitigate a physical $R_{ZZ}(\theta)$ gate, one could instead perform $I$, $ZZ$, or a noisy $R_{ZZ}(\pi/2)$ gate (which is Clifford, so the Type 4 elements can be twirled away) with appropriate probabilities, then use Pauli shaping to transform the resulting channel into a noiseless $R_{ZZ}(\theta)$ in the spirit of \cite{campbell:2019, koczor:2023, granet:2023, koczor:2024}. If $\theta$ is small, $R_{ZZ}( \pi/2)$ need only be performed with low probability, so the overall channel would contain little gate noise, and the resulting overhead could be reasonable. Another, more speculative approach, could be to approximately amplify non-Clifford gate noise (for ZNE) without ever learning the troublesome PTM elements. Consider, for example, a noisy $R_{ZZ}(\theta)$ gate twirled over the Paulis that commute with $ZZ$, whose PTM is therefore block-diagonal with $2\times 2$ blocks as in Eq.~\eqref{eq:G_color}. Consider one such block
\begin{equation}
B=
\begin{pmatrix}
\maroon{\boldsymbol{G}_{ii}} & \gray{\boldsymbol{G}_{ij}} \\
\gray{\boldsymbol{G}_{ji}} & \maroon{\boldsymbol{G}_{jj}}
\end{pmatrix},
\end{equation}
where $\maroon{\boldsymbol{G}_{ii}}$ and $\maroon{\boldsymbol{G}_{jj}}$ (called Type 1 elements in Sec.~\ref{sec:learning}) are known, and the Type 4 elements $\gray{\boldsymbol{G}_{ij}}$ and $\gray{\boldsymbol{G}_{ji}}$ are unknown but small. There are several possible notions of noise amplification in such a gate, some of which involve replacing $B$ with $B^{1+\alpha}$ through Pauli shaping, for different noise levels $1+\alpha \ge 1$. To first order in $\gray{\boldsymbol{G}_{ij}}$ and $\gray{\boldsymbol{G}_{ji}}$:
\begin{equation}
B^{1+\alpha}
\approx 
\begin{pmatrix}
\maroon{\boldsymbol{G}}^{\, 1+\alpha}_\maroon{ii} & \eta \, \gray{\boldsymbol{G}_{ij}} \\
\eta \, \gray{\boldsymbol{G}_{ji}} & \maroon{\boldsymbol{G}}^{\, 1+\alpha}_\maroon{jj}
\end{pmatrix}
=
B
\odot 
\overbrace{
\begin{pmatrix}
\maroon{\boldsymbol{G}}^{\, \alpha}_\maroon{ii} & \eta  \\
\eta & \maroon{\boldsymbol{G}}^{\, \alpha}_\maroon{jj}
\end{pmatrix}
}^C
\label{eq:B_1+alpha}
\end{equation}
for
\begin{equation}
\eta = \frac{
\maroon{\boldsymbol{G}}^{\, 1+\alpha}_\maroon{ii} - \maroon{\boldsymbol{G}}^{\, 1+\alpha}_\maroon{jj}
}{
\maroon{\boldsymbol{G}_{ii}} - \maroon{\boldsymbol{G}_{jj}}
}.
\label{eq:eta}
\end{equation}
Since $C$ depends only on $\maroon{\boldsymbol{G}_{ii}}$ and $\maroon{\boldsymbol{G}_{jj}}$, one could amplify the noise, up to an approximation error of order $O(\gray{\boldsymbol{G}}^{\, 2}_\gray{ij} ) + O(\gray{\boldsymbol{G}_{ij}} \gray{\boldsymbol{G}_{ji}}) + O(\gray{\boldsymbol{G}}^{\, 2}_\gray{ji})$, without knowing the exact value of $\gray{\boldsymbol{G}_{ij}}$ or $\gray{\boldsymbol{G}_{ji}}$. Doing so would entail a sampling overhead, although a potentially much smaller one than is required to cancel the noise (see Example 2 of Appendix~\ref{secA:examples}).

An additional practical complication with mitigating non-Clifford gates is that the noise must be learned separately for different angles $\theta$. But since noisy Cliffords on different qubits are already learned separately (because the noise can vary between qubits), and regularly (because of noise drift), the angle dependence does not seem to add much difficulty to any given quantum simulation. However, it makes the first-order Trotter formula preferable to higher-order product formulas \cite{suzuki:1991}, which typically involve multiple different angles for any pair of qubits. 

Similarly, while our mitigation techniques in Sec.~\ref{sec:mitigation} are very general, our noise learning techniques in Sec.~\ref{sec:learning} are deliberately specific to $R_{ZZ}$ gates (or rather $R_{PQ}$ gates, for Paulis $P,Q \in \{X,Y,Z\}$). Other 2-qubit gates, such as weakly-entangling iSWAP-like gates, can have a different structure which admits less twirling, leading to PTMs with larger nonzero blocks than in Eq.~\eqref{eq:G_color}. Whether and how our noise learning techniques generalize to such gates is an important open question to be answered case by case.

Finally, noise channels describing multiple 2-qubit gates performed in parallel have exponentially many parameters in general, for both Clifford and non-Clifford gates. Fortunately, the underlying device physics often allows one to approximate said noise using polynomially many parameters, thereby making the learning and mitigation tractable. The simplest approach is to approximate each noisy gate as being independent of (though not necessarily identical to) the others, meaning the techniques from Sections~\ref{sec:mitigation} and \ref{sec:learning} can be applied to each gate in parallel. In the Clifford case, sparse Pauli-Lindblad models are a generalization that has been empirically successful \cite{vandenberg:2023, kim:2023}. They model the twirled noise as arising from an effective Lindbladian $\mathcal{L}_\text{eff}$ that comprises only geometrically local terms matching the device connectivity (e.g., with 2-body terms only between coupled qubits), and learn its parameters through cycle benchmarking. We expect our learning techniques can be similarly generalized for parallel $R_{ZZ}$ gates twirled over $\mathbb{P}_\textsc{c}$. The main difference would be in allowing $\mathcal{L}_\text{eff}$ to include a slightly broader range of local terms (e.g., $ZZ$ Hamiltonian terms, and certain dissipative cross-terms) because the noise, while simplified from twirling, would remain non-Pauli. This extension of sparse Pauli-Lindblad models to non-Clifford gates seems conceptually simple (though perhaps tedious to work out), and would be a natural direction for subsequent research.

Whether or not non-Clifford error mitigation can outperform the Clifford variety remains to be seen. Ultimately, this comes down to whether reduced noise strength can outweigh increased noise complexity, which in turn, depends on specific techniques to handle this complexity, like those mentioned above. We expect such techniques to be a fruitful area for future research.

\acknowledgements \textit{Acknowledgments.} We wish to thank Lev Bishop, Andrew Eddins, Luke Govia, Seth Merkel, Kristan Temme, and Ewout van den Berg for helpful discussions.

\bibliographystyle{quantum}
\bibliography{references}

\begin{thebibliography}{10}

\bibitem{cai:2022}
Zhenyu Cai, Ryan Babbush, Simon~C. Benjamin, Suguru Endo, William~J. Huggins,
  Ying Li, Jarrod~R. McClean, and Thomas~E. O'Brien.
\newblock ``Quantum error mitigation''.
\newblock \href{https://dx.doi.org/10.1103/RevModPhys.95.045005}{Rev. Mod.
  Phys. {\bf 95}, 045005}~(2023).

\bibitem{temme:2017}
Kristan Temme, Sergey Bravyi, and Jay~M. Gambetta.
\newblock ``Error mitigation for short-depth quantum circuits''.
\newblock \href{https://dx.doi.org/10.1103/PhysRevLett.119.180509}{Phys. Rev.
  Lett. {\bf 119}, 180509}~(2017).

\bibitem{li:2017}
Ying Li and Simon~C. Benjamin.
\newblock ``Efficient variational quantum simulator incorporating active error
  minimization''.
\newblock \href{https://dx.doi.org/10.1103/PhysRevX.7.021050}{Phys. Rev. X {\bf
  7}, 021050}~(2017).

\bibitem{bravyi:2022}
Sergey Bravyi, Oliver Dial, Jay~M. Gambetta, Darío Gil, and Zaira Nazario.
\newblock ``{The future of quantum computing with superconducting qubits}''.
\newblock \href{https://dx.doi.org/10.1063/5.0082975}{Journal of Applied
  Physics {\bf 132}, 160902}~(2022).

\bibitem{lloyd:1996}
Seth Lloyd.
\newblock ``Universal quantum simulators''.
\newblock \href{https://dx.doi.org/10.1126/science.273.5278.1073}{Science {\bf
  273}, 1073--1078}~(1996).

\bibitem{childs:2018}
Andrew~M. Childs, Dmitri Maslov, Yunseong Nam, Neil~J. Ross, and Yuan Su.
\newblock ``Toward the first quantum simulation with quantum speedup''.
\newblock \href{https://dx.doi.org/10.1073/pnas.1801723115}{Proceedings of the
  National Academy of Sciences {\bf 115}, 9456--9461}~(2018).

\bibitem{childs:2021}
Andrew~M. Childs, Yuan Su, Minh~C. Tran, Nathan Wiebe, and Shuchen Zhu.
\newblock ``Theory of {T}rotter error with commutator scaling''.
\newblock \href{https://dx.doi.org/10.1103/PhysRevX.11.011020}{Phys. Rev. X
  {\bf 11}, 011020}~(2021).

\bibitem{clinton:2021}
Laura Clinton, Johannes Bausch, and Toby Cubitt.
\newblock ``Hamiltonian simulation algorithms for near-term quantum hardware''.
\newblock \href{https://dx.doi.org/10.1038/s41467-021-25196-0}{Nature
  Communications {\bf 12}, 4989}~(2021).

\bibitem{vandenberg:2023}
Ewout Van Den~Berg, Zlatko~K Minev, Abhinav Kandala, and Kristan Temme.
\newblock ``Probabilistic error cancellation with sparse {P}auli-{L}indblad
  models on noisy quantum processors''.
\newblock \href{https://dx.doi.org/10.1038/s41567-023-02042-2}{Nature
  Physics}~(2023).

\bibitem{kim:2023}
Youngseok Kim, Andrew Eddins, Sajant Anand, Ken~Xuan Wei, Ewout Van Den~Berg,
  Sami Rosenblatt, Hasan Nayfeh, Yantao Wu, Michael Zaletel, Kristan Temme,
  et~al.
\newblock ``Evidence for the utility of quantum computing before fault
  tolerance''.
\newblock \href{https://dx.doi.org/10.1038/s41586-023-06096-3}{Nature {\bf
  618}, 500--505}~(2023).

\bibitem{earnest:2021}
Nathan Earnest, Caroline Tornow, and Daniel~J. Egger.
\newblock ``Pulse-efficient circuit transpilation for quantum applications on
  cross-resonance-based hardware''.
\newblock \href{https://dx.doi.org/10.1103/PhysRevResearch.3.043088}{Phys. Rev.
  Res. {\bf 3}, 043088}~(2021).

\bibitem{stenger:2021}
John P.~T. Stenger, Nicholas~T. Bronn, Daniel~J. Egger, and David Pekker.
\newblock ``Simulating the dynamics of braiding of {M}ajorana zero modes using
  an {IBM} quantum computer''.
\newblock \href{https://dx.doi.org/10.1103/PhysRevResearch.3.033171}{Phys. Rev.
  Res. {\bf 3}, 033171}~(2021).

\bibitem{merkel:2013}
Seth~T. Merkel, Jay~M. Gambetta, John~A. Smolin, Stefano Poletto, Antonio~D.
  C\'orcoles, Blake~R. Johnson, Colm~A. Ryan, and Matthias Steffen.
\newblock ``Self-consistent quantum process tomography''.
\newblock \href{https://dx.doi.org/10.1103/PhysRevA.87.062119}{Phys. Rev. A
  {\bf 87}, 062119}~(2013).

\bibitem{blume:2013}
Robin Blume-Kohout, John~King Gamble, Erik Nielsen, Jonathan Mizrahi,
  Jonathan~D. Sterk, and Peter Maunz.
\newblock ``Robust, self-consistent, closed-form tomography of quantum logic
  gates on a trapped ion qubit''~(2013).
\newblock  \href{http://arxiv.org/abs/1310.4492}{arXiv:1310.4492}.

\bibitem{nielsen:2021}
Erik Nielsen, John~King Gamble, Kenneth Rudinger, Travis Scholten, Kevin Young,
  and Robin Blume-Kohout.
\newblock ``Gate {S}et {T}omography''.
\newblock \href{https://dx.doi.org/10.22331/q-2021-10-05-557}{{Quantum} {\bf
  5}, 557}~(2021).

\bibitem{endo:2018}
Suguru Endo, Simon~C. Benjamin, and Ying Li.
\newblock ``Practical quantum error mitigation for near-future applications''.
\newblock \href{https://dx.doi.org/10.1103/PhysRevX.8.031027}{Phys. Rev. X {\bf
  8}, 031027}~(2018).

\bibitem{greenbaum:2015}
Daniel Greenbaum.
\newblock ``Introduction to quantum gate set tomography''~(2015).
\newblock  \href{http://arxiv.org/abs/1509.02921}{arXiv:1509.02921}.

\bibitem{flammia:2020}
Steven~T Flammia and Joel~J Wallman.
\newblock ``Efficient estimation of {P}auli channels''.
\newblock \href{https://dx.doi.org/10.1145/3408039}{ACM Transactions on Quantum
  Computing {\bf 1}, 1--32}~(2020).

\bibitem{dur:2005}
W.~D\"ur, M.~Hein, J.~I. Cirac, and H.-J. Briegel.
\newblock ``Standard forms of noisy quantum operations via depolarization''.
\newblock \href{https://dx.doi.org/10.1103/PhysRevA.72.052326}{Phys. Rev. A
  {\bf 72}, 052326}~(2005).

\bibitem{dankert:2009}
Christoph Dankert, Richard Cleve, Joseph Emerson, and Etera Livine.
\newblock ``Exact and approximate unitary 2-designs and their application to
  fidelity estimation''.
\newblock \href{https://dx.doi.org/10.1103/PhysRevA.80.012304}{Phys. Rev. A
  {\bf 80}, 012304}~(2009).

\bibitem{knill:2004}
E.~Knill.
\newblock ``Fault-tolerant postselected quantum computation: Threshold
  analysis''~(2004).
\newblock
  \href{http://arxiv.org/abs/quant-ph/0404104}{arXiv:quant-ph/0404104}.

\bibitem{wallman:2016}
Joel~J. Wallman and Joseph Emerson.
\newblock ``Noise tailoring for scalable quantum computation via randomized
  compiling''.
\newblock \href{https://dx.doi.org/10.1103/PhysRevA.94.052325}{Phys. Rev. A
  {\bf 94}, 052325}~(2016).

\bibitem{santos:2024}
Jader P.~Santos, Ben Bar, and Raam Uzdin.
\newblock ``Pseudo twirling mitigation of coherent errors in non-clifford
  gates''.
\newblock \href{https://dx.doi.org/10.1038/s41534-024-00889-8}{npj Quantum
  Information {\bf 10}, 100}~(2024).

\bibitem{casella:2002}
G.~Casella and R.L. Berger.
\newblock ``Statistical inference''.
\newblock \href{https://dx.doi.org/10.1201/9781003456285}{Duxbury advanced
  series in statistics and decision sciences}. Thomson Learning. ~(2002).

\bibitem{harper:2021}
Robin Harper, Wenjun Yu, and Steven~T. Flammia.
\newblock ``Fast estimation of sparse quantum noise''.
\newblock \href{https://dx.doi.org/10.1103/PRXQuantum.2.010322}{PRX Quantum
  {\bf 2}, 010322}~(2021).

\bibitem{krantz:2019}
P.~Krantz, M.~Kjaergaard, F.~Yan, T.~P. Orlando, S.~Gustavsson, and W.~D.
  Oliver.
\newblock ``{A quantum engineer's guide to superconducting qubits}''.
\newblock \href{https://dx.doi.org/10.1063/1.5089550}{Applied Physics Reviews
  {\bf 6}, 021318}~(2019).

\bibitem{moses:2023}
S.~A. Moses, C.~H. Baldwin, M.~S. Allman, R.~Ancona, L.~Ascarrunz, C.~Barnes,
  J.~Bartolotta, B.~Bjork, P.~Blanchard, M.~Bohn, et~al.
\newblock ``A race-track trapped-ion quantum processor''.
\newblock \href{https://dx.doi.org/10.1103/PhysRevX.13.041052}{Phys. Rev. X
  {\bf 13}, 041052}~(2023).

\bibitem{evered:2023}
Simon~J Evered, Dolev Bluvstein, Marcin Kalinowski, Sepehr Ebadi, Tom Manovitz,
  Hengyun Zhou, Sophie~H Li, Alexandra~A Geim, Tout~T Wang, Nishad Maskara,
  et~al.
\newblock ``High-fidelity parallel entangling gates on a neutral atom quantum
  computer''.
\newblock \href{https://dx.doi.org/10.1038/s41586-023-06481-y}{Nature {\bf
  622}, 268–272}~(2023).

\bibitem{nielsen:2002}
Michael~A Nielsen.
\newblock ``A simple formula for the average gate fidelity of a quantum
  dynamical operation''.
\newblock \href{https://dx.doi.org/10.1016/S0375-9601(02)01272-0}{Physics
  Letters A {\bf 303}, 249--252}~(2002).

\bibitem{wallman:2015}
Joel Wallman, Chris Granade, Robin Harper, and Steven~T Flammia.
\newblock ``Estimating the coherence of noise''.
\newblock \href{https://dx.doi.org/10.1088/1367-2630/17/11/113020}{New Journal
  of Physics {\bf 17}, 113020}~(2015).

\bibitem{bravyi:2021}
Sergey Bravyi, Sarah Sheldon, Abhinav Kandala, David~C. McKay, and Jay~M.
  Gambetta.
\newblock ``Mitigating measurement errors in multiqubit experiments''.
\newblock \href{https://dx.doi.org/10.1103/PhysRevA.103.042605}{Phys. Rev. A
  {\bf 103}, 042605}~(2021).

\bibitem{vandenberg:2022}
Ewout van~den Berg, Zlatko~K. Minev, and Kristan Temme.
\newblock ``Model-free readout-error mitigation for quantum expectation
  values''.
\newblock \href{https://dx.doi.org/10.1103/PhysRevA.105.032620}{Phys. Rev. A
  {\bf 105}, 032620}~(2022).

\bibitem{erhard:2019}
Alexander Erhard, Joel~J Wallman, Lukas Postler, Michael Meth, Roman Stricker,
  Esteban~A Martinez, Philipp Schindler, Thomas Monz, Joseph Emerson, and
  Rainer Blatt.
\newblock ``Characterizing large-scale quantum computers via cycle
  benchmarking''.
\newblock \href{https://dx.doi.org/10.1038/s41467-019-13068-7}{Nature
  Communications {\bf 10}, 5347}~(2019).

\bibitem{chen:2022}
Senrui Chen, Sisi Zhou, Alireza Seif, and Liang Jiang.
\newblock ``Quantum advantages for {P}auli channel estimation''.
\newblock \href{https://dx.doi.org/10.1103/PhysRevA.105.032435}{Phys. Rev. A
  {\bf 105}, 032435}~(2022).

\bibitem{chen:2024}
Sitan Chen, Weiyuan Gong, and Qi~Ye.
\newblock ``Optimal tradeoffs for estimating {P}auli observables''.
\newblock In 2024 IEEE 65th Annual Symposium on Foundations of Computer Science
  (FOCS).
\newblock \href{https://dx.doi.org/10.1109/FOCS61266.2024.00072}{Pages
  1086--1105}.
\newblock ~(2024).

\bibitem{king:2024}
Robbie King, David Gosset, Robin Kothari, and Ryan Babbush.
\newblock ``Triply efficient shadow tomography''.
\newblock \href{https://dx.doi.org/10.1103/PRXQuantum.6.010336}{PRX Quantum
  {\bf 6}, 010336}~(2025).

\bibitem{wallman:2014}
Joel~J Wallman and Steven~T Flammia.
\newblock ``Randomized benchmarking with confidence''.
\newblock \href{https://dx.doi.org/10.1088/1367-2630/16/10/103032}{New Journal
  of Physics {\bf 16}, 103032}~(2014).

\bibitem{helsen:2019b}
Jonas Helsen, Joel~J. Wallman, Steven~T. Flammia, and Stephanie Wehner.
\newblock ``Multiqubit randomized benchmarking using few samples''.
\newblock \href{https://dx.doi.org/10.1103/PhysRevA.100.032304}{Phys. Rev. A
  {\bf 100}, 032304}~(2019).

\bibitem{wack:2021}
Andrew Wack, Hanhee Paik, Ali Javadi-Abhari, Petar Jurcevic, Ismael Faro,
  Jay~M. Gambetta, and Blake~R. Johnson.
\newblock ``Quality, speed, and scale: three key attributes to measure the
  performance of near-term quantum computers''~(2021).
\newblock  \href{http://arxiv.org/abs/2110.14108}{arXiv:2110.14108}.

\bibitem{flammia:2021}
Steven~T. Flammia.
\newblock ``{Averaged Circuit Eigenvalue Sampling}''.
\newblock In 17th Conference on the Theory of Quantum Computation,
  Communication and Cryptography (TQC 2022).
\newblock \href{https://dx.doi.org/10.4230/LIPIcs.TQC.2022.4}{Volume 232 of
  Leibniz International Proceedings in Informatics (LIPIcs), pages 4:1--4:10}.
\newblock Dagstuhl, Germany~(2022). Schloss Dagstuhl -- Leibniz-Zentrum f{\"u}r
  Informatik.

\bibitem{chen:2023}
Senrui Chen, Yunchao Liu, Matthew Otten, Alireza Seif, Bill Fefferman, and
  Liang Jiang.
\newblock ``The learnability of {P}auli noise''.
\newblock \href{https://dx.doi.org/10.1038/s41467-022-35759-4}{Nature
  Communications {\bf 14}, 52}~(2023).

\bibitem{kimmel:2014}
Shelby Kimmel, Marcus~P. da~Silva, Colm~A. Ryan, Blake~R. Johnson, and Thomas
  Ohki.
\newblock ``Robust extraction of tomographic information via randomized
  benchmarking''.
\newblock \href{https://dx.doi.org/10.1103/PhysRevX.4.011050}{Phys. Rev. X {\bf
  4}, 011050}~(2014).

\bibitem{helsen:2019}
Jonas Helsen, Francesco Battistel, and Barbara~M Terhal.
\newblock ``Spectral quantum tomography''.
\newblock \href{https://dx.doi.org/10.1038/s41534-019-0189-0}{npj Quantum
  Information {\bf 5}, 74}~(2019).

\bibitem{campbell:2019}
Earl Campbell.
\newblock ``Random compiler for fast {H}amiltonian simulation''.
\newblock \href{https://dx.doi.org/10.1103/PhysRevLett.123.070503}{Phys. Rev.
  Lett. {\bf 123}, 070503}~(2019).

\bibitem{koczor:2023}
B\'alint Koczor, John J.~L. Morton, and Simon~C. Benjamin.
\newblock ``Probabilistic interpolation of quantum rotation angles''.
\newblock \href{https://dx.doi.org/10.1103/PhysRevLett.132.130602}{Phys. Rev.
  Lett. {\bf 132}, 130602}~(2024).

\bibitem{granet:2023}
Etienne Granet and Henrik Dreyer.
\newblock ``Hamiltonian dynamics on digital quantum computers without
  discretization error''.
\newblock \href{https://dx.doi.org/10.1038/s41534-024-00877-y}{npj Quantum
  Information {\bf 10}, 82}~(2024).

\bibitem{koczor:2024}
B\'alint Koczor.
\newblock ``Sparse probabilistic synthesis of quantum operations''.
\newblock \href{https://dx.doi.org/10.1103/PRXQuantum.5.040352}{PRX Quantum
  {\bf 5}, 040352}~(2024).

\bibitem{suzuki:1991}
Masuo Suzuki.
\newblock ``General theory of fractal path integrals with applications to
  many‐body theories and statistical physics''.
\newblock \href{https://dx.doi.org/10.1063/1.529425}{Journal of Mathematical
  Physics {\bf 32}, 400--407}~(1991).

\end{thebibliography}

\appendix
\addtocontents{toc}{\protect\setcounter{tocdepth}{0}} 

\section{Pauli Shaping Produces the Intended Expectation Values}
\label{secA:exp_vals}

Suppose we wish to apply a channel $\mathcal{A}$ to an initial state $\rho$ on $n$ qubits, then estimate the expectation value
\begin{equation}
\langle O \rangle_\mathcal{A} = \tr \big[O \mathcal{A}(\rho) \big]
\end{equation}
of an observable $O = \sum_\lambda  \lambda \, \ket{\lambda} \! \bra{\lambda}$ for the resulting state. (Any gates preceding or following $\mathcal{A}$ in a quantum circuit can be absorbed into the definition of $\rho$ and $O$ respectively.) Suppose, however, that we can only implement a different channel $\mathcal{G}$ in place of $\mathcal{A}$, e.g., due to experimental imperfections. We will show that through Pauli shaping---that is, by inserting random $n$-qubit Paulis on either side of $\mathcal{G}$ with an appropriate distribution, then scaling the measurement outcomes $\{\lambda\}$---one can recover $\langle O \rangle_\mathcal{A}$ without needing to implement $\mathcal{A}$. The situation is summarized below, with the circuit on the left showing what we would like to implement, and that on the right showing what we will implement instead.
\vspace{-0.5em}
\begin{figure}[h!]
\centering 
\[ \raisebox{-1.75em}{
\Qcircuit @C=1em @R=.7em {
\lstick{\rho}  & \gate{\mathcal{A}} & \meterB{O} & \cw \\ 
\text{\small \hspace{5em} Desired circuit involving $\mathcal{A}$, giving $\langle O \rangle_\mathcal{A}$}
} }
\qquad \qquad \qquad \qquad \qquad \qquad 
\Qcircuit @C=1em @R=.7em {
& & \parbox{3cm}{\scriptsize \centering \vspace{-1ex} perform with \\[-0.2ex] probability $|\boldsymbol{Q}_{ij}|/\gamma$} &  \\
\lstick{\rho} & \gate{P_j} & \gate{\mathcal{G}} & \gate{P_i} & \meterB{O} & \cw & \rstick{ \hspace{-1em} \times \, \gamma \sgn(\boldsymbol{Q}_{ij})} 
{\gategroup{2}{2}{2}{4}{0.5em}{^\}}} \\
 & & & \text{\small \hspace{3em} Pauli shaping circuits involving $\mathcal{G}$, giving $\langle O \rangle_\textsc{ps}$}
} \]
\end{figure}
\vspace{0.5em}

We begin by defining the PTMs $\boldsymbol{A}$ and $\boldsymbol{G}$ of the channels $\mathcal{A}$ and $\mathcal{G}$, respectively, by 
\begin{equation}
\boldsymbol{A}_{ij} = 2^{-n} \tr[P_i \, \mathcal{A}(P_j)]
\qquad \text{and} \qquad 
\boldsymbol{G}_{ij} = 2^{-n} \tr[P_i \, \mathcal{G}(P_j)]
\label{eqA:AG_PTM}
\end{equation}
for $n$-qubit Paulis $P_i$ and $P_j$. We then take the characteristic matrix $\boldsymbol{C}$ to be any $4^n \times 4^n$ real matrix satisfying $\boldsymbol{A} = \boldsymbol{C} \odot \boldsymbol{G}$ as in Eq.~\eqref{eq:pauli_shaping} of the main text, where $\odot$ denotes a Hadamard/element-wise product. Finally, we define the corresponding quasi-probability matrix
\begin{equation}
\boldsymbol{Q} = 2^{-4n} \, \boldsymbol{WCW},
\end{equation}
where $\boldsymbol{W}$ is the Walsh matrix defined in Eq.~\eqref{eq:W} of the main text, as well as the normalizing factor
\begin{equation}
\gamma = \sum_{ij} |\boldsymbol{Q}_{ij}|.
\end{equation}

\textbf{Claim:} By inserting Paulis $P_j$ and $P_i$ before and after $\mathcal{G}$, respectively, with probability $|\boldsymbol{Q}_{ij}|/\gamma$ in each shot independently, and multiplying the measurement outcomes (i.e., the recorded eigenvalues of $O$) by $\gamma \, \sgn(\boldsymbol{Q}_{ij})$, the resulting expectation value $\langle O \rangle_\textsc{ps}$ equals the desired one $\langle O \rangle_\mathcal{A}$.

\textbf{Proof:} We begin by finding $\langle O \rangle_\textsc{ps}$. There are two types of randomness involved in Pauli shaping: that of the measurement outcomes $\{\lambda\}$  for a given quantum circuit (i.e., for fixed Paulis $P_i$ and $P_j$), and that from the choice of circuit. For a fixed circuit in which $\mathcal{G}$ is flanked by Paulis $P_j$ and $P_i$ as shown above, the probability of observing $\lambda$ when measuring $O$ is
\begin{equation}
\text{Pr}(\lambda \,|\, i,j ) = \tr \Big[ \ket{\lambda}\!\bra{\lambda} \,  P_i \, \mathcal{G} (P_j \, \rho P_j ) P_i \Big].
\end{equation}
One then records $\lambda \times \gamma \sgn(\boldsymbol{Q}_{ij})$ in place of $\lambda$, as described above. The probability of running this random circuit in the first place is
\begin{equation}
\text{Pr}(i,j) = |\boldsymbol{Q}_{ij}|/\gamma.
\end{equation}
Therefore, the overall expectation value from Pauli shaping, combining both sources of randomness and invoking the chain rule (for probabilities), is:
\begin{align}
\langle O \rangle_\textsc{ps} 
&=
\sum_{\lambda i j}
\lambda \gamma \sgn(\boldsymbol{Q}_{ij}) \, 
\underbrace{ \text{Pr}(\lambda \,|\, i,j ) \, \text{Pr}(i,j) }_{\text{Pr}(\lambda, \, i, \, j) }
=
\sum_{ij}
 \gamma \sgn(\boldsymbol{Q}_{ij}) \, 
 \tr \big[O P_i \, \mathcal{G} (P_j \,\rho P_j ) P_i \big] \, \frac{|\boldsymbol{Q}_{ij}|}{\gamma} \\
 &=
 \sum_{ij} \boldsymbol{Q}_{ij} \tr \big[O P_i \, \mathcal{G} (P_j \, \rho P_j ) P_i \big]. \nonumber
\end{align}
To show that this expression equals $\langle O \rangle_\mathcal{A}$, we then decompose both $\rho = 2^{-n} \sum_\ell s_\ell P_\ell$ and $O = \sum_k r_k P_k$ in the Pauli basis using appropriate coefficients $\{ s_\ell \}$ and $\{r_k\}$ to get
\begin{align}
\langle O \rangle_\textsc{ps}
&= 
2^{-n} \sum_{ijk\ell}
\boldsymbol{Q}_{ij} \, r_k \, s_\ell \tr \big[ P_k P_i \, \mathcal{G} (P_j P_\ell P_j) P_i \big]
=
2^{-n} \sum_{k \ell} \, r_k \, s_\ell 
\left( \sum_{ij} \boldsymbol{W}_{k i} \boldsymbol{Q}_{ij} \boldsymbol{W}_{j \ell} \right)\tr[P_k \, \mathcal{G}(P_\ell)] \nonumber \\ 
&=
\sum_{k\ell} \, r_k \, s_\ell  \, \boldsymbol{C}_{k\ell}  \, \boldsymbol{G}_{k\ell},
\end{align}
where we've used Eq.~\eqref{eqA:AG_PTM} together with the facts that $P_i P_k P_i = \boldsymbol{W}_{ki} P_k$ and $\boldsymbol{WQW}=\boldsymbol{C}$. Finally, since $\boldsymbol{A}_{k\ell} = \boldsymbol{C}_{k\ell} \, \boldsymbol{G}_{k\ell}$ by definition of $\boldsymbol{C}$, we can use Eq.~\eqref{eqA:AG_PTM} again to conclude that:
\begin{align}
\langle O \rangle_\textsc{ps}
&= 
\sum_{k \ell}   r_k \, s_\ell  \,  \boldsymbol{A}_{k \ell}
=
2^{-n} \sum_{k \ell}   r_k \, s_\ell  \, \tr[P_k \, \mathcal{A}(P_\ell)] 
=
\tr[O \mathcal{A} (\rho)]
=
\langle O \rangle_\mathcal{A}. \quad \square
\end{align}

Of course, the variance of the recorded outcomes is generally larger with Pauli shaping than it would be if we could implement $\mathcal{A}$ rather than $\mathcal{G}$. Given access to $\mathcal{A}$, the variance would be
\begin{equation}
\Delta O^2_\mathcal{A}
=
\tr \big[ O^2 \mathcal{A} (\rho) \big] - \langle O \rangle_\mathcal{A}^2,
\end{equation}
which depends on $\rho$ and $\mathcal{A}$, of course, but can be upper-bounded as 
\begin{align}
\Delta O^2_\mathcal{A}  &\le 
\tr \big[ O^2 \mathcal{A} (\rho) \big]
\le \sum_i \sigma_i(O^2) \, \sigma_i \big[\mathcal{A} (\rho) \big] \\
&\le 
\sigma_1(O^2) \sum_i \sigma_i \big[\mathcal{A} (\rho) \big]
= 
\| O^2 \| \, \tr \big[\mathcal{A} (\rho) \big]
=
\| O \|^2 \nonumber
\end{align}
using the von Neumann trace inequality, where $\sigma_i(M)$ denotes the $i^\text{th}$ largest singular value of a matrix $M$ and $\| \cdot \|$ denotes the operator/spectral/2 norm. (We use the standard notation $\Delta O^2$ to denote the variance in measurement outcomes for a quantum observable $O$. It has no relation to the quantity $\Delta$ in Eq.~\eqref{eq:Var_mu'} of the main text. The letter $\sigma$ is similarly overloaded: its meaning above has no relation to the function in Eq.~\eqref{eq:Clifford_Q}, which describes the action of a generic Clifford gate on Paulis.) Due to the extra randomness inherent in Pauli shaping, its recorded outcomes are generally less concentrated, having a variance of
\begin{align}
\Delta O^2_\textsc{ps}
& =
\sum_{\lambda i j}
\Big[ \lambda \gamma \sgn(\boldsymbol{Q}_{ij}) \Big]^2 \, 
\text{Pr}(\lambda \,|\, i,j ) \, \text{Pr}(i,j) 
-
\langle O \rangle_\textsc{ps}^2 \\
&  = \gamma \sum_{ij} |\boldsymbol{Q}_{ij}| \tr \big[O^2 P_i \, \mathcal{G} (P_j \, \rho P_j ) P_i \big] - \langle O \rangle_\mathcal{A}^2. \nonumber
\end{align}
This quantity also depends on $\rho$, $\mathcal{A}$, and $\mathcal{G}$, but it can be similarly upper-bounded as
\begin{equation}
\Delta O^2_\textsc{ps}
\le 
\gamma \left( \sum_{ij} |\boldsymbol{Q}_{ij}| \right) \max_{k\ell} \tr \big[O^2 P_k \, \mathcal{G} (P_\ell \, \rho P_\ell) P_k \big]
\le
\gamma^2 \| O \|^2.
\end{equation}
Since the standard error of the mean in estimating $\langle O \rangle_\mathcal{A}$ using $N$ shots (i.e., $N$ circuit executions) with access to $\mathcal{A}$ is $\Delta O_\mathcal{A}/\sqrt{N}$, whereas that from Pauli shaping is $\Delta O_\textsc{ps}/\sqrt{N}$, Pauli shaping incurs roughly a $\gamma^2$ sampling overhead for estimating expectation values to within a given statistical error.

\section{Pauli Shaping Reduces to Clifford PEC/ZNE}
\label{secA:clifford_reduction}

Consider a $2^n \times 2^n$ Clifford unitary $U$. Using the same notation as in the main text, we define an invertible function $\sigma : \{0, 4^n-1\} \rightarrow \{0, 4^n-1\}$ such that $P_{\sigma(i)} \propto U^\dag P_i U$ for every $n$-qubit Pauli $P_i$. Moreover, we use the notation $k = i \oplus j$ when $P_k \propto P_i P_j$. We begin with two lemmas about the Walsh matrix elements $\boldsymbol{W}_{ij}$ defined in Eq.~\eqref{eq:W} of the main text.

\noindent \textbf{Lemma B1:} $\boldsymbol{W}_{ij} \boldsymbol{W}_{ik} = \boldsymbol{W}_{i, \, j\oplus k}$.

\noindent \textbf{Proof:} By definition, $P_{j\oplus k} = z \, P_j P_k$ for some $z \in \mathbb{C}$. Then:
\[
\boldsymbol{W}_{i,\, j\oplus k} \; P_{j\oplus k}
=
P_i P_{j\oplus k} P_i
=
z P_i P_j P_k P_i
=
z \, \boldsymbol{W}_{ij} \boldsymbol{W}_{ik} \; P_j (P_i)^2 P_k
=
\boldsymbol{W}_{ij} \boldsymbol{W}_{ik} \; P_{j\oplus k}. \quad \square
\]

\noindent \textbf{Lemma B2:} $\boldsymbol{W}_{ij} = \boldsymbol{W}_{\sigma(i), \, \sigma(j)}$. 

\noindent \textbf{Proof:} By definition, $P_{\sigma(i)} = v_i \, U^\dag P_i U$ and $P_{\sigma(j)} = v_j \, U^\dag P_j U$ for $v_i, v_j \in \{-1,1\}$. Then:
\[
\boldsymbol{W}_{\sigma(i), \, \sigma(j)} \; P_{\sigma(j)}
=
P_{\sigma(i)} P_{\sigma(j)} P_{\sigma(i)}
=
v_i^2 \, v_j \, (U^\dag P_i U) (U^\dag P_j U) (U^\dag P_i U)
=
v_j \, U^\dag P_i P_j P_i U
=
\boldsymbol{W}_{ij} \; P_{\sigma(j)}.
\quad \square
\]

Suppose we implement a channel $\mathcal{G}$ in place of an ideal gate $\mathcal{U}(\rho) = U \rho \, U^\dag$, which we want to transform (in expectation) into $\mathcal{A}$ from Eq.~\eqref{eq:A_clifford} of the main text through Pauli shaping---without invoking the formalism of Clifford PEC/ZNE. That is, we want to realize an aggregate PTM of
\begin{equation}
\boldsymbol{A} = \boldsymbol{U} \bar{\boldsymbol{N}}^{1+\alpha}
\end{equation}
as in Eq.~\eqref{eq:clifford_mitigation}, for some desired $\alpha$, where $\boldsymbol{U}$ is the PTM of $\mathcal{U}$ and $\bar{\boldsymbol{N}} = \text{diag}(\vec{f}\,)$  is the PTM of the twirled noise channel $\bar{\mathcal{N}}$ (which comes from Pauli-twirling $\mathcal{N} = \mathcal{U}^{-1} \mathcal{G}$) with Pauli fidelities $f_i = \tr[P_i \, \mathcal{N}(P_i)]/2^n$. The elements of $\boldsymbol{U}$ are give by
\begin{equation}
\boldsymbol{U}_{ij} = \tr( P_i U P_j U^\dag)/2^n =  v_i \tr( P_{\sigma(i)} P_j )/2^n = v_i \; \delta_{\sigma(i), \, j},
\end{equation}
using the same notation of $P_{\sigma(i)} = v_i \, U^\dag P_i U$ for $v_i = \pm 1$ as in the proof of Lemma B2. The elements of $\boldsymbol{A}$ are therefore
\begin{equation}
\boldsymbol{A}_{ij}
=
\sum_k \boldsymbol{U}_{ik} \left( \bar{\boldsymbol{{N}}}^{1+\alpha} \right)_{kj}
=
\sum_k v_i \; \delta_{\sigma(i), \, k} \; f_k^{1+\alpha} \; \delta_{kj}
=
v_i \, f_j^{1+\alpha} \; \delta_{\sigma(i), \, j}.
\end{equation}
It follows that, for generic noise, we need a characteristic matrix $\boldsymbol{C}$ with elements
\begin{equation}
\boldsymbol{C}_{ij} = f_j^\alpha \; \delta_{\sigma(i), \, j}
\end{equation}
to satisfy Eq.~\eqref{eq:pauli_shaping} of the main text (that is, to get $\boldsymbol{A} = \boldsymbol{C} \odot \boldsymbol{G}$, where $\boldsymbol{G}$ is the PTM of $\mathcal{G}$). The associated quasi-probability matrix $\boldsymbol{Q} = \boldsymbol{WCW}/4^{2n}$ has elements
\begin{align}
\boldsymbol{Q}_{ij}
&= 
4^{-2n} \sum_{k\ell} \boldsymbol{W}_{ik} \boldsymbol{C}_{k\ell} \boldsymbol{W}_{\ell j}
\quad = \quad 
4^{-2n} \sum_{k\ell} \boldsymbol{W}_{ik} \Big( f_\ell^\alpha \;\delta_{\sigma(k), \, \ell} \Big)\boldsymbol{W}_{\ell j}
\quad = \quad 
4^{-2n} \sum_k \boldsymbol{W}_{ik} \boldsymbol{W}_{\sigma(k),\,j} \; f^\alpha_{\sigma(k)} \nonumber\\
& \hspace{-1.2em} \stackrel{\text{\tiny Lemma B2}}{=}
4^{-2n} \sum_k \boldsymbol{W}_{\sigma(i),\, \sigma(k)} \boldsymbol{W}_{\sigma(k),\,j} \; f^\alpha_{\sigma(k)}  \\
& \hspace{-1.2em} \stackrel{\text{\tiny Lemma B1}}{=}
4^{-2n} \sum_k \boldsymbol{W}_{\sigma(i)\oplus j, \, \sigma(k)} \; f^\alpha_{\sigma(k)} \nonumber \\
&=
4^{-2n} \big (\boldsymbol{W} \vec{f}^{\,\alpha} \big)_{\sigma(i) \oplus j}, \nonumber
\end{align}
or, written in terms of the vector of quasi-probabilities $\vec{q}$ from Sec.~\ref{sec:mitigation_clifford} of the main text, defined by $\boldsymbol{W} \vec{q} = \vec{f}^{\, \alpha}$:
\begin{equation}
\boldsymbol{Q}_{ij}
=
4^{-n} \, \vec{q}_{\sigma(i) \oplus j},
\end{equation}
which is precisely Eq.~\eqref{eq:Clifford_Q} from the main text. That is, Step 1 of Clifford PEC/ZNE (from Sec.~\ref{sec:mitigation_clifford}) can be described as inserting Paulis $P_{\sigma(i)}$ and $P_i$ before and after $\mathcal{G}$ respectively, where $P_i \sim \text{unif}(\mathbb{P})$. Step 2 can be described as inserting an extra Pauli $P_k$ before $P_{\sigma(i)}$ with quasi-probability $q_k$. Both steps are shown separately in the left circuit below. Combining the two adjacent Paulis (as in the middle circuit), and relabeling $k \oplus \sigma(i)$ as $j$ (as in the right circuit), we arrive at the previous equation. In other words, Pauli shaping is operationally identical to Clifford PEC/ZNE when the target gate $U$ is Clifford, as claimed in the main text.
\[
\Qcircuit @C=0.6em @R=1em { 
& \gate{P_k} & \qw & \gate{P_{\sigma(i)}} & \gate{\mathcal{G}} & \gate{P_i} & \qw 
& & = & & & \gate{P_{k \oplus \sigma(i)}} & \gate{\mathcal{G}} & \gate{P_i} & \qw 
& & = & & & \gate{P_j} & \gate{\mathcal{G}} & \gate{P_i} & \qw \\
& \parbox{3cm}{\scriptsize \centering \vspace{2ex} quasi-prob. \\[-0.2ex] $=q_k$} & & \text{\hspace{4em} \scriptsize prob.\ $=4^{-n}$} & & & & & & & & \text{\hspace{4em} \scriptsize quasi-prob.\ $=4^{-n} q_k$} & & & & & & & & & \parbox{4cm}{\scriptsize \centering \vspace{2ex} quasi-prob. $=4^{-n} q_{\sigma(i) \oplus j} $ \\[0ex] (setting $j=k\oplus \sigma(i)$)}
\gategroup{1}{2}{1}{2}{.5em}{_\}}
\gategroup{1}{4}{1}{6}{.5em}{_\}}
\gategroup{1}{12}{1}{14}{.5em}{_\}}
\gategroup{1}{20}{1}{22}{.5em}{_\}}
}
\]

\section{Mathematical Details of Pauli Shaping Examples}
\label{secA:examples}

Before getting into the details of the Pauli shaping examples from Sec.~\ref{sec:examples} of the main text, we begin this section with a useful lemma. Consider two characteristic matrices $\boldsymbol{C}^{(1)}$ and $\boldsymbol{C}^{(2)}$, which correspond respectively to quasi-probability matrices
\begin{equation}
\boldsymbol{Q}^{(1)} = 2^{-4n} \, \boldsymbol{WC}^{(1)} \boldsymbol{W}
\qquad \qquad 
\boldsymbol{Q}^{(2)} = 2^{-4n} \, \boldsymbol{WC}^{(2)} \boldsymbol{W}.
\end{equation}
Suppose we apply Pauli shaping to a channel $\mathcal{G}$ (with PTM $\boldsymbol{G}$) according to $\boldsymbol{C}^{(1)}$ to create an aggregate channel with PTM $\boldsymbol{C}^{(1)} \odot \boldsymbol{G}$ at the cost of a sampling overhead $\gamma_1^2$, where $\gamma_1 = \sum_{ij}|\boldsymbol{Q}_{ij}^{(1)}|$. Then, treating this aggregate channel as a black box, suppose we do a second, outer layer of Pauli shaping according to $\boldsymbol{C}^{(2)}$, thus producing an aggregate channel with PTM
\begin{equation}
\boldsymbol{C}^{(2)} \odot \big( \boldsymbol{C}^{(1)} \odot \boldsymbol{G} \big)
=
\big( \boldsymbol{C}^{(2)} \odot \boldsymbol{C}^{(1)} \big) \odot \boldsymbol{G},
\end{equation}
at the cost of a total sampling overhead $(\gamma_1 \gamma_2)^2$, where $\gamma_2 = \sum_{ij}|\boldsymbol{Q}_{ij}^{(2)}|$. Alternatively, we could realize the same aggregate channel by applying a single layer of Pauli shaping with characteristic matrix $\boldsymbol{C} = \boldsymbol{C}^{(2)} \odot \boldsymbol{C}^{(1)}$, with a corresponding quasi-probability matrix $\boldsymbol{Q} = \boldsymbol{WCW}/4^{2n}$, incurring a potentially different sampling overhead of $\gamma^2$ where $\gamma = \sum_{ij} |\boldsymbol{Q}_{ij}|$.

\noindent \textbf{Lemma C1:} $\gamma \le \gamma_1 \gamma_2$.

\noindent \textbf{Proof:} We begin by rewriting the relation $\boldsymbol{C} = \boldsymbol{C}^{(2)} \odot \boldsymbol{C}^{(1)}$ in terms of the quasi-probability matrices associated with each characteristic matrix:
\begin{equation}
\boldsymbol{WQW} = \big( \boldsymbol{WQ}^{(2)}\boldsymbol{W} \big)
\odot
\big( \boldsymbol{WQ}^{(1)}\boldsymbol{W} \big),
\end{equation}
or equivalently:
\begin{equation}
\boldsymbol{Q} = 4^{-2n} \, \boldsymbol{W} \Big[ \big( \boldsymbol{WQ}^{(2)}\boldsymbol{W} \big)
\odot
\big( \boldsymbol{WQ}^{(1)}\boldsymbol{W} \big) \Big] \boldsymbol{W}.
\end{equation}
The elements of $\boldsymbol{Q}$ can therefore be expressed as
\begin{align}
\boldsymbol{Q}_{ij}
& =
4^{-2n}
\sum_{abcdk\ell}
\boldsymbol{W}_{ia} 
\Big[
\big( \boldsymbol{W}_{ak} \boldsymbol{Q}_{k \ell}^{(2)} \boldsymbol{W}_{\ell b} \big)
\big( \boldsymbol{W}_{ac} \boldsymbol{Q}_{cd}^{(1)} \boldsymbol{W}_{db} \big)
\Big]
\boldsymbol{W}_{bj} \nonumber \\
& \hspace{-1.2em} \stackrel{\text{\tiny Lemma B1}}{=} \;
4^{-2n}
\sum_{abcdk\ell} \big( \boldsymbol{W}_{i\oplus k, \, a} \boldsymbol{W}_{ac} \big)
\big( \boldsymbol{W}_{j\oplus \ell, \,b} \boldsymbol{W}_{bd} \big)
\boldsymbol{Q}_{k \ell}^{(2)} \, \boldsymbol{Q}_{cd}^{(1)} \\
&= 
\sum_{cdk\ell} \delta_{i \oplus k, \, c} \; \delta_{j\oplus \ell,\,d} \; \boldsymbol{Q}_{k \ell}^{(2)}\, \boldsymbol{Q}_{cd}^{(1)} \nonumber\\
&=
\sum_{k\ell}  \boldsymbol{Q}_{k \ell}^{(2)} \, \boldsymbol{Q}_{i\oplus k, \, j \oplus \ell}^{(1)}. \nonumber
\end{align}
In other words, $\boldsymbol{Q}$ is a convolution of $\boldsymbol{Q}^{(1)}$ and $\boldsymbol{Q}^{(2)}$. It follows immediately that
\begin{align}
\gamma = \sum_{ij} \left|\boldsymbol{Q}_{ij} \right|
= 
\sum_{ij} 
\left| 
\sum_{k\ell}  \boldsymbol{Q}_{k \ell}^{(2)} \, \boldsymbol{Q}_{i\oplus k, \, j \oplus \ell}^{(1)}
\right|
\le 
\sum_{k\ell}
\left| \boldsymbol{Q}_{k\ell}^{(2)} \right| 
\sum_{ij}  \left| \boldsymbol{Q}_{i\oplus k, \, j \oplus \ell }^{(1)} \right| 
=
\sum_{k\ell}
\left| \boldsymbol{Q}_{k\ell}^{(2)} \right|  \gamma_1
= \gamma_1 \gamma_2.  \quad \square
\end{align}

An important consequence of this lemma is that, in Examples 1 and 2 from Sec.~\ref{sec:examples} of the main text, we only need to consider block-diagonal characteristic matrices. More precisely, suppose we want to transform a noisy gate $\mathcal{G}$ into an a channel $\mathcal{A}$ through Pauli shaping, and that we do so using a generic characteristic matrix $\boldsymbol{C}^{(2)}$ such that $\boldsymbol{A} =\boldsymbol{C}^{(2)} \odot \boldsymbol{G}$. Now consider the operation of twirling $\mathcal{G}$ over the 8 Paulis that commute with $ZZ$. In the ordered basis \eqref{eq:pauli_order} from the main text, this twirling is described by
\begin{equation}
\boldsymbol{Q}^{(1)} = \diag \left( \frac{1}{8}, \, \frac{1}{8}, \, \frac{1}{8}, \,
 \frac{1}{8}, \, \frac{1}{8}, \, \frac{1}{8}, \,\frac{1}{8}, \, \frac{1}{8}, \, 
 0, \,  0, \,  0, \,  0, \,  0, \,  0, \,  0, \,  0
\right),
\end{equation}
whose corresponding characteristic matrix is block-diagonal with identical $2\times 2$ blocks:
\begin{equation}
\boldsymbol{C}^{(1)}
=
\boldsymbol{WQ}^{(1)}\boldsymbol{W}
=
\begin{pmatrix}
1 & 1 & & & & & 0 \\
1 & 1 \\
& & 1 & 1 \\ 
& & 1 & 1 \\ 
& & & & \ddots \\
& & & & & 1 & 1 \\ 
0 & & & & & 1 & 1
\end{pmatrix}.
\end{equation}
In both examples, $\boldsymbol{G}$ is block-diagonal with $2\times 2$ blocks, so $\boldsymbol{G} = \boldsymbol{C}^{(1)} \odot \boldsymbol{G}$. Therefore, 
\begin{equation}
\boldsymbol{A}
=
\boldsymbol{C}^{(2)} \odot \boldsymbol{G}
=
\boldsymbol{C}^{(2)} \odot \big( \boldsymbol{C}^{(1)} \odot \boldsymbol{G} \big)
= 
\boldsymbol{C} \odot \boldsymbol{G},
\end{equation}
where $\boldsymbol{C} = \boldsymbol{C}^{(2)} \odot \boldsymbol{C}^{(1)}$ is also block-diagonal with $2\times 2$ blocks, and $\gamma \le \gamma_1 \gamma_2$ from Lemma C1. Therefore, in both examples from Sec.~\ref{sec:examples} of the main text, the optimal (i.e., lowest possible) sampling overhead is achieved by a block-diagonal characteristic matrix with $2 \times 2$ blocks.

\subsection*{Example 1}

\begin{figure}
\centering
\includegraphics[]{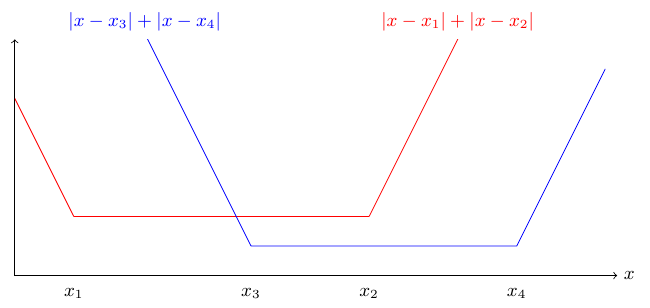}
\caption{The components of Eq.~\eqref{eq:gamma_coherent_reduced} grouped into two functions which sum to $4 \gamma$.}
\label{fig:gamma_coherent}
\end{figure}

To correct a coherent over- or under-rotation by $\epsilon$, as in the first example, we consider a block-diagonal characteristic matrix $\boldsymbol{C}$ given by Eqs.~\eqref{eq:T_coherent}--\eqref{eq:Ta_coherent} of the main text. The bottom four blocks are fixed by Eq.~\eqref{eq:pauli_shaping}, whereas we choose the top four to be identical in order to simplify the calculations. The associated $\gamma = \frac{1}{256}\sum_{ij}|(\boldsymbol{WCW})_{ij}|$ is given by
\begin{equation}
8 \gamma =
8 \big|x-y \big| + 
\big| 2 + x + y + 2 c + 2s \big| + 
\big| 2 + x + y - 2 c - 2s \big| + 
\big| 2 - x - y - 2 c + 2s \big| + 
\big| 2 - x - y + 2 c - 2s \big|,
\label{eq:gamma_coherent_full}
\end{equation}
where $x$ and $y$ are the free coefficients introduced in the main text, and
\begin{equation}
c = \frac{\cos (\theta)}{\cos (\theta + \epsilon)}
\qquad \qquad 
s = \frac{\sin (\theta)}{\sin (\theta + \epsilon)}.
\end{equation}
By inspection, Eq.~\eqref{eq:gamma_coherent_full} is minimized when $x=y$, which we will consider from now on. Notice also that Eq.~\eqref{eq:gamma_coherent_full} is invariant under the transformations $(c,s) \mapsto (s,c)$ and $(c,s) \mapsto (-c, -s)$. We can simplify the equation by removing these symmetries and expressing $\gamma$ as
\begin{equation}
4 \gamma = 
\big | x  - \underbrace{( -1 - M - m )}_{x_1} \big | + 
\big |x - \underbrace{(-1 + M + m )}_{x_2} \big | + 
\big | x - \underbrace{( 1  - M + m )}_{x_3} \big | + 
\big | x - \underbrace{( 1  + M - m  )}_{x_4} \big |
\label{eq:gamma_coherent_reduced}
\end{equation}
for 
\begin{equation}
M = \max\big \{ |c|, |s| \big \}
\qquad \qquad 
m = \text{sgn}(cs) \, \min\big \{ |c|, |s| \big \}.
\end{equation}
Note that 
\begin{equation}
\sin^2 (\theta) - \sin^2 (\theta + \epsilon )
=
- \big[ 
\cos^2 (\theta) - \cos^2 (\theta + \epsilon )
\big]
\end{equation}
for any $\theta$ and $\epsilon$, and that the sign of the left (and right) hand side indicates whether $|s|$ (and $|c|$) is greater or smaller than 1. It follows that $M \ge 1$ and $|m| \le 1$, which implies $x_1 \le x_3 \le x_2 \le x_4$. It is helpful to visualize $\gamma$ as a sum of two functions, one comprising the first two terms in Eq.~\eqref{eq:gamma_coherent_reduced} and one comprising the last two, as shown in Fig.~\ref{fig:gamma_coherent}. It is then clear, geometrically, that the minimum value of $\gamma$ is achieved at
\begin{equation}
    x = \frac{1}{2} (x_2 + x_3) = m,
\end{equation}
which gives $\gamma = M$, as in Eq.~\eqref{eq:gamma_overrot} of the main text.

\subsection*{Example 2}

The second example of Pauli shaping from Sec.~\ref{sec:examples} of the main text can be derived from a Lindblad equation. A general Lindblad equation on $n$ qubits can be written as
\begin{equation}
\frac{d \rho}{dt}
=
\mathcal{L}(\rho)
=
-i [H, \rho]
+
\sum_{j,k=1}^{4^n-1} \Gamma_{jk} \Big(
P_j \rho P_k - \frac{1}{2} \big \{ P_k P_j, \, \rho \big \} \Big),
\label{eqA:lindblad}
\end{equation}
where $H=H^\dag$ is a Hamiltonian, $\{ P_j \}$ are $n$-qubit Paulis (excluding the identity), and $\Gamma = (\Gamma_{jk})_{j,k=1}^{4^n-1} = \Gamma^\dag$ is a positive semi-definite matrix. When all terms are time-independent, the solution to Eq.~\eqref{eqA:lindblad} for an initial state $\rho$ is 
\begin{equation}
\rho' = e^{\mathcal{L} t}(\rho);
\end{equation}
in other words, it is described by the channel $\mathcal{G} = e^{\mathcal{L} t}$. To simplify the notation, we will absorb $t$ into the definition of $\mathcal{L}$, i.e., set $t=1$ and consider $\mathcal{G} = e^\mathcal{L}$. The second example in the main text comes from picking $H=\frac{\theta}{2} ZZ$, which describes the effect of the ideal gate, together with a nonzero dissipative term with a coefficient matrix
\begin{equation}
\Gamma 
=
-\frac{1}{2}\ln(1- 2\epsilon)
\left( 
\begin{array}{*{11}c}
0 &&&&&&&&&& 0 \\[-1ex]
& \, \ddots \\[-0.5ex]
&& 0 \\
&&& 1 & i \\
&&& -i & 1 \\
&&&&& 1 & i \\
&&&&& -i & 1 \\
&&&&&&& 1 & i \\
&&&&&&& -i & 1 \\
&&&&&&&&& 1 & i \\
0 &&&&&&&&& -i & 1 \\
\end{array}
\right),
\end{equation}
expressed in the ordered basis \eqref{eq:pauli_order} from the main text. Computing the PTM of $\mathcal{L}$ and taking its matrix exponential then gives Eq.~\eqref{eq:bad_PTM}. We can cancel this noise through Pauli shaping using the characteristic matrix $\boldsymbol{C}$ from Eq.~\eqref{eq:T_bad} of the main text, which corresponds to a $\gamma$ of
\begin{align}
\gamma 
= &  
\frac{1}{16} \left| 1-x+ \frac{1}{1-2\epsilon} + \frac{6}{1-\epsilon} \right| \nonumber \\
& +
\frac{1}{32} \left( \left| 1+x + \frac{1}{1-2\epsilon} - \frac{16}{\sqrt{1-2\epsilon}} + \frac{6}{1-\epsilon} \right| 
+
\left| 1+x+\frac{1}{1-2\epsilon} + \frac{16}{\sqrt{1-2 \epsilon}} + \frac{6}{1-\epsilon} \right| \right)  \\
& +
\frac{1}{4} \left( \left|1-x + \frac{1}{2\epsilon-1} \right| 
+
\left| 1+x + \frac{1}{2\epsilon-1} \right| \right)
+
\frac{3}{16} \left( \left|x  - \frac{2 \epsilon^2}{1-3 \epsilon + 2 \epsilon^2}\right|
+
\left|x  + \frac{2 \epsilon^2}{1-3 \epsilon + 2 \epsilon^2}\right| \right),  \nonumber
\end{align}
where $x$ is a free parameter that we can choose so as to minimize $\gamma$. Taking the $\epsilon \rightarrow 0$ limit of this expression gives Eq.~\eqref{eq:gamma_limit} of the main text, which is lower-bounded by 3/2.

It is possible, however, to approximately amplify the noise at much lower cost. Specifically, notice that we can factor the PTM $\boldsymbol{G}$ of the noisy gate from Eq.~\eqref{eq:bad_PTM} as either $\boldsymbol{G}=\boldsymbol{UN}$ or $\boldsymbol{G}=\boldsymbol{NU}$ here, for the same noise channel
\begin{equation}
\resizebox{0.92\textwidth}{!}{$
\boldsymbol{N}
= \!
\begin{pmatrix}
\!\begin{pmatrix}
1 & 0 \\
2 \epsilon & 1 - 2\epsilon
\end{pmatrix} &&&&&&& 0\\ 
& \!\!\! \begin{pmatrix}
1 - \epsilon & -\epsilon \\
-\epsilon & 1- \epsilon
\end{pmatrix} \\
&& \!\!\! \begin{pmatrix}
1 - \epsilon & \epsilon \\
\epsilon & 1- \epsilon
\end{pmatrix} \\
&&& \!\!\! \begin{pmatrix}
1 - \epsilon & \epsilon \\
\epsilon & 1- \epsilon
\end{pmatrix}\\ 
&&&& \sqrt{1\!-\!2\epsilon} \, I\\
&&&&& \sqrt{1\!-\!2\epsilon} \, I \\
&&&&&& \sqrt{1\!-\!2\epsilon} \, I \\ 
0 &&&&&&& \sqrt{1\!-\!2\epsilon} \, I
\end{pmatrix}
$}
\end{equation}
either way, where $I$ is a $2\times 2$ matrix and $\boldsymbol{U}$ is the PTM of the ideal $R_{ZZ}(\theta)$ gate from Eq.~\eqref{eq:U_PTM}. There are several reasonable notions of what it means to amplify the noise in $\boldsymbol{G}$ by a factor of $1+\alpha$, as discussed in the main text, but one is to use Pauli shaping to replace $\boldsymbol{G}$ with 

\begin{equation}
\resizebox{0.93\textwidth}{!}{$
\boldsymbol{UN}^{1+\alpha} = \boldsymbol{N}^{1+\alpha} \, \boldsymbol{U}
= \!
\begin{pmatrix}
\!\begin{pmatrix}
1 & 0 \\
2 \epsilon & 1 - 2\epsilon
\end{pmatrix}^{1+\alpha} \!\!\!\!\!\!\!\!\!\! &&&&&& 0\\ 
& \!\!\! \begin{pmatrix}
1 - \epsilon & -\epsilon \\
-\epsilon & 1- \epsilon
\end{pmatrix}^{1+\alpha} \!\!\!\!\!\!\!\!\!\! \\
&& \!\!\! \begin{pmatrix}
1 - \epsilon & \epsilon \\
\epsilon & 1- \epsilon
\end{pmatrix}^{1+\alpha} \!\!\!\!\!\!\!\!\!\! \\
&&& \!\!\! \begin{pmatrix}
1 - \epsilon & \epsilon \\
\epsilon & 1- \epsilon
\end{pmatrix}^{1+\alpha} \!\!\!\!\!\!\! \\ 
&&&& (1\!-\!2\epsilon)^{\frac{1+\alpha}{2}} \; R^{(2)} \\
&&&&& \ddots \\
0 &&&&&& (1\!-\!2\epsilon)^{\frac{1+\alpha}{2}} \; R^{(2)}  \\ 
\end{pmatrix}.
$}
\end{equation}

Consider a $2 \times 2$ block from the top-left of $\boldsymbol{G}$:
\begin{equation}
B = 
\begin{pmatrix}
\boldsymbol{G}_{ii} & \boldsymbol{G}_{ij} \\
\boldsymbol{G}_{ji} & \boldsymbol{G}_{jj}
\end{pmatrix},
\label{eqA:example2PEA}
\end{equation}
where $i \in \{0,2,4,6\}$ and $j=i+1$. In order to Pauli-shape $\boldsymbol{G}$ into $\boldsymbol{UN}^{1+\alpha}$, we need to effectively replace each such block $B$ by $B^{1+\alpha}$. In general, if the off-diagonal elements of $B$ are small (as they are in this example) we can hope to instead approximate $B^{1+\alpha}$ with a simple expression that differs from the exact $B^{1+\alpha}$ by errors of order $O(\boldsymbol{G}_{ij}^2)$, $O(\boldsymbol{G}_{ji}^2)$ and $O(\boldsymbol{G}_{ij} \boldsymbol{G}_{ji})$. To keep track of error orders, we will rewrite the off-diagonal terms of $B$ as $\boldsymbol{G}_{ij} = \zeta \boldsymbol{G}_{ij}'$ and $\boldsymbol{G}_{ji} = \zeta \boldsymbol{G}_{ji}'$ for some small bookkeeping parameter $\zeta$ that we will use for series expansions. To (quickly) approximate $B^{1+\alpha}$ to first order in $O(\zeta)$, we will use the standard identity 
\begin{equation}
\frac{d}{d\zeta} \, e^{F(\zeta)}
=
\int_0^1 e^{z F(\zeta)} \, F'(\zeta) \, e^{(1-z) F(\zeta)} \, dz
\label{eqA:exp_derivative}
\end{equation}
for the derivative of a matrix exponential, where $F$ is a suitably well-behaved matrix-valued function. Using Eq.~\eqref{eqA:exp_derivative}, one can formally show that
\begin{equation}
B = 
\underbrace{\begin{pmatrix}
\boldsymbol{G}_{ii} & 0 \\
0 & \boldsymbol{G}_{jj}  
\end{pmatrix}}_{B_D}
+ \zeta
\underbrace{\begin{pmatrix}
0 & \boldsymbol{G}_{ij}' \\
\boldsymbol{G}_{ji}' & 0 
\end{pmatrix}}_{B_A}
=
e^{F(\zeta)}
\end{equation}
for some function
\begin{equation}
F(\zeta)
=
\ln(B_D) + \zeta  \left[ \frac{\ln(\boldsymbol{G}_{ii}/\boldsymbol{G}_{jj})}{\boldsymbol{G}_{ii} - \boldsymbol{G}_{jj}} \right] B_A  +  O(\zeta^2),
\end{equation}
assuming $\boldsymbol{G}_{ii}, \boldsymbol{G}_{jj} > 0$. Then, using Eq.~\eqref{eqA:exp_derivative} again, we can formally expand $B^{1+\alpha} = e^{(1+\alpha)F(\zeta)}$ in powers of $\zeta$ as
\begin{align}
B^{1+\alpha}
& =
e^{(1+\alpha)F(0)} + \zeta \frac{d}{d\zeta} \Big|_{\zeta=0} \, e^{(1+\alpha) F(\zeta)} + O(\zeta^2) \nonumber  \\ 
& =
B_D^{1+\alpha}
+ 
\zeta \left[ \frac{ (1+\alpha) \ln(\boldsymbol{G}_{ii}/\boldsymbol{G}_{jj})}{\boldsymbol{G}_{ii} - \boldsymbol{G}_{jj}} \right] \int_0^1 B_D^{z(1+\alpha)} B_A B_D^{(1-z)(1+\alpha)} dz + O(\zeta^2) \\
& =
B_D^{1+\alpha} + \zeta \, \eta  B_A + O(\zeta^2) \nonumber  \\ 
&=
\begin{pmatrix}
 \boldsymbol{G}_{ii}^{1+\alpha} & \eta \, \boldsymbol{G}_{ij} \nonumber \\    
 \eta \, \boldsymbol{G}_{ji} &  \boldsymbol{G}_{jj}^{1+\alpha}
\end{pmatrix}
+ 
O \big(\boldsymbol{G}_{ij}^2 \big) + O\big(\boldsymbol{G}_{ij} \boldsymbol{G}_{ji}\big) + O\big(\boldsymbol{G}_{ji}^2\big) \nonumber
\end{align}
for $\eta$ in Eq.~\eqref{eq:eta} of the main text. (Note that $\eta \rightarrow (1+\alpha) \boldsymbol{G}_{ii}^\alpha$ in the limit of $\boldsymbol{G}_{jj} \rightarrow \boldsymbol{G}_{ii}$.) Therefore, to first order in the off-diagonal elements, $B^{1+\alpha}$ can be written as the Hadamard product of $B$ with a matrix that only depends on the diagonal terms of $B$, as in Eq.~\eqref{eq:B_1+alpha}.

This means that by Pauli-shaping $\boldsymbol{G}$ with a characteristic matrix of
\begin{equation}
\boldsymbol{C}
=
\text{diag} (C_1, \, C_2,\,  C_2,\,  C_2,\,  C_3,  \, C_3,  \, C_3,\,   C_3 ),
\end{equation}
where 
\begin{equation}
C_1 = 
\begin{pmatrix}
1 & x \\
1 + \alpha & 1 - 2 \alpha \epsilon
\end{pmatrix}
\qquad \qquad 
C_2 = 
\begin{pmatrix}
1-\alpha \epsilon & 1 + \alpha \\ 1 + \alpha  & 1 - \alpha \epsilon 
\end{pmatrix}
\qquad \qquad 
C_3 = (1-\alpha \epsilon) 
\begin{pmatrix}
1 & 1 \\ 1 & 1
\end{pmatrix}
\end{equation}
for any $x$ of our choice, we get an aggregate channel of $ \boldsymbol{A} = \boldsymbol{C} \odot \boldsymbol{G} = \boldsymbol{UN}^{1+\alpha} + O(\epsilon^2)$. In other words, using a characteristic matrix $\boldsymbol{C}$ that does not require knowledge of the off-diagonal elements $\boldsymbol{G}_{ij}$ in the top-left of $\boldsymbol{G}$ (which are called Type 4 elements in Sec.~\ref{sec:learning} of the main text), we can effectively implement a channel that is very close to the desired $\boldsymbol{UN}^{1+\alpha} = \boldsymbol{N}^{1+\alpha} \, \boldsymbol{U}$ when the gate noise is weak. (That is, we constructed $C_1$ and $C_2$ without needing to know the values of the Type 4 elements.) Unlike in Clifford ZNE, this noise amplification still comes at a sampling overhead, albeit a much smaller one than is required to cancel the noise. Specifically, picking $x = 1+\alpha$ for convenience in $C_1$ leads to
\begin{equation}
\gamma = \frac{1}{256} \sum_{ij} \big| (\boldsymbol{WCW})_{ij} \big|
=
\frac{(3+7 \epsilon)|\alpha|}{4}
+
\frac{1}{4} \big| 4 + \alpha ( 1-3\epsilon  ) \big|,
\end{equation}
which reduces to $\gamma = 1 + \alpha (1+\epsilon)$ for $\alpha > 0$ and small $\epsilon$. That is, we can approximately amplify weak noise by a factor of $1+\alpha$ at the cost of a sampling overhead that is approximately linear in $\alpha$ and that vanishes as $\alpha \rightarrow 0$.

This example is unusual in that the associated noise channel does not depend on the factorization order: if we write $\mathcal{G} = \mathcal{U N} = \mathcal{N}' \mathcal{U}$ as in the main text, then $\mathcal{N} = \mathcal{N}'$. This is because the noise contributes only an overall damping factor to Paulis that anti-commute with $ZZ$ (i.e., in the bottom-right of $\boldsymbol{G}$). While this will not generally be the case in experiments, the trick above with $B^{1+\alpha}$ nonetheless applies more broadly. To see how, let $\mathcal{G}$ denote an $R_{ZZ}(\theta)$ gate with arbitrary noise. If we twirl $\mathcal{G}$ over $\mathbb{P}_\textsc{c}$, the set of Paulis that commute with $ZZ$, the resulting PTM $\bar{\boldsymbol{G}}_\textsc{c}$ will be $2\times 2$ block-diagonal, as in Eq.~\eqref{eq:G_color}. Because the ideal PTM $\boldsymbol{U}$ also has this block structure, so does the PTM of the resulting (partially twirled) noise channel $\bar{\boldsymbol{N}}_\textsc{c} = \boldsymbol{U}^{-1} \bar{\boldsymbol{G}}_\textsc{c}$. And since the top-left blocks of $\boldsymbol{U}$ are identity matrices, the top-left blocks of the noise-amplified PTM $\bar{\boldsymbol{N}}_\textsc{c}^{1+\alpha} \, \boldsymbol{U}$ are just those of $\bar{\boldsymbol{G}}$ raised to the power of $\alpha$, as in the example above. This may make it possible to approximately amplify generic noise on $R_{ZZ}(\theta)$ gates without learning the troublesome Type 4 PTM elements. In general, of course, the factorizations $\bar{\mathcal{G}}_\textsc{c} = \mathcal{U} \bar{\mathcal{N}}_\textsc{c} = \bar{\mathcal{N}}_\textsc{c}' \, \mathcal{U}$ will lead to different noise channels $\bar{\mathcal{N}}_\textsc{c} \neq \bar{\mathcal{N}}_\textsc{c}'$, and it is not clear at present which one---if either---provides a useful notion of noise amplification.

\section{Readout Twirling Details}
\label{secA:readout_twirling}

Consider estimating the expectation value $\langle P_i \rangle = \tr(\rho P_i)$ of an $n$-qubit Pauli $P_i$ with respect to a state $\rho$, subject to measurement/readout errors. In this section, we show that readout twirling \cite{vandenberg:2022} gives an estimate that is proportional to the ground truth $\langle P_i \rangle$, on average, with a proportionality coefficient that depends on the statistics of the readout errors, but not on $\rho$.

We begin by defining some notation. Let 
\begin{equation}
W = \begin{pmatrix}
1 & 1 \\ 
1 & -1
\end{pmatrix} ^{\otimes n}
\end{equation}
be a $2^n$-dimensional Walsh-Hadamard matrix, then $W^\top = W$ and $W^2 = 2^n I$. Also, for any $j \in \{0,1\}^n$ let 
\begin{equation}
\mathcal{X}_j = X^{j_1} \otimes \cdots \otimes X^{j_n}
\qquad \qquad 
\mathcal{Z}_j = Z^{j_1} \otimes \cdots \otimes Z^{j_n}
\end{equation}
be $n$-qubit Paulis comprising $X$'s and $I$'s, or $Z$'s and $I$'s, respectively, as labeled by $j$. Define the vector $\vec{z}_j \in \mathbb{R}^{2^n}$ such that $\mathcal{Z}_j = \diag(\vec{z}_j)$, then evidently $\vec{z}_j = W \ket{j}$, where $\ket{j} = \ket{j_1} \otimes \cdots \otimes \ket{j_n}$ is a standard basis vector. Also, note that $\mathcal{X}_j = 2^{-n} W \mathcal{Z}_j W$. Finally, define the vectors $\vec{p}, \vec{v} \in \mathbb{R}^{2^n}$ of ideal probabilities and ideal expectation values, respectively, with elements
\begin{equation}
p_j = \bra{j} \rho \ket{j} 
\qquad \qquad 
v_j = \tr(\mathcal{Z}_j \, \rho) = \langle \mathcal{Z}_j \rangle.
\end{equation}
Then $\vec{v} = W \vec{p}$, since
\begin{equation}
\sum_k W_{jk} \, p_k
=
\sum_k W_{jk} \tr \big( \ket{k} \! \bra{k} \rho \big)
=
\tr \Big[ 
\Big(
\sum_k W_{jk} \ket{k} \! \bra{k}
\Big) \rho
\Big]
=
\tr \Big[ 
\diag \!  \big ( W\ket{j} \big) \, \rho
\Big]=
\tr(\mathcal{Z}_j \, \rho) = v_j.
\end{equation}

The typical way to estimate $\langle P_i \rangle$ is to apply single-qubit gates to rotate $P_i$ into a matrix that is diagonal in the computational basis, then measure each qubit in this same basis (i.e., the $Z$ eigenbasis). Since we are modeling single-qubit gates as being noiseless, it therefore suffices to consider a diagonal $P_i$ from the start, i.e., we can take $P_i = \mathcal{Z}_j$ for some $j \in \{0,1\}^n$ in this analysis. Each measurement returns a random $n$-bit string $k \in \{0,1\}^n$ distributed according to $\vec{p}$, in the absence of readout error. From each measured $k$,  we can record $\bra{k} \mathcal{Z}_j \ket{k} = \pm 1$, the corresponding eigenvalue of $\mathcal{Z}_j$. If we perform $N_\text{tot}$ such measurements on identical copies of $\rho$, and estimate $\langle Z_j \rangle$ by $\hat{\mu}$ defined in Eq.~\eqref{eq:mu_hat} of the main text, then
\begin{equation}
\mathbb{E}(\hat{\mu})
=
\frac{N_\text{tot} \, \text{Pr}(+1) - N_\text{tot} \, \text{Pr}(-1)}{N_\text{tot}}
=
\sum_k W_{jk} \, p_k 
=
(W  \vec{p})_j
=
v_j
=
\langle \mathcal{Z}_j \rangle
\end{equation}
as expected. That is, without readout errors, $\hat{\mu}$ is an unbiased estimate of $\langle \mathcal{Z}_j \rangle$.

Readout errors can be described by a stochastic matrix $A$ whose elements $A_{\ell k}$ give the probability that a $k$ measurement outcome gets misreported as an $\ell$, for any $k,\ell \in \{0,1\}^n$ \cite{bravyi:2021}. Under such errors, the distribution of measured bit-strings becomes $\vec{p}\,' = A\vec{p}$ (according to the law of total probability), rather than $\vec{p}$. Therefore:
\begin{equation}
\mathbb{E}(\hat{\mu})
=
(W  \vec{p} \,' )_j
=
(W A \vec{p} \, )_j
=
2^{-n} (WAW \vec{v} \,)_j
=
2^{-n} \sum_\ell (WAW)_{j \ell}  \, \langle \mathcal{Z}_\ell \rangle,
\end{equation}
which generally does not equal $\langle \mathcal{Z}_j \rangle$, nor does it have any simple functional relation to it. 

Readout twirling implements a random $\mathcal{X}_m$ operation both before and after the noisy readout, for $m$ drawn uniformly from $\{0,1\}^n$ in each shot. In effect, this replaces $A$ with 
\begin{align}
A'
&= 
2^{-n} \sum_m \mathcal{X}_m A \mathcal{X}_m
=
2^{-3n} \sum_m (W \mathcal{Z}_m W) A (W \mathcal{Z}_m W)
=
2^{-3n} W \Big[ \Big(
\sum_m \vec{z}_m \vec{z}_m^\top \Big) \odot 
(WAW)
\Big] W \\ 
&= 
2^{-2n} W \big[ I \odot (WAW) \big] W, \nonumber 
\end{align}
where $\odot$ denotes a Hadamard/element-wise product. To derive this expression, we used the identity 
\begin{equation}
\diag(\vec{x}) \, M \diag(\vec{x}) = (\vec{x} \vec{x}^\top) \odot M
\end{equation}
for any matrix $M$ and vector $\vec{x}$ of compatible sizes, and the fact that 
\begin{equation}
\sum_m \vec{z}_m \vec{z}_m^\top
=
\sum_m W \ket{m} \! \bra{m} W^\top
=
2^n I,
\end{equation}
where $\{\ket{m}\}$ are standard basis vectors. This means that with noisy, twirled readout, the distribution of measured bit-strings is $\vec{p}\, '' = A' \vec{p}$. The average value of $\hat{\mu}$ therefore becomes
\begin{equation}
\mathbb{E}(\hat{\mu})
=
(W \vec{p} \, '')_j
=
2^{-n} (WA'W \vec{v})_j
=
2^{-n} \Big( \big[ I \odot (WAW) \big] \vec{v} \Big)_j
=
m_j \langle \mathcal{Z}_j \rangle,
\end{equation}
where $m_j = 2^{-n} (WAW)_{jj}$ is independent of the measured state $\rho$, as claimed in the main text.

\section{Sensitivity of Learning Schemes}
\label{secA:sensitivity}

Suppose we want to estimate the expectation value of a Pauli observable $P_m$ with respect to a state $\rho$ by repeating many identical measurements in the eigenbasis of $P_m$ on copies of $\rho$. Let the random variable $\mathsf{Y} = \pm 1$ denote the outcome of one such measurement, where the value of $\mathsf{Y}$ indicates the observed eigenvalue. Following the notation from the main text, we will denote the expectation value of $\mathsf{Y}$ by $\mu = \mathbb{E}(\mathsf{Y})$. In the absence of readout error, $\mu$ is equal to the ideal Pauli expectation value of $\langle P_m \rangle = \tr(P_m \, \rho)$, but with readout error, $\mu$ need not equal $\langle P_m \rangle$ in general, as explained in Appendix~\ref{secA:readout_twirling}.

The probability mass function of $\mathsf{Y}$ is 
\begin{equation}
p(y) =
\text{Pr}(\mathsf{Y}=y) = 
\begin{cases}
 (1+\mu)/2, & y=1 \\ 
 (1-\mu)/2, & y=-1.
\end{cases}
\end{equation}
Suppose the state $\rho$ is produced by a noise learning scheme, and that $\mu$ depends on some parameter $\phi$ that we wish to learn. The corresponding (classical) Fisher information is
\begin{equation}
\mathcal{I}(\phi) = \mathbb{E} \left[ 
\left( 
\frac{\partial}{\partial \phi} 
\ln p(\mathsf{Y})
\right)^2
\right]
=
\sum_{y \in \{1, -1\}}
p(y) \left(
\frac{y}{2 p(y)} \frac{\partial \mu}{\partial \phi}
\right)^2
=
\frac{1}{1-\mu^2} \left(  \frac{\partial \mu}{\partial \phi} \right)^2,
\label{eqA:general_FI}
\end{equation}
as in Eq.~\eqref{eq:fisher_info} of the main text. Now consider the case where $\mu$ decays exponentially with circuit depth $d$, as in cycle benchmarking (CB), modified CB, and correlated-twirl benchmarking (see Eqs.~\eqref{eq:exp_decay} and \eqref{eq:exp_decay_correlated}). That is, suppose 
\begin{equation}
\mu = A r^d,
\end{equation}
where we wish to learn the decay rate $0 < r \le 1$ (so we will write $r$ instead of the general placeholder $\phi$ in the equations below), and where the amplitude $A$ depends on SPAM errors. The corresponding Fisher information is
\begin{equation}
\mathcal{I}(r)
=
\frac{\mu^2}{1-\mu^2} \frac{d^2}{r^2}
=
\frac{A^2 \, r^{2d}}{1-A^2 \, r^{2d}} \frac{d^2}{r^2},
\label{eqA:CB_FI}
\end{equation}
as in Eq.~\eqref{eq:fisher_info}. Since we aim to characterize the Fisher information in the noiseless limit, we begin by taking $A \rightarrow 1$ in the expression above, which corresponds to the limit of ideal state preparation and measurement. The result still depends on the circuit depth $d$. To find the maximum Fisher information, we will optimize over $d$, treating it as a continuous parameter for the moment and demanding that $\frac{\partial}{\partial d} \mathcal{I}(r) \big|_{A=1} = 0$. We know of no closed-form expression for this optimum, so it is convenient to parameterize the Fisher information in terms of  $x = 2d \ln(1/r)$ instead:
\begin{equation}
\mathcal{I}(r) \big|_{A=1}
=
\frac{1}{r^{-2d}-1} \frac{d^2}{r^2}
=
\frac{1}{r^2 \ln(r)^2} \underbrace{ \frac{x^2}{4(e^x-1)}}_{g(x)} ,
\label{eq:FI_g(x)}
\end{equation}
as in Eq.~\eqref{eq:fisher_info_max} of the main text. This way all dependence on $d$ is captured by $g(x)$, plotted in Fig.~\ref{fig:g(x)}, which can be shown numerically to achieve a maximum value of $g(x_\star) \approx 0.162$ at $x_\star \approx 1.59$. This means the maximum Fisher information is approximately
\begin{equation}
\mathcal{I}(r) \big|_{ \! \! \substack{A=1\\ \;\; d=d_\star}} \approx \frac{0.162}{r^2 \ln(r)^2}
\label{eqA:max_I(r)}
\end{equation}
at  $d_\star \approx \frac{1.59}{2 \ln(1/r)}$, as in Eq.~\eqref{eq:fisher_info_max}. Of course, $d_\star$ is not generally an integer, so the true maximum Fisher information can be slightly lower due to rounding. (The impact of this rounding on $\mathcal{I}(r)$ is negligible for $r \approx 1$ since $\frac{\partial x}{\partial d} \approx 0$ in this regime, but it can be substantial when $r$ is small. Eq.~\eqref{eqA:max_I(r)} should therefore not be used in the latter regime.) If $r \rightarrow 1$ in the noiseless limit then $\mathcal{I}(r) \big|_{A=1, \, d=d_\star} \rightarrow \infty$, meaning that the measurement outcomes can be informative about the value of $r$, so relatively few shots are needed. This is the case in Clifford CB, modified CB (for Type 1 elements) and correlated-twirl benchmarking (since $\blue{\boldsymbol{G}}^2_\blue{ii} - \green{\boldsymbol{G}_{ij} \boldsymbol{G}_{ji}} \rightarrow \sin(\theta)^2 + \cos(\theta)^2 =1$ in the noiseless limit). However, if we were to use modified CB to learn the Type 2 elements, we would have $r \rightarrow \cos(\theta)$ in the noiseless limit, leading to a much lower Fisher information when $\theta \not \approx 0$, and therefore requiring substantially more shots. As a result, we recommend using partial-twirl benchmarking and correlated-twirl benchmarking to learn the Type 2 elements instead, in general.

\begin{figure}
\centering
\begin{tikzpicture}[domain=0.01:8, xscale=0.7, yscale=15]

  \draw[->] (0, 0) -- (8.2, 0) node[right] {$x$};
  \draw[->] (0, 0) -- (0, 0.19) node[above] {$g(x)$};
  \draw[color=blue, samples=1000]    plot (\x, {\x^2 / ( 4*( exp(\x) - 1))} );
  \draw[-, densely dotted] (1.59, 0) -- (1.59, 0.16);
  \draw[-, densely dotted] (0, 0.162) -- (1.59, 0.162);

  \node [] at (0, -0.015) {\footnotesize 0};
  \node [] at (1.59, -0.015) {\footnotesize $x_\star \approx 1.59$};
  \node [] at (4, -0.015) {\footnotesize 4};
  \node [] at (6, -0.015) {\footnotesize 6};
  \node [] at (8, -0.015) {\footnotesize 8};

  \node [] at (-0.5, 0.00) {\footnotesize 0};
  \node [] at (-0.5, 0.05) {\footnotesize 0.05};
  \node [] at (-0.5, 0.10) {\footnotesize 0.10};
  \node [] at (-0.8, 0.162) {\footnotesize $\approx 0.162$};
\end{tikzpicture}
\caption{The function $g(x)$ from Eq.~\eqref{eq:FI_g(x)}.}
\label{fig:g(x)}
\end{figure}

Finally, consider the case where 
\begin{equation}
\mu = A r^d \cos(\omega d - \delta),
\label{eqA:oscillating_decay}
\end{equation}
as in partial-twirl benchmarking (see Eq.~\eqref{eq:oscillating_decay} of the main text). Before proceeding, we note that for any $x \in (0,1)$ and $\varphi \in \mathbb{R}$,
\begin{equation}
\frac{\cos(\varphi)^2}{1 - x \cos(\varphi)^2}
\le 
\frac{1}{1-x}
\qquad \qquad 
\frac{\sin(\varphi)^2}{1 - x \cos(\varphi)^2}
\le 
1.
\label{eqA:cos_sin_bounds}
\end{equation}
The Fisher information for the decay rate $r$ from Eq.~\eqref{eqA:oscillating_decay} is
\begin{equation}
\mathcal{I}(r)
=
\left( \frac{Ar^d d}{r} \right)^2
\frac{\cos(\omega d - \delta)^2}{1 - A^2 r^{2d} \cos(\omega d - \delta)^2 },
\end{equation}
which follows from plugging Eq.~\eqref{eqA:oscillating_decay} into Eq.~\eqref{eqA:general_FI}. It is difficult to maximize this expression over $d$ in general, since it can behave differently depending on how the oscillations line up with the exponential decay. (E.g., the optimal depth $d_\star$ used in Eq.~\eqref{eqA:max_I(r)} could happen to give $\cos(\omega d_\star - \delta)=0$.) In the noiseless limit, however, the exponential decay will be slow compared to the oscillations, which motivates the upper bound
\begin{equation}
\mathcal{I}(r) \le 
\left( \frac{Ar^d d}{r} \right)^2
\max_\varphi 
\frac{\cos(\varphi)^2}{1 - A^2 r^{2d} \cos(\varphi)^2 }.
\end{equation}
We can bound this expression further using Eq.~\eqref{eqA:cos_sin_bounds} provided $A r^d < 1$, which is guaranteed whenever 
\begin{equation}
d > d_0 := -\frac{\ln(A)}{\ln(r)}.
\end{equation}
Consider a $2\times 2$ PTM block $B$ (from the diagonal of $\boldsymbol{G}_\textsc{c}$) which deviates from the ideal expression in Eq.~\eqref{eq:R2} of the main text by an arbitrary perturbation:
\begin{equation}
B = \begin{pmatrix}
    \cos\theta  & -\sin\theta \\ 
    \sin\theta & \cos\theta 
\end{pmatrix}
+ 
\varepsilon 
\begin{pmatrix}
    x_{00} & x_{01} \\ 
    x_{10} & x_{11}
\end{pmatrix},
\end{equation}
for some small $\varepsilon>0$, then one can show that $d_0 = O(\varepsilon)$ through Eq.~\eqref{eq:params_to_PTM}. In other words, we can use Eq.~\eqref{eqA:cos_sin_bounds} for any non-trivial depth $d > d_0 \rightarrow 0$ in the noiseless limit to find
\begin{equation}
\mathcal{I}(r) \le \left( \frac{Ar^d d}{r} \right)^2 \frac{1}{1 - A^2 r^{2d}},
\end{equation}
which is identical to Eq.~\eqref{eqA:CB_FI}, and therefore also diverges at $d_\star$ as $r \rightarrow 1$. Similarly, now using $\omega$ in place of the generic parameter $\phi$:
\begin{equation}
\mathcal{I}(\omega)
=
\big( Ar^d d \big)^2
\frac{\sin(\omega d - \delta)^2}{1 - A^2 r^{2d} \cos(\omega d - \delta)^2 }
\le 
\big( Ar^d d \big)^2
\le \frac{A^2}{e^2 \ln(r)^2}
\end{equation}
in the noiseless limit, which also diverges as $r \rightarrow 1$. In other words, the measurement outcomes from partial-twirl benchmarking are informative about both $r$ and $\omega$. The phase $\delta$, however, is harder to learn---while $\partial \mu/\partial r$ and $\partial \mu/\partial \omega$ both pick up a factor of $d$, $\partial \mu/\partial \delta$ does not. More precisely, the second inequality in \eqref{eqA:cos_sin_bounds} gives
\begin{equation}
\mathcal{I}(\delta)
=
\big( Ar^d \big)^2
\frac{\sin(\omega d - \delta)^2}{1 - A^2 r^{2d} \cos(\omega d - \delta)^2 }
\le 
\big( Ar^d \big)^2
\le A^2
\end{equation}
in the noiseless limit, which is much smaller than the maximum values of both $\mathcal{I}(r)$ and $\mathcal{I}(\omega)$.

\section{Concentration in Learning Schemes}
\label{secA:concentration}

Suppose we would like to execute random quantum circuits from some particular family (e.g., defined by one of the learning or mitigation schemes discussed in the main text) then estimate some Pauli expectation value $\langle P_m \rangle$. Ideally, we would pick a new random circuit for every shot and record $\pm 1$ based on the observed eigenvalue of $P_m$, repeating this process $N_\text{tot}$ times and averaging the results. (We denote this average as $\hat{\mu}$ in the main text.) However, it is convenient in many experiments to instead pick a small number $N_\text{c}$ of random circuits and to run each one $N_\text{s/c}$ times, leading to a total number of shots $N_\text{tot} = N_\text{c} N_\text{s/c}$, as described in the main text.

Consider a random quantum channel $\mathcal{T}$ describing a random (potentially noisy) circuit which is chosen from some prescribed family of circuits with probability $\Pr(\mathcal{T})$. Let the random variable $\mathsf{Y}(\mathcal{T}) = \pm 1$ denote a measurement outcome (more precisely, the observed eigenvalue of $P_m$) from running $\mathcal{T}$. Define the conditional expectation
\begin{equation}
\mu(\mathcal{T})
=
\mathbb{E} \big[ \mathsf{Y}(\mathcal{T}) \big| \mathcal{T} \big],
\end{equation}
which is the expected value of the measurement outcomes from a given circuit $\mathcal{T}$ (i.e., the average outcome in the limit of infinitely many executions of $\mathcal{T}$). Since $\mathcal{T}$ is random, $\mu(\mathcal{T})$ is also a random variable, with some mean
\begin{equation}
\mu 
=
\mathbb{E} \big[ \mu(\mathcal{T}) \big]
\end{equation}
and some variance $\Delta^2$, as illustrated in Fig.~\ref{fig:T_variance} of the main text. We are interested in the case where there are $N_\text{c}$ different random circuits, all drawn independently from the same distribution, and each one is executed $N_\text{s/c}$ times. Let the random variable $\mathcal{T}_j$ denote the $j^\text{th}$ random circuit, and the random variable $\textsf{Y}_i(\mathcal{T}_j) = \pm 1$ denote the measurement outcome from the $i^\text{th}$ time circuit $\mathcal{T}_j$ is run, for $1 \le j \le N_\text{c}$ and $1 \le i \le N_\text{s/c}$. As described in the main text, we will use the random variable
\begin{equation}
\hat{\mu}'
=
\frac{1}{N_\text{tot}} \sum_{i=1}^{N_\text{s/c}} \sum_{j=1}^{N_\text{c}} \mathsf{Y}_i (\mathcal{T}_j)
\end{equation}
as an estimator for $\mu$. It is unbiased, since
\begin{equation}
\mathbb{E}(\hat{\mu}')
=
\frac{1}{N_\text{tot}} \sum_{i=1}^{N_\text{s/c}} \sum_{j=1}^{N_\text{c}}
\mathbb{E} \Big\{ 
\mathbb{E} \big[ \mathsf{Y}_i (\mathcal{T}_j) \big| \mathcal{T}_j \big]
\Big\}
=
\frac{1}{N_\text{tot}} \sum_{i=1}^{N_\text{s/c}} \sum_{j=1}^{N_\text{c}}
\mathbb{E} \big[
\mu(\mathcal{T})
\big]
=
\mu
\end{equation}
using the law of total expectation (we follow the standard notation in which the inner $\mathbb{E}$ is over $\mathsf{Y}_i$ for fixed $\mathcal{T}_j$, and the outer one is over $\mathcal{T}_j$), and it has a variance of
\begin{equation}
\Var(\hat{\mu}')
=
\mathbb{E}(\hat{\mu}'^2)
-
\mathbb{E}(\hat{\mu}')^2
=
\mathbb{E}(\hat{\mu}'^2)
-
\mu^2.
\end{equation}
There are three distinct types of terms in the sum
\begin{equation}
\mathbb{E}(\hat{\mu}'^2)
=
\frac{1}{N_\text{tot}^2}
\sum_{i=1}^{N_\text{s/c}} \sum_{j=1}^{N_\text{c}} 
\sum_{k=1}^{N_\text{s/c}} \sum_{\ell=1}^{N_\text{c}} 
\mathbb{E} \Big[
\mathsf{Y}_i (\mathcal{T}_j) \mathsf{Y}_k (\mathcal{T}_\ell) \Big],
\label{eq:E_mu_hat'}
\end{equation}
which we will analyze separately. 
\begin{description}
\item[Same shot from the same circuit ($\boldsymbol{i=k}$ and $\boldsymbol{j = \ell}$)] There are $N_\text{tot}$ such terms in Eq.~\eqref{eq:E_mu_hat'}, for which
\begin{equation}
\mathbb{E} \Big[ \mathsf{Y}_i (\mathcal{T}_j) \mathsf{Y}_i (\mathcal{T}_j) \Big]
=
\mathbb{E} \big[ (\pm 1)^2 \big]  = 1.
\end{equation}
\item[Different circuits ($\boldsymbol{j \neq \ell}$)] There are $N_\text{c}( N_\text{c}-1) N_\text{s/c}^2$ such terms, for which $\mathsf{Y}_i (\mathcal{T}_j)$ and $\mathsf{Y}_k (\mathcal{T}_\ell)$ are independent since they come from different random circuits $\mathcal{T}_j$ and $\mathcal{T}_\ell$, which are themselves independent:
\begin{equation}
\mathbb{E} \Big[ \mathsf{Y}_i (\mathcal{T}_j) \mathsf{Y}_k (\mathcal{T}_\ell) \Big]
=
\mathbb{E} \Big\{ 
\mathbb{E} \big[  \mathsf{Y}_i (\mathcal{T}_j) \mathsf{Y}_k (\mathcal{T}_\ell) \big| \mathcal{T}_j, \mathcal{T}_\ell \big ]
\Big \}
=
\mathbb{E} \big[ 
\mu(\mathcal{T}_j) \mu(\mathcal{T}_\ell)
\big]
=
\mu^2.
\end{equation}
\item[Different shots from the same circuit ($\boldsymbol{i \neq k}$ and $\boldsymbol{j = \ell}$)] There are $ N_\text{s/c} (N_\text{s/c}-1) N_\text{c}$ such terms, which are conditionally independent given $\mathcal{T}_j$:
\begin{equation}
\mathbb{E} \Big[ \mathsf{Y}_i (\mathcal{T}_j) \mathsf{Y}_k (\mathcal{T}_j) \Big]
=
\mathbb{E} \Big\{ 
\mathbb{E} \big[  \mathsf{Y}_i (\mathcal{T}_j) \mathsf{Y}_k (\mathcal{T}_j) \big| \mathcal{T}_j \big ]
\Big \}
=
\mathbb{E} \big[ \mu(\mathcal{T}_j)^2 \big]
=
\mu^2 + \Delta^2.
\end{equation}
\end{description}
Therefore
\begin{align}
\mathbb{E}(\hat{\mu}'^2)
& =
\frac{1}{N_\text{tot}^2}
\Big[
N_\text{tot} 
+ 
N_\text{c} (N_\text{c}-1) N_\text{s/c}^2 \, \mu^2
+
N_\text{s/c} (N_\text{s/c}-1) N_\text{c} \, (\mu^2 + \Delta^2)
\Big] \\
&=
\mu^2 + \frac{1-\mu^2}{N_\text{tot}} +
\left( \frac{N_\text{s/c}-1}{N_\text{s/c}} \right)
\frac{\Delta^2}{N_\text{c}},
\end{align}
so the standard error $\sqrt{\Var(\hat{\mu}')}$ is given by Eq.~\eqref{eq:Var_mu'} of the main text, which reduces to Eq.~\eqref{eq:Var_mu} when $N_\text{s/c}=1$ (in which case we denote the estimator as $\hat{\mu}$ rather than $\hat{\mu}'$).

We now examine the quantity $\Delta$ for the different learning schemes discussed in the main text, which are meant to extract certain elements of a PTM $\boldsymbol{G}$ describing a noisy $R_{ZZ}(\theta)$ gate. In general, $\Delta$ will depend on the very elements of $\boldsymbol{G}$ we wish to measure, so its exact value cannot be known \textit{a priori}. Instead, we approximate it by calculating $\Delta$ in the noiseless limit (where it is tractable) for various learning schemes, as discussed in the main text. We use the expression for $\Delta$ in Eq.~\eqref{eq:circ_variance} as a starting point. In general, the random channel $\mathcal{T}$ in that equation should describe not just the quantum gates being characterized and the random single-qubit gates surrounding them, but also any state-prep twirling, readout error, and readout twirling. (Readout error can be described by a stochastic matrix $A$ that acts after an ideal measurement, as in Appendix~\ref{secA:readout_twirling}, or equivalently here, as a quantum channel that acts before an ideal measurement, as in Fig.~\ref{fig:T_variance}.) However, these latter elements vanish in the noiseless limit, so we need only consider $\mathcal{T}$ comprising repeated ideal $R_{ZZ}(\theta)$ gates (described by a unitary channel $\mathcal{U}$ with PTM $\boldsymbol{U}$ from Eqs.~\eqref{eq:U_PTM} and \eqref{eq:R2}) surrounded by random Pauli gates.

We begin with modified cycle benchmarking, where $\mathcal{U}$ is repeated $d$ times, and each occurrence is twirled independently over the full set of 2-qubit Paulis. This means
\begin{equation}
\mathcal{T} = \mathcal{T}^{(d)} \mathcal{T}^{(d-1)} \cdots \mathcal{T}^{(1)}
\end{equation}
for
\begin{equation}
\mathcal{T}^{(k)}(\rho) = P_\ell \, \mathcal{U} (P_\ell \rho P_\ell ) P_\ell
=
\begin{cases}
\, \, \mathcal{U}(\rho) = U \rho \, U^\dag, & [P_\ell, \, Z \!\otimes \! Z]=0 \\
\mathcal{U}^\dag(\rho) = U^\dag \rho \, U,& \{ P_\ell, \, Z \!\otimes \! Z \}=0,
\end{cases}
\label{eqA:T_layer}
\end{equation}
where $P_\ell \sim \text{unif}(\mathbb{P})$ is sampled independently in each layer $k$. Therefore, when estimating the expectation value of some Pauli $P_i \in \mathbb{P}$:
\begin{equation}
\Delta^2 
= 
\mathbb{E} \big \{ \tr[ P_i \mathcal{T}(\rho)]^2 \big \} - \mu^2
=
\tr \Big[ 
P_i^{\otimes 2} \,
\mathbb{E} \big( \mathcal{T}^{\otimes 2} \big)
( \rho^{\otimes 2} )
\Big] - \mu^2
=
\tr \Big \{
P_i^{\otimes 2} \,
\Big[
\mathbb{E} \big( \mathcal{T}^{(k) \, \otimes 2} \big) 
\Big]^d
( \rho^{\otimes 2} )
\Big \} - \mu^2,
\label{eqA:Delta_sq_tr}
\end{equation}
where we used the identity $\tr(M)^2 = \tr(M \otimes M)$ for any square matrix $M$ to bring the expectation into the trace, then the fact that each layer is twirled independently to distribute the overall expectation into each layer. Since $P_\ell$ in Eq.~\eqref{eqA:T_layer} commutes or anti-commutes with $Z \! \otimes \! Z$ with equal probabilities,
\begin{equation}
\mathbb{E} \big( \mathcal{T}^{(k) \, \otimes 2} \big) 
=
\frac{1}{2} \, \mathcal{U}^{\otimes 2}
+
\frac{1}{2} \, \mathcal{U}^{\dag \, \otimes 2} 
=:
\mathcal{V}.
\end{equation}
So for a generic initial state $\rho = \frac{1}{4} \sum_k s_k P_k$,
\begin{equation}
\Delta^2 
=
\tr \big[ P_i^{\otimes 2} \, (\mathcal{V}^d) (\rho) \big] - \mu^2
=
 \sum_{k \ell} s_k s_\ell 
 \underbrace{ 
 \frac{1}{16} \tr \big[ P_i^{\otimes 2} \, \mathcal{V}^d (P_k \! \otimes P_\ell ) \big]}_{(\boldsymbol{V}^d)_{i \! i \!, k \! \ell}}
 - \mu^2,
\end{equation}
where $(\boldsymbol{V}^d)_{ii , k \ell}$ are the PTM elements of $\mathcal{V}^d$. Moreover, because $\text{span}\{P_i, P_j\}$ is an invariant subspace of $\mathcal{U}$ for $P_j \propto (Z \! \otimes \! Z) P_i$, $\text{span}(\mathbb{S})$ is invariant under $\mathcal{V}$ for
\begin{equation}
\mathbb{S}
=
\big ( P_i \otimes P_i, \; P_i \otimes P_j, \; P_j \otimes P_i, \; P_j \otimes P_j  \big ).
\end{equation}
In other words, because the PTM of $\mathcal{U}$ is block-diagonal with $2\times 2$ blocks, the PTM of $\mathcal{V}$ is block-diagonal with $4\times 4$ blocks (in an appropriate ordered basis). We can therefore evaluate $\Delta$ by calculating one of these blocks, denoted $\boldsymbol{V}_{\!\mathbb{S}}$, and taking the $d^\text{th}$ power of it. Concretely, 
\begin{align}
\boldsymbol{V}_{\!\mathbb{S}}
&=
\left(
\frac{1}{16}
\tr \big[ S_k \mathcal{V} (S_\ell) \big] 
\right)_{1 \le k,\ell \le 4}
=
\frac{1}{2} 
\begin{pmatrix}
\boldsymbol{U}_{ii} & \boldsymbol{U}_{ij} \\ 
\boldsymbol{U}_{ji} & \boldsymbol{U}_{jj}
\end{pmatrix}^{\otimes 2} 
+ 
\frac{1}{2} 
\begin{pmatrix}
\boldsymbol{U}_{ii} & -\boldsymbol{U}_{ij} \\ 
-\boldsymbol{U}_{ji} & \boldsymbol{U}_{jj}
\end{pmatrix}^{\otimes 2}  \\
& =
\begin{pmatrix}
\boldsymbol{U}_{ii}^2 & 0 & 0 & \boldsymbol{U}_{ij}^2 \\ 
0 & \boldsymbol{U}_{ii} \boldsymbol{U}_{jj} & \boldsymbol{U}_{ij} \boldsymbol{U}_{ji} & 0 \\ 
0 & \boldsymbol{U}_{ij} \boldsymbol{U}_{ji} & \boldsymbol{U}_{ii} \boldsymbol{U}_{jj} & 0 \\ 
\boldsymbol{U}_{ji}^2 & 0 & 0 & \boldsymbol{U}_{jj}^2
\end{pmatrix}, \nonumber
\end{align}
since $\boldsymbol{U}_{ij} = -\boldsymbol{U}_{ji}$, where $S_k$ denotes the $k^\text{th}$ element of $\mathbb{S}$. Then 
\begin{equation}
\Delta^2 = 
\begin{pmatrix}
1 & 0 & 0 & 0
\end{pmatrix}
\begin{pmatrix}
\boldsymbol{U}_{ii}^2 & 0 & 0 & \boldsymbol{U}_{ij}^2 \\ 
0 & \boldsymbol{U}_{ii} \boldsymbol{U}_{jj} & \boldsymbol{U}_{ij} \boldsymbol{U}_{ji} & 0 \\ 
0 & \boldsymbol{U}_{ij} \boldsymbol{U}_{ji} & \boldsymbol{U}_{ii} \boldsymbol{U}_{jj} & 0 \\ 
\boldsymbol{U}_{ji}^2 & 0 & 0 & \boldsymbol{U}_{jj}^2
\end{pmatrix}^d
\begin{pmatrix}
s_i s_i \\ s_i s_j \\ s_j s_i \\ s_j s_j
\end{pmatrix} - \mu^2,
\end{equation}
where 
\begin{equation}
\mu = \tr \Big[ P_i \, \Big(\frac{1}{2} \, \mathcal{U} + \frac{1}{2} \, \mathcal{U}^\dag \Big)^d (\rho)
\Big]
= \boldsymbol{U}_{ii}^d \, s_i.
\end{equation}
If $[P_i, \, Z \otimes  Z]$, which is the case when learning Type 1 elements through modified CB, then $\boldsymbol{U}_{ii}=\boldsymbol{U}_{jj}=1$ and $\boldsymbol{U}_{ij}=\boldsymbol{U}_{ji}=0$, so $\boldsymbol{V}_{\!\mathbb{S}}=I$ and therefore $\Delta = 0$ in the noiseless limit. The same is true for standard CB on Clifford gates. Note however that this property does not arise automatically: if instead $\{ P_i, \, Z \!\otimes \! Z \} = 0$, then $\boldsymbol{U}_{ii} = \boldsymbol{U}_{jj} = \cos(\theta)$ and $\boldsymbol{U}_{ij} = -\boldsymbol{U}_{ji} = \pm \sin(\theta)$, so
\begin{equation}
\Delta^2 = \frac{1}{2}
\begin{pmatrix}
1, &  \!\!\!\! 0, & \!\!\!\! 0, & \!\!\!\! 0
\end{pmatrix} \!
\begin{pmatrix}
1+\cos(2 \theta)^d & 0 & 0 & 1-\cos(2 \theta)^d \\ 
0 & 1+\cos(2 \theta)^d & -1+\cos(2 \theta)^d & 0 \\ 
0 & -1+\cos(2 \theta)^d & 1+\cos(2 \theta)^d & 0 \\ 
1-\cos(2\theta)^d & 0 & 0 & 1+\cos(2\theta)^d
\end{pmatrix} \!\!
\begin{pmatrix}
s_i s_i \\ s_i s_j \\ s_j s_i \\ s_j s_j
\end{pmatrix} - \cos(\theta)^{2d} \, s_i^2.
\end{equation}
Since $(s_i, s_j)\rightarrow (1,0)$ in the noiseless limit, the resulting $\Delta$ is given by Eq.~\eqref{eq:Delta} of the main text. In other words, the random circuits arising in modified CB lead to expectation values that are highly spread out when measuring Paulis that anti-commute with $Z \! \otimes \! Z$. This is another reason why modified CB is generally impractical for learning Type 2 elements.

In contrast, partial-twirl benchmarking (PTB) and correlated-twirl benchmarking (CTB) both have $\Delta=0$ in the noiseless limit since, for any given depth $d$, the random circuits they use are all logically equivalent (i.e., they describe identical unitaries). More precisely, in PTB each layer $\mathcal{T}^{(k)}$ has the same form as in Eq.~\eqref{eqA:T_layer}, but $P_\ell \sim \text{unif}(\mathbb{P}_\textsc{c})$, so $\mathcal{T}^{(k)}=\mathcal{U}$ in the noiseless limit, which immediately gives $\Delta=0$ using Eq.~\eqref{eqA:Delta_sq_tr}. The situation is similar with CTB, although to analyze that scheme it is convenient to decompose $\mathcal{T} = \mathcal{T}^{(d/2)} \cdots \mathcal{T}^{(1)}$ with (even) depth $d$ into $d/2$ independent random layers 
\begin{equation}
\mathcal{T}^{(k)}(\rho) = \mathcal{P}_m \, \mathcal{U} \mathcal{P}_m \mathcal{P}_\ell \, \mathcal{U} \mathcal{P}_\ell,
\end{equation}
each of which comprises two $R_{ZZ}(\theta)$ gates, where we use the notation $\mathcal{P}_\ell(\rho) = P_\ell \rho P_\ell$. Then $P_\ell \sim \text{unif}(\mathbb{P})$ and $P_m \sim \text{unif}(\mathbb{P}_\textsc{a})$ if $P_\ell \in \mathbb{P}_\textsc{c}$, or $P_m \sim \text{unif}(\mathbb{P}_\textsc{c})$ if $P_\ell \in \mathbb{P}_\textsc{a}$.  In either case, $\mathcal{T}^{(k)}$ reduces to the identity channel in the noiseless limit, which immediately implies $\Delta=0$.

\section{Partial-Twirl Benchmarking Details}
\label{secA:partial_twirl_benchmarking}

In this section we derive Eqs.~\eqref{eq:block_powers}, \eqref{eq:params_to_PTM}, and \eqref{eq:type3_prod}--\eqref{eq:G_jj} from the main text. Consider the $2\times 2$ matrix 
\begin{equation}
B 
=
\begin{pmatrix}
\blue{\boldsymbol{G}_{ii}} & \green{\boldsymbol{G}_{ij}} \\ 
\green{\boldsymbol{G}_{ji}} & \blue{\boldsymbol{G}_{jj}}
\end{pmatrix},
\end{equation}
which forms a block on the bottom-right of the diagonal of $\boldsymbol{G}_\textsc{c}$, the PTM describing a noisy $R_{ZZ}(\theta)$ gate twirled over $\mathbb{P}_\textsc{c}$, the set of Paulis that commute with $ZZ$. The blue and green colors serve to denote Type 2 and Type 3 elements respectively. As in the main text, we will assume that $B$ has complex eigenvalues, i.e., that its eigenvalues satisfy condition \eqref{eq:eig_assumption}. To avoid any ambiguity with complex square roots being multi-valued (e.g., $+i$ and $-i$ are both square roots of $-1$), we will write these eigenvalues as 
\begin{equation}
\lambda_\pm = x \pm i y = r e^{\pm i \omega}
\end{equation}
here, where the real and imaginary parts
\begin{equation}
x = \frac{1}{2} \big(
\blue{\boldsymbol{G}_{ii}} + \blue{\boldsymbol{G}_{jj}}
\big)
\qquad \qquad 
y = 
\frac{1}{2} \sqrt{-(\blue{\boldsymbol{G}_{ii}}-\blue{\boldsymbol{G}_{jj}})^2 
-
4 \, \green{\boldsymbol{G}_{ij} \boldsymbol{G}_{ji}}}
\end{equation}
are both well-defined, and the magnitude $r$ and argument $\omega$ are defined in the usual way in Eq.~\eqref{eq:params_to_PTM} of the main text. Likewise, we will write the corresponding (unnormalized) eigenvectors of $B$ as 
\begin{equation}
\vec{v}_\pm
=
\begin{pmatrix}
\blue{\boldsymbol{G}_{ii}} - \blue{\boldsymbol{G}_{jj}}
\pm 
2 i y \\
2 \, \green{ \boldsymbol{G}_{ji} }
\end{pmatrix}.
\end{equation}
It follows that
\begin{equation}
\begin{pmatrix}
1 \\ 0
\end{pmatrix}
=
\frac{-i}{4 y}
\big(\vec{v}_+ - \vec{v}_- \big),
\end{equation}
so for any circuit depth $d$, we can express the top-left element of $B^d$ as
\begin{align}
\begin{pmatrix}
1 & 0
\end{pmatrix}
\begin{pmatrix}
\blue{\boldsymbol{G}_{ii}} & \green{\boldsymbol{G}_{ij}} \\ 
\green{\boldsymbol{G}_{ji}} & \blue{\boldsymbol{G}_{jj}}
\end{pmatrix}^d
\begin{pmatrix}
    1 \\ 0
\end{pmatrix}
&=
\frac{-i}{4y}
\begin{pmatrix}
1 \\ 0
\end{pmatrix} \!\cdot\! 
\Big( \lambda_+^d \vec{v}_+ -  \lambda_-^d \vec{v}_- \Big)
=
\frac{-i}{4y}
\Big[ 
(\blue{\boldsymbol{G}_{ii}} - \blue{\boldsymbol{G}_{jj}}) (\lambda_+^d - \lambda_-^d)
+ 2iy (\lambda_+^d + \lambda_-^d)
\Big] \nonumber \\ 
&=
r^d
\left[ \left(
\frac{\blue{\boldsymbol{G}_{ii}} - \blue{\boldsymbol{G}_{jj}}}{2 y}
\right)
\sin(\omega d) 
+ 
\cos(\omega d)
\right]. 
\end{align}
We can combine both terms in the square brackets using the identity
\begin{equation}
    a \cos(\omega d - \delta) = a \big[ \sin(\delta) \sin(\omega d)  + \cos(\delta) \cos(\omega d) \big ]
\end{equation}
by demanding that the amplitude $a$ and phase $\delta$ satisfy 
\begin{equation}
    a \sin(\delta) = \frac{\blue{\boldsymbol{G}_{ii}} - \blue{\boldsymbol{G}_{jj}}}{2 y} 
    \qquad \qquad 
    a \cos(\delta) = 1,
    \label{eq:delta_trig}
\end{equation}
which immediately gives the expressions for $a$ and $\delta$ in Eq.~\eqref{eq:params_to_PTM} of the main text. These definitions then yield
\begin{equation}
\begin{pmatrix}
1 & 0
\end{pmatrix}
\begin{pmatrix}
\blue{\boldsymbol{G}_{ii}} & \green{\boldsymbol{G}_{ij}} \\ 
\green{\boldsymbol{G}_{ji}} & \blue{\boldsymbol{G}_{jj}}
\end{pmatrix}^d
\begin{pmatrix}
    1 \\ 0
\end{pmatrix}
=
a r^d \cos (\omega d - \delta),
\end{equation}
as in Eq.~\eqref{eq:block_powers}.

We now derive Eqs.~\eqref{eq:type3_prod}--\eqref{eq:G_jj} of the main text, which give $\blue{\boldsymbol{G}_{ii}}$, $\blue{\boldsymbol{G}_{jj}}$ and $\green{\boldsymbol{G}_{ij} \boldsymbol{G}_{ji}}$ as functions of $r$, $\omega$ and $\delta$, which can be estimated experimentally. Note, from the definitions above, that
\begin{equation}
r \cos \omega = x
\qquad \qquad 
r \sin \omega = y
\qquad \qquad 
a y = \sqrt{- \green{\boldsymbol{G}_{ij} \boldsymbol{G}_{ji}} }.
\end{equation}
Combining these equations with Eq.~\eqref{eq:delta_trig} gives
\begin{equation}
\cos (\delta) = \frac{1}{a} = \frac{y}{\sqrt{-\green{\boldsymbol{G}_{ij} \boldsymbol{G}_{ji}}}}
= \frac{r \sin(\omega) }{\sqrt{-\green{\boldsymbol{G}_{ij} \boldsymbol{G}_{ji}}}},
\end{equation}
which immediately leads to Eq.~\eqref{eq:type3_prod} of the main text. Note that since $\cos(\delta)^{-2} = 1 + O(\delta^2)$ for $\delta$ near 0, $\green{\boldsymbol{G}_{ij} \boldsymbol{G}_{ji}}$ depends only weakly on the fitted value of $\delta$ when the noise is weak (meaning $\delta$ is small). Similarly, 
\begin{equation}
\tan(\delta) = \frac{\blue{\boldsymbol{G}_{ii}} - \blue{\boldsymbol{G}_{jj}}}{2 y} 
=
\frac{\blue{\boldsymbol{G}_{ii}} - \blue{\boldsymbol{G}_{jj}}}{2 r \sin(\omega)},
\end{equation}
so 
\begin{equation}
\frac{\blue{\boldsymbol{G}_{ii}} - \blue{\boldsymbol{G}_{jj}}}{2}
=
r \sin(\omega) \tan(\delta).
\end{equation}
Adding this quantity to $x$, or subtracting it from $x$, gives the expressions for $\blue{\boldsymbol{G}_{ii}}$ and $\blue{\boldsymbol{G}_{jj}}$ from Eqs.~\eqref{eq:G_ii} and \eqref{eq:G_jj} respectively. However, since $\tan(\delta) = O(\delta)$, these expressions depend strongly on the fitted values of $\delta$, so we do not advise using them. Instead, we recommend correlated-twirl benchmarking.

\end{document}